\documentclass[aps,pre,preprint]{revtex4-1}
\usepackage{blindtext}
\usepackage{graphicx,epsfig,fullpage}
\usepackage{epstopdf}
\usepackage{subfigure}
\usepackage{wrapfig,color,enumitem}
\usepackage{amsmath,amssymb,latexsym,float}
\usepackage{mathrsfs,varioref,wrapfig,threeparttable}
\usepackage{dcolumn}   
 \newcolumntype{d}{D{.}{.}{-1}}
\usepackage{textcomp}
\usepackage{lineno}
\usepackage{epsfig}
\usepackage{float}
\usepackage{tikz}
\usepackage{indentfirst}
\usepackage{hyperref}
\usepackage{setspace}
\begin{document}
\title{Kinetic modeling of three-dimensional electrostatic-solitary and surface waves in beam neutralization}
\author{Nakul Nuwal$^\#$}
\email{nuwal2@illinois.edu}
\author{Deborah A. Levin$^\#$}
\author{Igor D. Kaganovich$^\dagger$}
\affiliation{$^\#$University of Illinois Urbana-Champaign, IL, USA, $^\dagger$Princeton Plasma Physics Lab, NJ, USA}


\begin{abstract}

 This work studies the fundamental plasma processes involved in the neutralization of an ion beam’s  space-charge by electrons emitted by a filament using Particle-in-Cell simulations. While filament neutralization is economical, previous experiments have shown that a variety of waves become excited in this process that limit the space-charge neutralization. In this work, the formation and movement of electrostatic solitary waves (ESWs), which have low dissipation rates, are characterized for 2D planar and 3D cylindrical beams and are observed to generate waves that survive for a long time and slow the process of beam neutralization. Further, through a 1D Bernstein-Greene-Kruskal (BGK) analysis, we find that the non-Maxwellian nature of the beam electrons gives rise to large-sized ESWs that are not  predicted by theory which assumes that the electrons may be described by a Maxwellian distribution.  Our PIC simulations are sufficiently sensitive to be able to resolve important three-dimensional effects in a 3D cylindrical geometry that lead to the excitation of Trivelpiece-Gould surface waves due to high energy electrons present at the beginning of neutralization.
\end{abstract}
\maketitle
\setlength\parindent{24pt}
\section{Introduction}
Ion beams are used in various engineering applications such as particle accelerators, ion-thrusters\cite{korkut2015,jambunathan2018chaos,nuwal2020kinetic}, nanopantography\cite{chen2019nearly}, and ion-implantation\cite{conrad1989plasma}. The electrons are introduced to neutralize the beam and reduce its divergence along the axis. The source of electrons can be from a background plasma or through an external source such as filaments or cathodes. Although better neutralization is achieved using the background cold electron plasma\cite{vay2001intense,vay1998charge,kaganovich2001nonlinear,berdanier2015intense}, this method is expensive as it requires the production of plasma in the vicinity of the ion beam. Filaments, on the other hand, are easier to install and control, however, experiments have found multiple challenges in their use in the ion beam neutralization. For example, the rate of neutralization is affected by the energy and location of the electron source relative to the ion beam\cite{lan2020neutralization,kaganovich2009physics}. Additionally, in recent numerical works by Lan and Kaganovich\cite{lan2020neutralization,lan2020neutralization2}, electrostatic-solitary-waves (ESWs), also known as electron solitons, were observed to form when electrons were injected into a 2D ion beam via a filament source. This is problematic since a solitary wave is a non-linear wave structure which sustains itself for a long time with low dissipation\cite{bernstein1957exact}, thus delaying neutralization. 

An ESW forms when enough electrons are trapped in a local electric potential-well. ESWs can be multidimensional structures, and their movement along the beam depends on the electric potential profile\cite{lan2020neutralization2} and electron velocity distribution of the beam. ESWs were first described in one-dimension by Bernstein, Green, and Kruskal\cite{bernstein1957exact} where a spatially symmetric electric potential structure was shown to be stable and stationary when it trapped a finite fraction of electrons. A number of theoretical solutions of ESW or electron holes have been found by other authors, such as Sagdeev, and Schamel\cite{schamel1979theory}, where the Vlasov-Poisson equation system was solved for the electric potential and electron velocity phase distribution, assuming a stationary solitary wave. Schamel\cite{schamel1979theory} showed an analytical solution of a finite (small) amplitude slow moving electron hole by solving the Vlasov-Poisson system of equations using the Classical-potential method\cite{schamel1979theory,bernstein1957exact,hutchinson2017electron}. Since their theoretical prediction, these waves have been observed in experiments\cite{saeki1979formation} and in the geomagnetic tail\cite{matsumoto1994electrostatic}, which is comprised of charged particles. While such theoretical analysis provided insights into the properties of individual solitary waves in a Maxwellian far-field electron distribution, we will show how they appear with a different shape and size in a neutralizing beam where the electron distribution is non-Maxwellian. Further, we will show how the ESWs in a 3D cylindrical beam are different from a 2D planar beam configuration and demonstrate why the 1D BGK analysis is inadequate due to more complex electron trajectories in an unmagnetized plasma beam. 
   
\begin{figure}
\centering
        \includegraphics[trim = 0.14cm 0.0cm 0.0cm 0.0cm, clip,width = 0.65\textwidth]{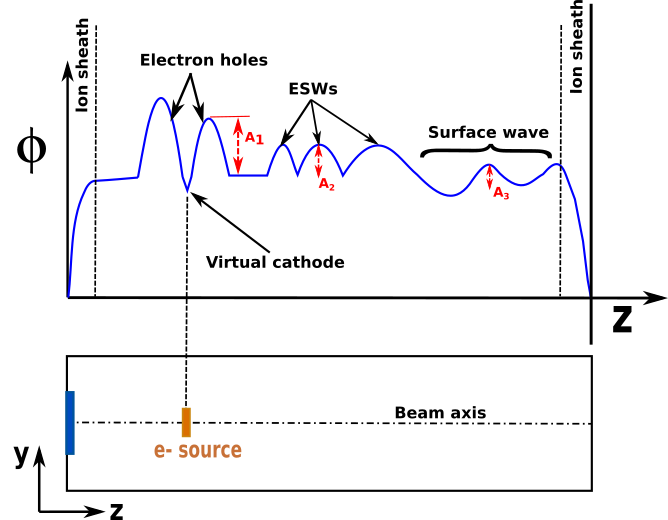}
\caption{Schematic of physical features that appear along the beam axis during beam neutralization, where $A_1, A_2$, and $A_3$ are the amplitudes of the electron holes that form near the electron source, solitary waves, and surface wave, respectively, and $A_1>A_2>A_3$.  }
\label{fig:Phys_schem}
\end{figure}

ESWs affect the beam neutralization because they create local pockets of positive potential, as shown in the Fig. \ref{fig:Phys_schem} schematic, thereby preventing the beam from achieving a perfect neutralization\cite{lan2020neutralization2}. Figure \ref{fig:Phys_schem} shows a schematic of the electric potential profile along the axis of a typical beam in which an electron source is placed along its axis. As will be shown later, the continuous emission of electrons from the electron source results in the formation of a virtual cathode and two electrons holes on its either sides, which can be seen in the electron potential-wells labeled by amplitude $A_1$ in Fig. \ref{fig:Phys_schem}. Further downstream of the electron source, ESWs appear in the beam, shown by the waves with the amplitude $A_2$ in Fig. \ref{fig:Phys_schem}, and they move along the beam with speeds ranging from zero to 0.92$v_{te}$\cite{hutchinson2017electron}, where $v_{te}$ is the electron thermal speed. \textcolor{black}{The difference between an electron hole and an ESW is that an electron hole does not move while an ESW does.} As will be shown later, many of these wave features depend on the shape and geometry of the beam.  Since for a number of applications\cite{conrad1989plasma} a beam is cylindrical in shape, performing a numerical study for a 3D beam helps us to understand the effects of geometry on solitary and other plasma waves. 
{In this work, we aim to understand the beam neutralization in 2D planar and 3D cylindrical beams and the formation of ESWs, and their movement along the beam axis using the Particle-in-Cell (PIC) method.} 

\textcolor{black}{Depending on the geometry of the plasma, \textcolor{black}{small disturbances at the plasma-vacuum interface} can result in the excitation of surface waves, as was theorized by Trivelpiece and Gould\cite{trivelpiece1959space}. These waves were studied because of their ability to interact with the electromagnetic waves of certain frequencies that cause interference with the passive measurement methods used in plasma beam diagnostics\cite{krall1973principles}. The amplitude of these waves, designated as $A_3$ in Fig. \ref{fig:Phys_schem}, are lower than ESWs. These waves are especially likely to appear in systems such as ours, where the electrons neutralize a high-electric-potential beam and the average electron temperature of the beam declines with time. Such surface waves, if excited, will not be damped by high energy electrons. }{As will be shown later, we found Trivelpiece-Gould\cite{trivelpiece1959space} surface waves in our 3D plasma beam simulations. However, such waves did not appear in the 2D planar beam case} \textcolor{black}{because a higher-phase speed is required to excite them in a planar beam geometry, as will be discussed further in Sec. \ref{sec:surface_waves}.}

In this paper, we first discuss the numerical approach of the PIC calculations in Sec. \ref{sec:Num_app}. Then, in Sec. \ref{sec:soliton_excitation}, we describe the excitation, movement, and structure of the ESWs that form along the beam. In Sec. \ref{sec:surface_waves}, we discuss the excitation and propagation of surface waves in the beam, where we show a comparison of our PIC results with the theoretical disperson relations for 2D and 3D beams. Finally, in Sec. \ref{sec:conclusions}, we list the key findings and conclusions of our numerical study. 


\begin{figure}
\centering
\subfigure[2D planar beam]{\label{fig:Dom_schem_2D}\includegraphics[trim = 0.0cm 0.0cm 0.0cm 0.0cm, clip,width = 0.40\textwidth]{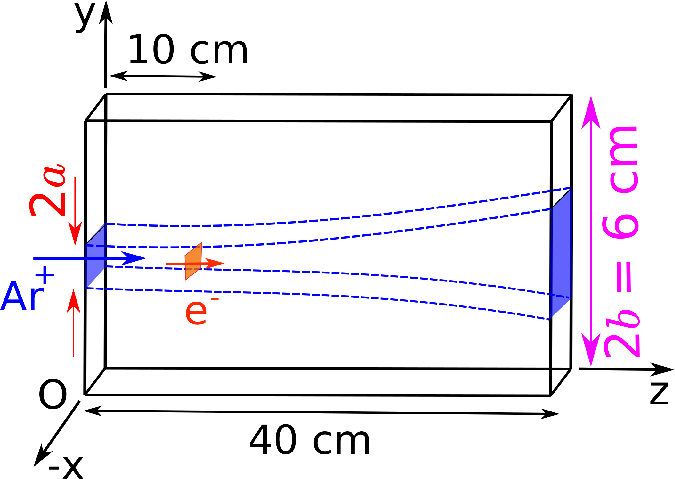}}
       \subfigure[3D cylindrical beam]{\label{fig:Dom_schem_3D}\includegraphics[trim = 0.0cm 0.0cm 0.0cm 0.0cm, clip,width = 0.40\textwidth]{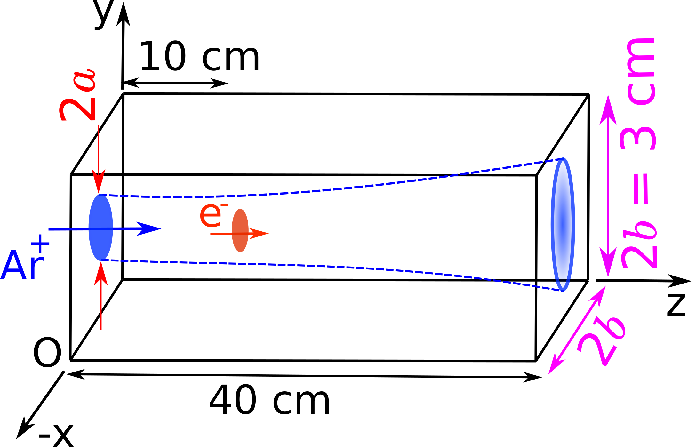}}
\caption{Schematic (not drawn to scale) of the (a) 2D and (b) 3D numerical domains. Here, the channel width is $2b$ and the ion-source width or diameter for 2D planar and 3D cylindrical beams is $2a$. }
\label{fig:Dom_schem}
\end{figure}

\begin{table*}
		\begin{center}
		\caption{Numerical case parameters for 2D and 3D beams$^{1-7}$}
		\label{tab:Numparam_2D}
	
		\begin{tabular}{l | c | c c c c c c c c c c c c}
            \hline \hline
            \multicolumn{1}{c} {Numerical parameter}  & \multicolumn{1}{c}{2D planar} & \multicolumn{3}{c}{3D cylindrical} \\
              & $a = 2.5$ mm &  $a = 2.5$ mm & $5.0$ mm & $7.5$ mm \\  
            \hline
      $\Delta x$ (mm)  & 0.46 & 0.46 & 0.46 & 0.46\\    
      Number of GPUs (Nvidia Tesla-K20)  & 128 & 128 & 128 & 128\\ 
      Particles-per-cell & 400-650 & 450-800 & 250-500 & 200-420\\
      $F_{\rm num}$ & 40 & 30 & 60 & 80 \\
      Particles per species (Millions) & 33 & 46 & 92 & 154\\  
      No. of PIC grid cells (Millions) & 0.9 & 3.7 & 3.7 & 3.7 \\
   \hline \hline
    \end{tabular}
\begin{tablenotes}
\item $^1$ $\omega_{pe} = 7.46\times10^8$ rad/s,   $n_{\rm 0}$ = $1.75\times 10^{14} \ {\rm m^{-3}}$, and $\Delta t = 80 \times10^{-12}$ s.
\item $^2$ $n_{\rm Ar^+}$ = $1.75\times 10^{14} \ {\rm m^{-3}}$ at the ion source; electron current, $I_{e-} = I_{\rm Ar^+}/3$, $I_{\rm Ar^+}/3$ being the ion current.
\item $^3$ Half-width ion source in 2D planar beam case, $a = 2.5$ mm.
\item $^4$ Radius of ion source, $r_{\rm Ar^+} \equiv a = 2.5$, $5.0$, and $7.5$ mm for different 3D cylindrical cases.
\item $^5$ $b = 30$ mm for the 2D planar and $b = 15$ mm for all 3D cylindrical cases.
\item $^5$ $T_e = 2$ eV at the electron source for all cases.  
\item $^6$ All ions are monoenergetic and introduced with velocity, $v_z = 427$ km/s and no thermal velocity.
\item $^7$ Debye length, $\lambda_{\rm D_0} = 0.78$ mm, at plasma density $1.75\times 10^{14} \ {\rm m^{-3}}$ and $T_e = 2$ eV. 

\end{tablenotes}		
		\end{center}
\end{table*}

\section{Numerical approach and general plasma conditions}\label{sec:Num_app}
The numerical study in this paper uses our in-house time-explicit CPU-GPU based electrostatic Particle-in-Cell (PIC) solver CHAOS (Cuda-based Hybrid Approach for Octree Simulations)\cite{Jambunathan_CnF}. We solve Poisson's equation to compute the electric potential scalar and electric fields to move charged particles in the domain. To satisfy the numerical validity of the time-explicit nature of the coupling between Poisson's equation and particle movement, we use a time-step, $\Delta t < 0.1 \omega_{pe}^{-1}$ and grid cell-size, $\Delta x < \lambda_{\rm D}$\cite{godunov1959difference}, where $\omega_{pe}$ and $\lambda_{\rm D}$ are the local electron plasma frequency and Debye length, respectively. The code CHAOS has previously been used in plume-plasma\cite{jambunathan2018chaos,nuwal2020kinetic}, and plasma-surface interaction\cite{nuwal2021influence} studies, and provides the multi-GPU scalability necessary for this study.

In this work, we performed calculations of 2D planar and 3D cylindrical beams in a metal cavity. For the 2D planar beam calculations,  we used a numerical domain of $0.375\times6\times40$ cm, where $2b =6$ cm is the channel width, as shown in Fig. \ref{fig:Dom_schem_2D}. The size of our ion source and the domain length is of the same order as other plasma wave experiments\cite{lynov1979observations,ikezi1971electron,stepanov2016dynamics} and nanopantography\cite{chen2019nearly} applications. Since CHAOS is a 3D code, the third dimension of the domain was suppressed for the 2D case by not moving particles in the $x$ direction, applying a zero potential gradient boundary along the $x$ direction ($\partial \phi / \partial x = 0$), and setting the $x$ direction thickness to be $0.375$ cm, which is much thinner than the other two dimensions of the simulation domain. All other parameters are given in Tab. \ref{tab:Numparam_2D}. We also performed three cases for 3D cylindrical beams, each with a different ion beam radius. For all 3D cylindrical beam cases, a domain of size $1.5\times1.5\times40$ cm, shown in Fig. \ref{fig:Dom_schem_3D}, was used with a uniform grid of cell size, $\Delta x = 0.46$ mm. \textcolor{black}{The domain width was reduced from $2b = $ 6 cm in the 2D planar beam to $2b = $ 3 cm in the 3D cylindrical beam case to reduce the computational expense while keeping the same domain length.}  All walls of the 3D domain were considered charge-absorbing with $\phi = 0$~V, similar to other numerical\cite{lan2020neutralization,lan2020neutralization2} and experimental work\cite{stepanov2016dynamics,lynov1979observations}, where Poisson's equation was solved with Dirichlet boundary conditions at all domain walls. The numerical parameters for all 3D cases are also listed in Tab. \ref{tab:Numparam_2D}.


In both 2D and 3D cases, ions were introduced in the domain with a constant axial velocity of 427 km/s\cite{lan2020neutralization,stepanov2016dynamics} and zero initial velocity in the transverse direction from an ion source of width (diameter for 3D cases) $2a = 5$ mm at $z = 0$ mm, as shown in the Fig. \ref{fig:Dom_schem} schematic. As discussed before and also shown in Lan et al.\cite{lan2019neutralization}, the beam diverges along the z-axis because of self-repulsive space-charge among ions. Electron injection starts when the ions reach the end of domain, $z = 40$ cm, at $t = 1 \ \mu s$.  Electrons in both 2D planar and 3D cylindrical beams were injected from the axial location of $z =10 $ cm with a Maxwellian velocity distribution function of electron temperature, $T_e = 2$ eV, in all three directions. Similar to Lan et al.\cite{lan2020neutralization}, the injected electron current, $I_{e^-}$, was one third of the ion current $I_{\rm Ar^+}$ and the electrons were introduced from a rectangular source of width 2 mm for the 2D planar beam. For all the cases of a 3D cylindrical beam, the electrons were introduced from a circular source of radius $r_{e^-}$, such that the ratio of ion and electron source radii, $a / r_{e^-} = 2.5$, which is consistent with the ion to electron source width ratio for the 2D planar beam case. Unlike the 2D case, the charged particles were allowed to move in all three directions for the 3D beam cases.  The number of particles-per-cell ranges from 200 to 800 for all cases, and a convergence study of the 3D case with $a = 5.0$ mm is presented in Appendix~\ref{sec:Num_conv}.

\begin{figure}
\centering
\subfigure[Electric potential at $z = 20$ cm comparison with Lan et al.\cite{lan2019neutralization}.]{\label{fig:Probe_z20cm}
        \includegraphics[trim = 0.14cm 0.14cm 0.14cm 0.14cm, clip,width = 0.48\textwidth]{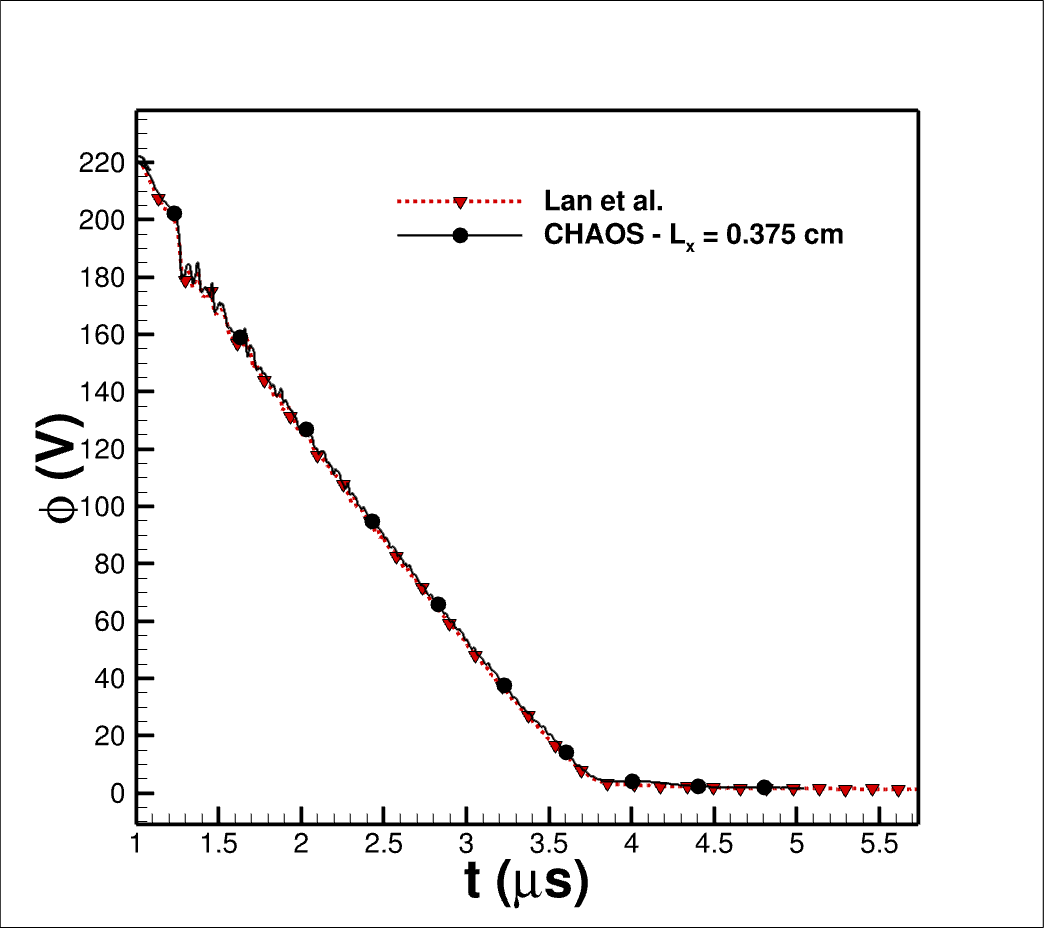}}
        \hspace{0.3mm}    
 \subfigure[Electron and ion density profiles at $z = 20$ cm compared at $t = 2 $ and $4.3 \ \mu s$ ]{\label{fig:nelec_profile_comp}
        \includegraphics[trim = 0.14cm 0.14cm 0.14cm 0.14cm, clip,width = 0.48\textwidth]{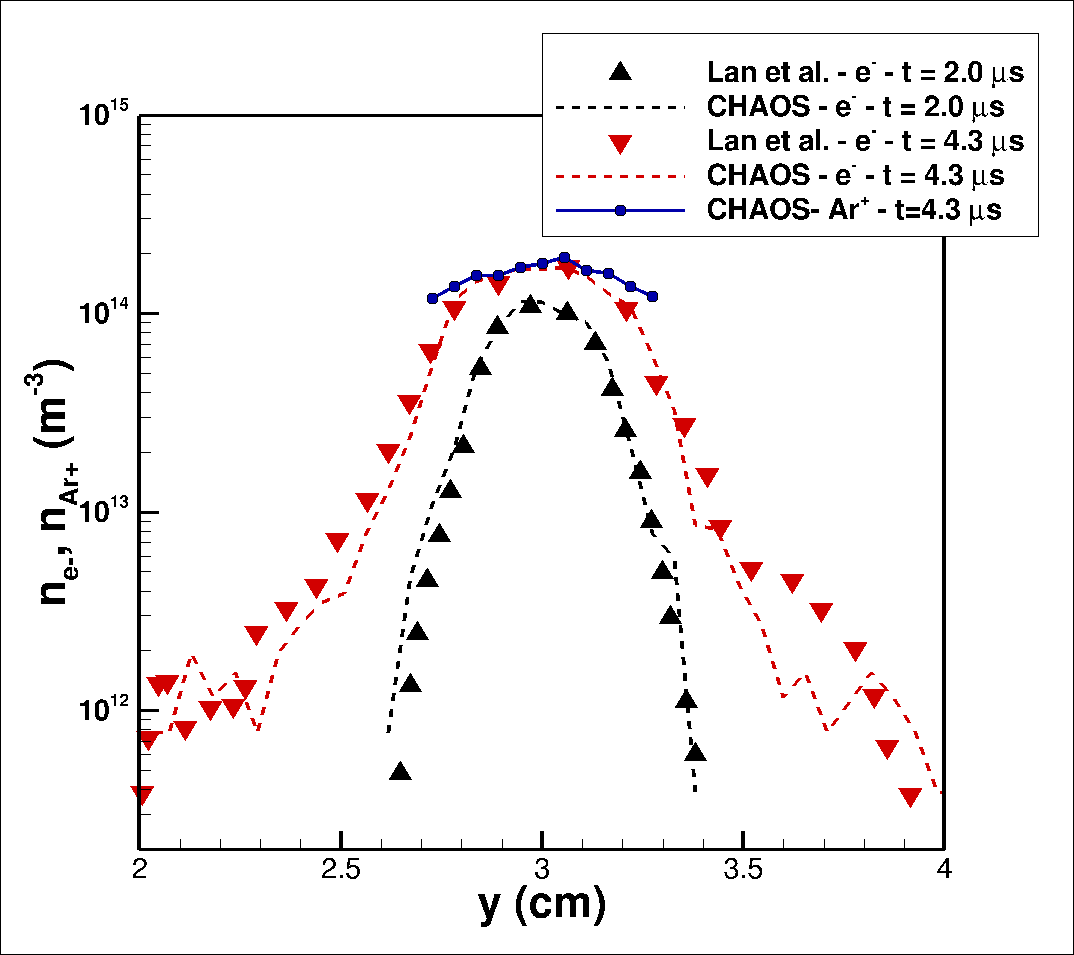}}        
 \subfigure[Evolution of electric potential profile along the beam axis with time. ]{\label{fig:phi_profile_2D_sheath}
        \includegraphics[trim = 0.14cm 0.14cm 0.14cm 0.14cm, clip,width = 0.8\textwidth]{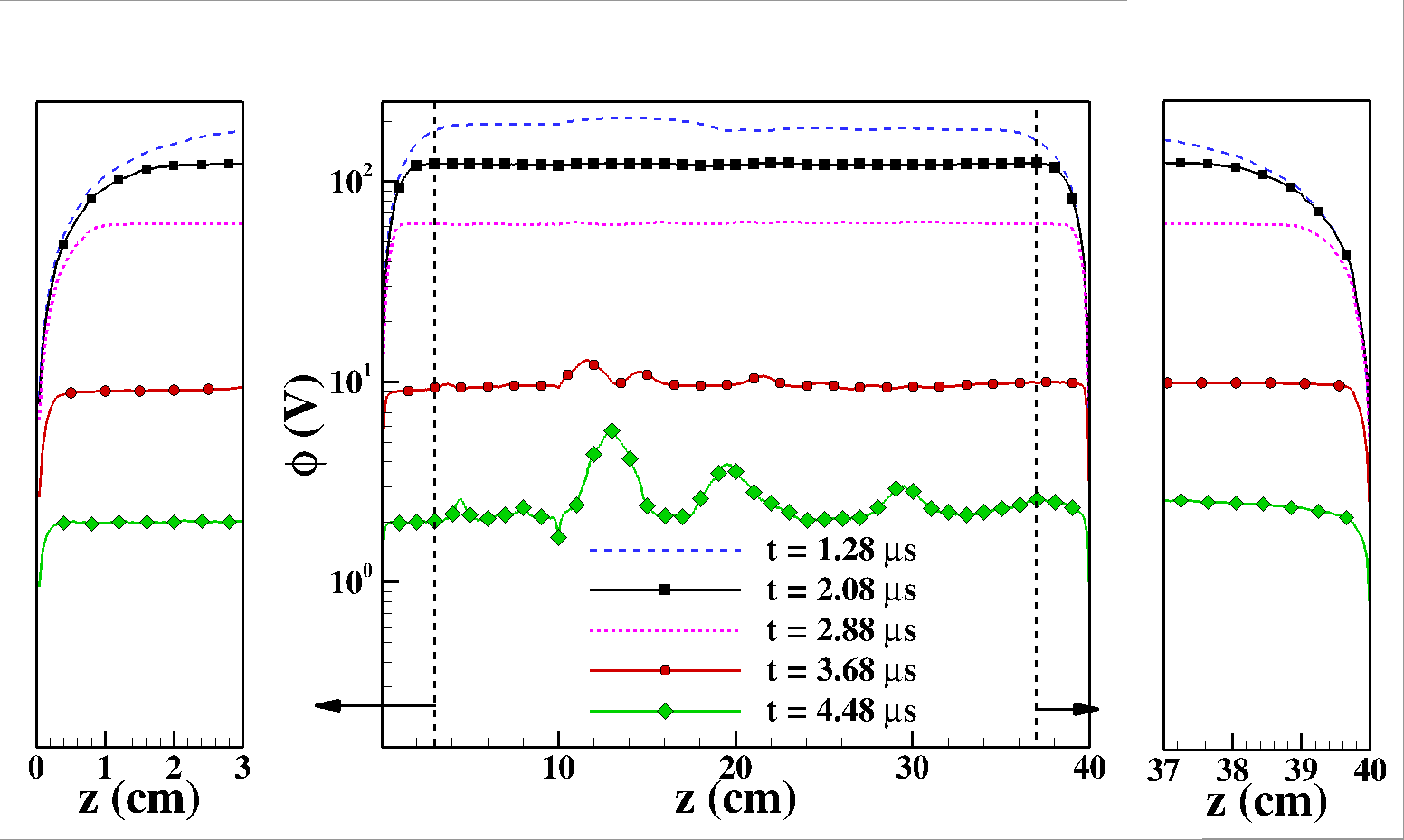}}                
\caption{Time evolution of (a) probe electric potential, (b) electron density profiles, and (c) axial electric potential profile for a 2D planar beam.  }
\label{fig:lan_comp}
\end{figure}

 In general, regardless of the specific geometry, there are two phases of beam neutralization\cite{lan2020neutralization}: (1) when the rate of neutralization of the beam, $\partial \phi/\partial t$, is high, and (2) when the beam has reached a quasi-steady state value and $\partial \phi/\partial t$ is very small. In all our cases, the first phase lasts from $t = 1$ to $3.5 \ \mu s$ and the second phase starts at $t = 4.0 \ \mu s$, as seen in Fig. \ref{fig:Probe_z20cm}. In a neutralizing beam, ESWs form in the first phase and then dissipate very slowly in the second phase\cite{lan2020neutralization2} of neutralization. In this work, we focus only on the first phase of neutralization. \textcolor{black}{As more electrons are introduced, the beam electric potential decreases, and the electron density profile in the beam transverse direction widens and assumes a shape that is similar to the ion density profile, as shown by its evolution in Fig. \ref{fig:nelec_profile_comp}. The electron density profiles at different times, shown in Fig. \ref{fig:nelec_profile_comp}, also compare well with the results of Lan et al.\cite{lan2020neutralization} with less than 1\% difference. Because of the conducting walls biased at $\phi = 0$ at $z_{\rm min}$ and $z_{\rm max}$, ion plasma sheaths form on both these boundaries, as shown in Fig. \ref{fig:phi_profile_2D_sheath}. These plasma sheaths reflect low energy electrons and result in the formation of two electron streams in the beam, which create plasma waves, such as ESWs\cite{omura1996electron}.}  

 
\begin{figure}
\centering
\subfigure[$t = 1.2 \ \mu s$.]{\label{fig:2D_3D_t=10000}
        \includegraphics[trim = 0.14cm 0.14cm 0.14cm 0.14cm, clip,width = 0.48\textwidth]{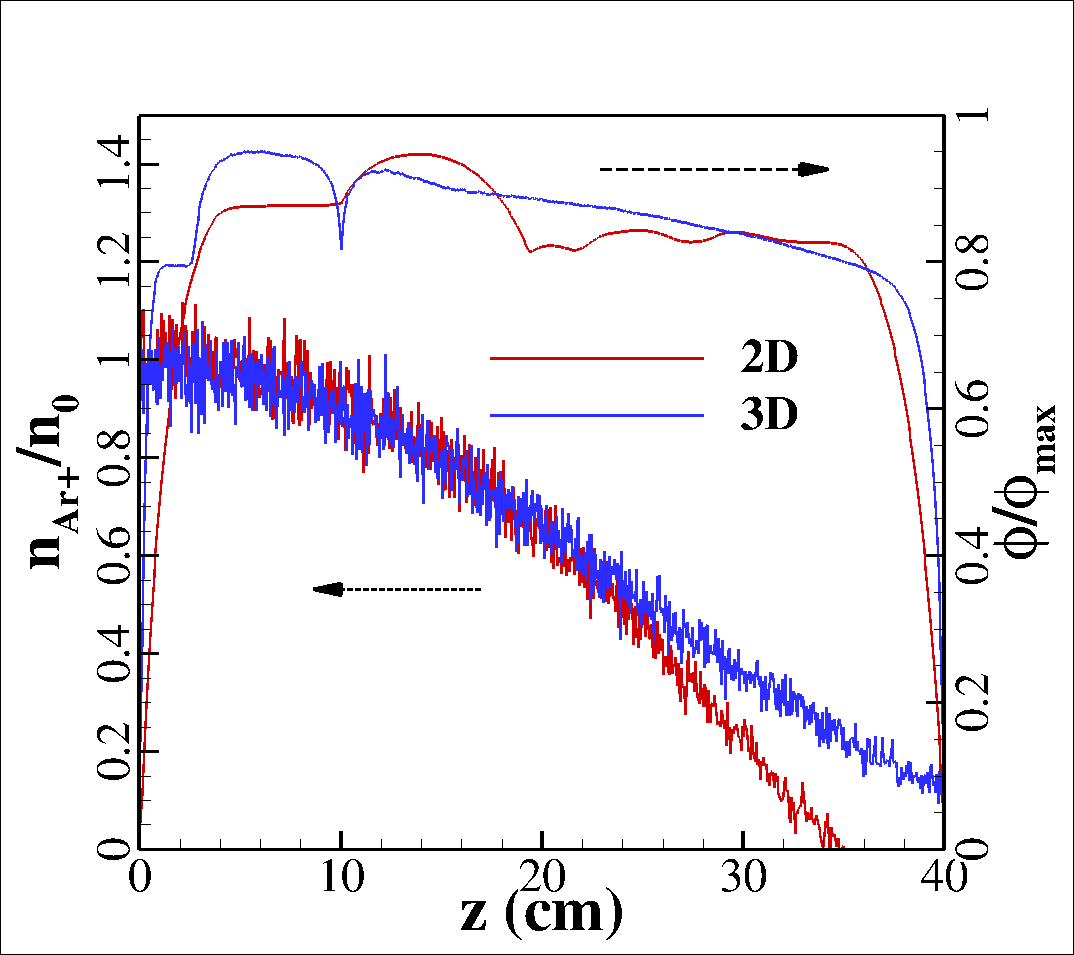}}
\hspace{0.3mm}        
  \subfigure[$t = 4.0 \ \mu s$.]{\label{fig:2D_3D_t=50000}
        \includegraphics[trim = 0.14cm 0.14cm 0.14cm 0.14cm, clip,width = 0.48\textwidth]{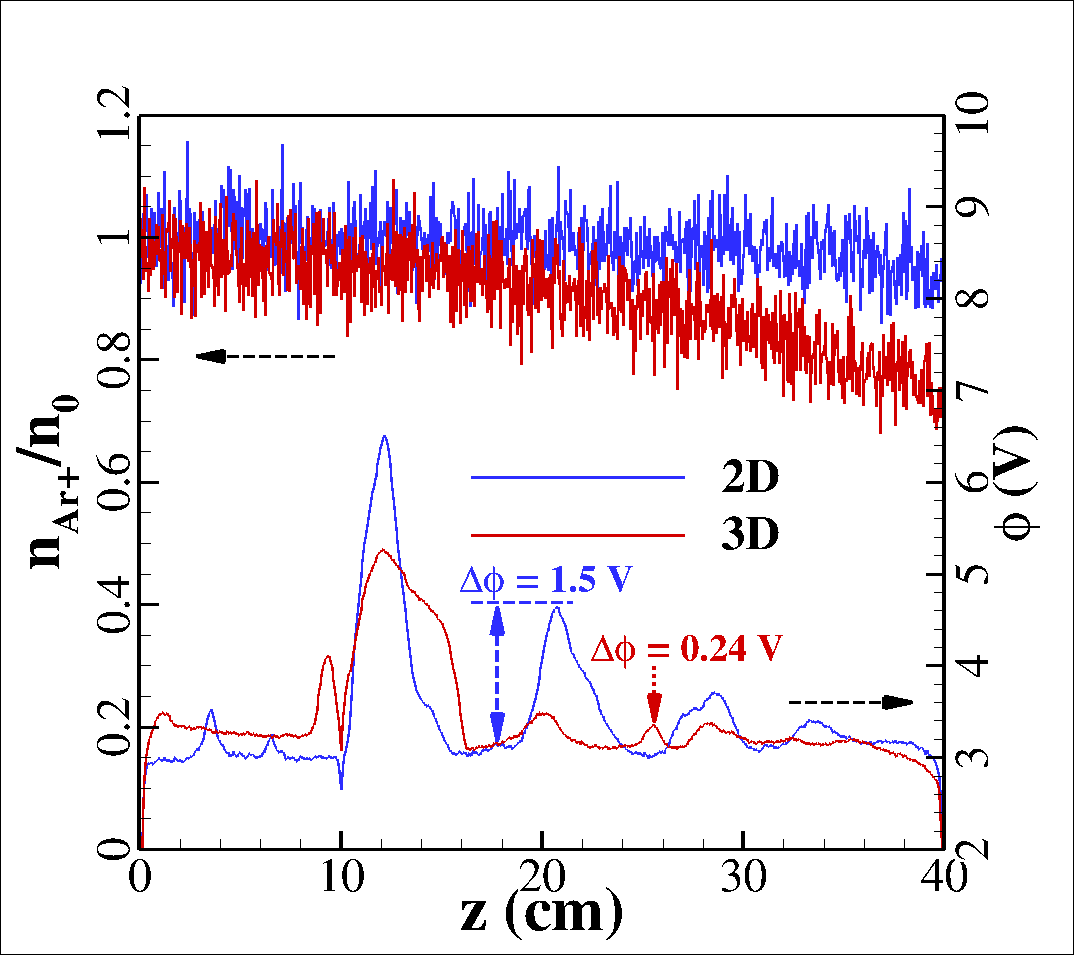}}
\caption{ Comparison of time evolution of plasma parameters along the beam axis for 2D and 3D cases. Here, $n_0 = 1.75\times10^{14} \ {\rm m^{-3}}$ and $\phi_{\rm max} = 220$ and 14 V for 2D and 3D cases, respectively. In (b),  $\Delta \phi$ is the amplitude of the ESW, i.e. the difference between the peak and base potential. }
\label{fig:2D-3D}
\end{figure}

Since we will be discussing differences in ESWs and surface wave formations for 2D vs 3D geometries, we provide a comparison of the evolution of their electric potential and ion density profiles with time in the first phase of neutralization.
Figure \ref{fig:2D_3D_t=10000} shows that the normalized electric potential and ion density profiles have a similar shape for both 2D planar and 3D cylindrical beams at $t = 1.2 \ \mu s$, right after the electron injection starts at $t = 1.0 \ \mu s$.  \textcolor{black}{The electric potential in Fig. \ref{fig:2D_3D_t=10000} is normalized by $\phi_{\rm max} = 220$ and 14 V for 2D planar and 3D cylindrical beams, respectively. Because of higher electric potential in the 2D planar case, the ion beam is initially more expanded along the beam axis compared to the 3D cylindrical case at $t = 1.2 \ \mu  s$, as shown in Fig. \ref{fig:2D_3D_t=10000}. However, with time the ion beam becomes more coloumnated in the 2D planar beam case, compared to the 3D cylindrical beam case, as shown in Fig. \ref{fig:2D_3D_t=50000}, where the ion density value drops by factors of about 1.1 and 2.82 from $z = 0$ cm to $z = 30$ cm in the 2D planar and 3D cylindrical beam cases, respectively at $t = 4.0 \ \mu s$. This comparison shows that the beam-neutralization process is slower in the 3D cylindrical beam case than the 2D planar beam case because of the lower beam potential of the initially un-neutralized beam. 
The electric potential `bumps' in Fig. \ref{fig:2D_3D_t=50000} provide clues about the presence of electrostatic solitary waves (ESWs) which are excited in both 2D and 3D cases with $a = 2.5$ mm. In Fig. \ref{fig:2D_3D_t=50000}, the amplitude of the ESWs near the end of the first phase of neutralization is higher for the 2D than 3D case, likely because of the higher electric potential of the initially un-neutralized 2D planar beam, which enables it to trap more electrons than the 3D beam.   } 

\section{Excitation of ESWs in beam neutralization}\label{sec:soliton_excitation}

\begin{figure}
\centering
\subfigure[Electron phase distribution, $f_{e-}$, is computed by normalizing number density with $n_0 = 1.75\times10^{14} \ {\rm m^{-3}}$, and the electric potential profile at $t = 3.12 \ {\rm \mu s}$. The black line shows electric potential on the right vertical axis. The red dots show the position of the ESW along the dashed black line in (b).]{\label{fig:2D_soliton_schematic_t312}
        \includegraphics[trim = 0.14cm 0.14cm 0.14cm 0.14cm, clip,width = 0.65\textwidth]{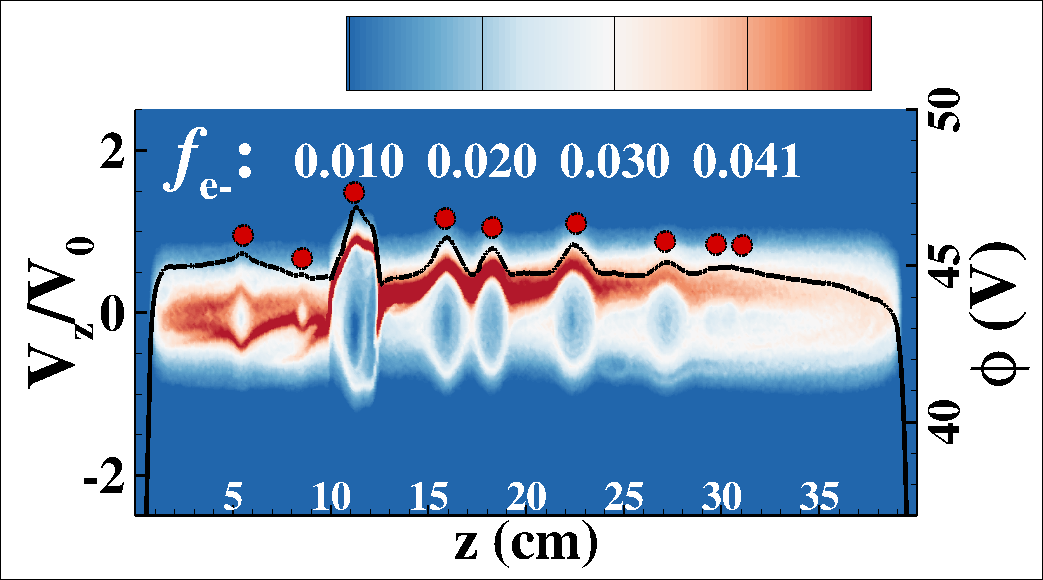}}
  \subfigure[Movement of ESWs along the beam axis with time overlayed on $\phi (z)$ vs $t$.]{\label{fig:2D_soliton_movement_timeevol}
        \includegraphics[trim = 0.14cm 0.14cm 0.14cm 0.14cm, clip,width = 0.7\textwidth]{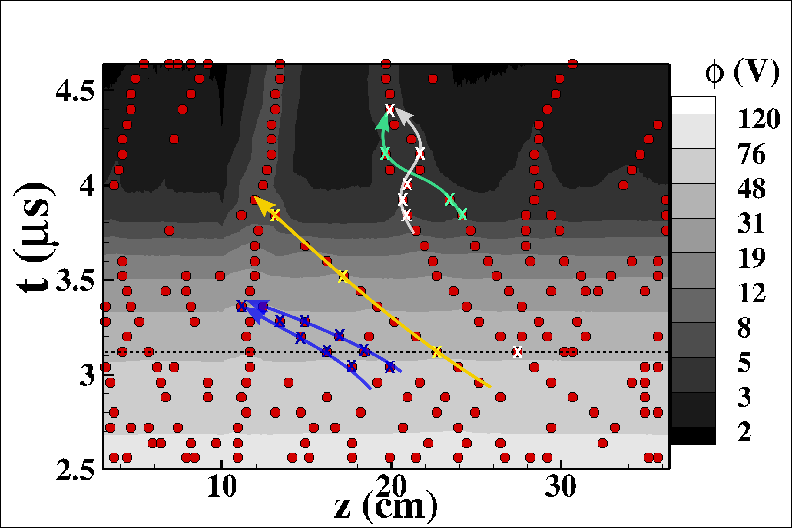}}
\caption{ Time evolution of ESW positions along the beam axis, $(x,y) = (3.0,3.0)$ cm, for the 2D planar beam case, $a = 2.5$ mm and $b = 30$ mm. In (b), the ESW shown by blue \textbf{`X'}s merge with the large ESW near the electron source at $z = 10$ cm, and ESWs shown by white and green \textbf{`X'}s merge with each other. The ESW shown by yellow \textbf{`X'} moves with a constant speed from about $t = 2.75$ to $3.75 \ \mu s$.  }
\label{fig:2D_soliton_movement_timeevol_all}
\end{figure}

During the neutralization of a beam, a fraction of electrons are trapped in regions of the beam which results in the formation of non-linear waves called ESWs, electron holes, or solitons\cite{hutchinson2017electron,bernstein1957exact}, as mentioned in the previous section.  
Figure \ref{fig:2D_soliton_schematic_t312} shows a time-snapshot of the axial velocity phase space for electrons at $t = 3.12 \ \mu s$ where multiple regions of trapped electrons, i.e. ESWs, appear along the beam. These ESWs always appear at the local maxima in the electric potential, as shown in Fig. \ref{fig:2D_soliton_schematic_t312}, and their spatial location can be estimated by numerically differentiating the electric potential profile (see Appendix \ref{sec:Num_conv}). The $z$ locations of these local maxima are shown by the red bubbles in Fig. \ref{fig:2D_soliton_schematic_t312} for $t = 3.12 \ \mu s$. The location of these ESWs extracted over multiple discrete time-steps is shown in Fig. \ref{fig:2D_soliton_movement_timeevol} where their positions, shown by the red bubbles, is overlayed on the electric potential variation in time along the beam axis. 

\subsection{Movement of ESWs along beam axis}\label{sec:move_ESWs}
Figure \ref{fig:2D_soliton_movement_timeevol} shows the movement, collisions, and merging of various ESWs with time and, as will be shown later, can be used to estimate the speed of an ESW to determine its properties, such as the trapped electron particle population and their energies. Figure \ref{fig:2D_soliton_movement_timeevol} also shows a gradual decrease in electron potential in the beam from about 120 V at $t = 2.5 \ \mu s$ to about 5 V at $t = 4.5 \ \mu s$. \textcolor{black}{In Fig. \ref{fig:2D_soliton_movement_timeevol}, we can see \textit{finger-like} structures in the $\phi$ contours on the  $z$ $vs$ $t$ plot for $t>4 \  \mu s$ which coincide with the red dots. These structures represent the slow moving ESWs that survive in the beam at the end of the first phase of beam neutralization. These structures then take a long time to dissipate in the second phase of the beam neutralization, as was shown in Lan et al.\cite{lan2020neutralization2}. } The continuous nature of these points shows that ESWs travel along the beam with velocities ranging from nearly zero to $20 \ \rm cm / \mu s$ with one of the fastest ESWs shown by blue `X's in Fig. \ref{fig:2D_soliton_movement_timeevol}. \textcolor{black}{These speeds of ESWs are in the same range of zero to $0.92\sqrt{(2k_bT_e)/m_e}$ shown in Hutchinson\cite{hutchinson2017electron}, where $k_b$ and $m_e$ are Boltzmann constant and the mass of electron, respectively.} Since the slope of the curves passing through these ESW locations gives their velocities, the lines that are nearly parallel to the  $t$ axis show ESWs that are stationary, whereas the lines that are `flatter' represent ESWs moving with very high speeds in Fig. \ref{fig:2D_soliton_movement_timeevol}.   

\begin{figure}
\centering
  \subfigure[Snapshots of ESWs for $3.04<t<3.36 \ {\rm \mu s}$.]{\label{fig:time_snaps_2D_merge}
        \includegraphics[trim = 9cm 0.14cm 10cm 0.14cm, clip,width = 0.47\textwidth]{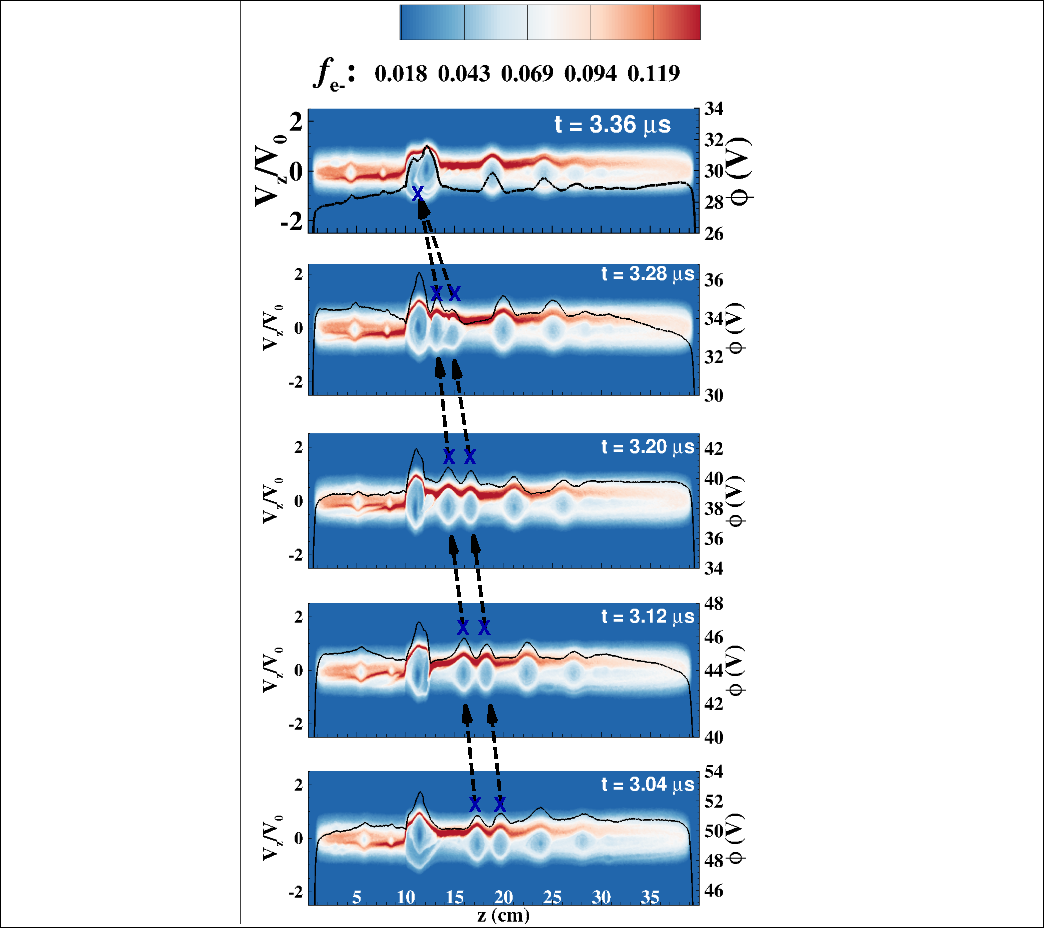}}
        \hspace{0.1mm}
\subfigure[Snapshots of ESWs for $3.84<t<4.40 \ {\rm \mu s}$.]{\label{fig:time_snaps_2D_coll}
        \includegraphics[trim = 9cm 0.1cm 10cm 0.14cm, clip,width = 0.485\textwidth]{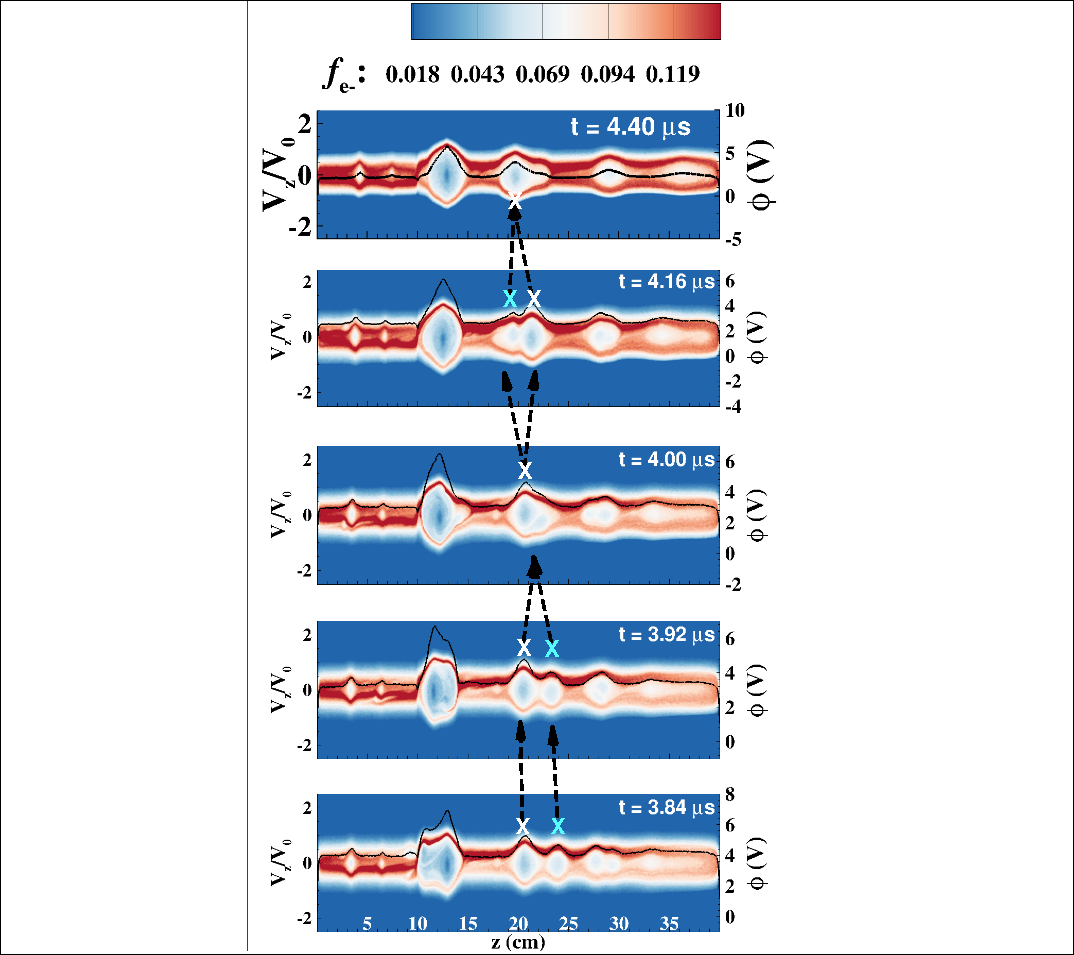}}
\caption{ Movement of ESWs in 2D planar beam case, represented by the \textbf{X} marks in Fig. \ref{fig:2D_soliton_movement_timeevol}. In (a), two blue ESWs move in the $-z$ direction and merge with the attached electron hole located at $z \approx 12$ cm. In (b), the two ESWs, shown by white and green \textbf{X} marks, merge into a single ESW. Here, the electron distribution $f_{e-}$ is computed by normalizing number density with $n_0 = 1.75\times10^{14} \ {\rm m^{-3}}$.   }
\label{fig:2D_soliton_movement_timesnaps}
\end{figure}

Similar to Lan and Kaganovich\cite{lan2020neutralization2}, these ESWs merge with the other ESWs, as shown by the merging trajectories of a number of ESWs in Fig. \ref{fig:2D_soliton_movement_timeevol}. An example of this merging can be seen by following two ESWs marked in blue \textbf{`X'} in Figs. \ref{fig:2D_soliton_movement_timeevol} and \ref{fig:time_snaps_2D_merge} from $t = 3.04$ to $3.96 \ \mu s$, in which both the ESWs move towards the electron source at $z = 10$ cm and finally merge with the stationary electron hole at $z \approx 11$ cm at $t = 3.36 \ \mu$s. Both the ESWs retain their approximate shape, i.e. electric potential and length, before merging, as shown in Fig. \ref{fig:time_snaps_2D_merge}. Similarly, the ESWs marked in white and green \textbf{`X'} in Figs. \ref{fig:2D_soliton_movement_timeevol} and \ref{fig:time_snaps_2D_coll} collide and merge between $t = 4.0 \ \mu s$ and $4.4 \ \mu s$. As seen in Fig. \ref{fig:time_snaps_2D_coll}, the smaller ESW, shown by green \textbf{`X'} tries to pass through the larger ESW, shown by white \textbf{`X'}. However, because of a low relative velocity between these ESWs, they merge together to now move at a speed lower than either of those individual ESWs, as seen in Fig. \ref{fig:2D_soliton_movement_timeevol}, where the merged ESW, marked by white \textbf{`X'} at $t = 4.40 \ \mu s$, has a much smaller velocity.  
In both cases, the merging of ESWs results in an ESW of larger length to accommodate a larger number of trapped particles (as discussed further in Sec. \ref{sec:ESW_BGK}). Such inelastic collisions between ESWs have also been previously observed in experiments where they merge when their relative speed is small\cite{saeki1979formation}. The electron hole with the largest length that forms next to the electron source does not significantly move and remains attached to the electron source and therefore cannot be considered an ESW.       

\begin{figure}
\centering
\subfigure[Time evolution of electric potential along beam axis and movement of ESWs.]{\label{fig:Phi_z_t_3D-A}
        \includegraphics[trim = 3.2cm 2.2cm 3.0cm 4.0cm, clip,width = 0.48\textwidth]{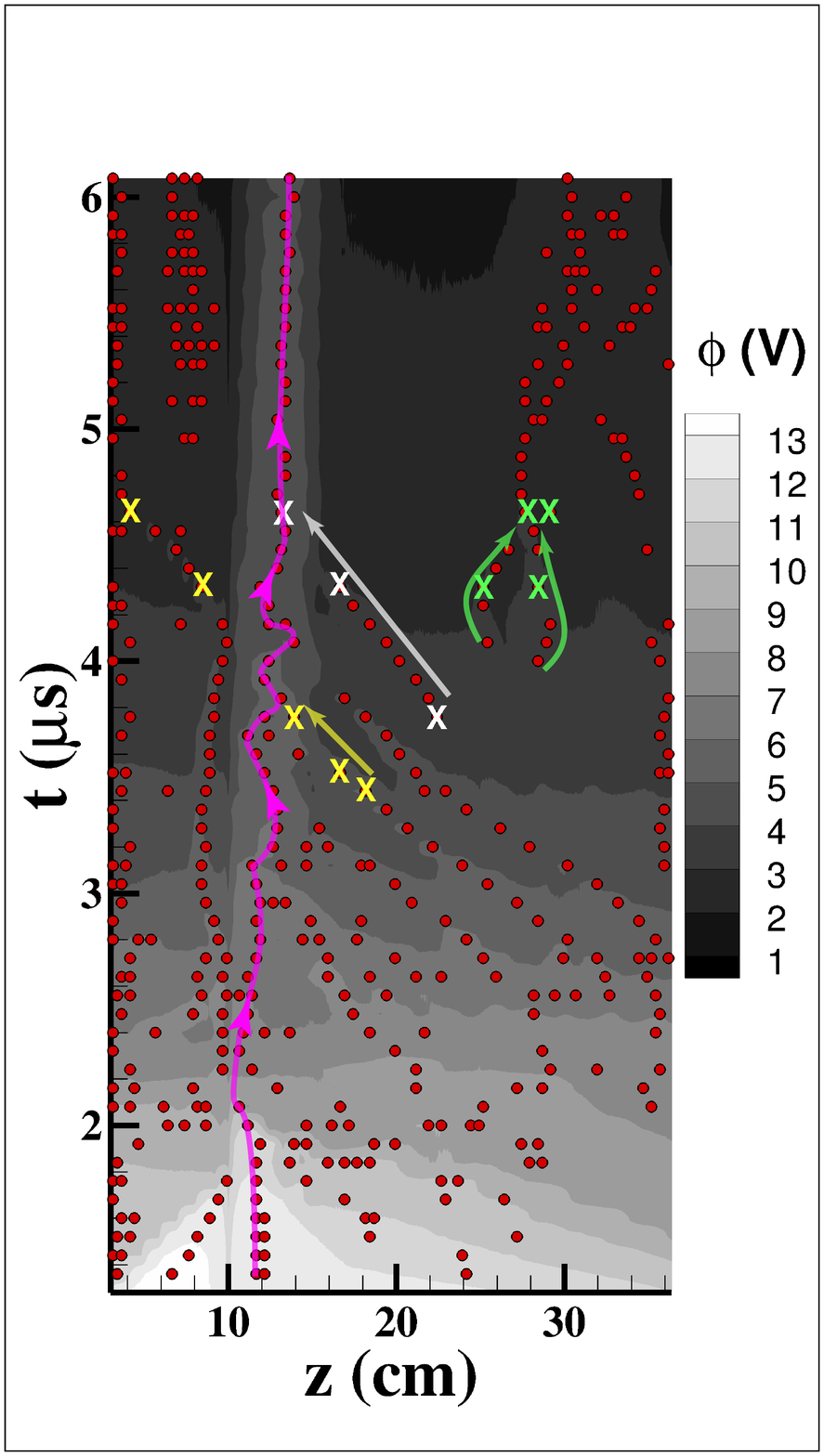}}
\hspace{0.3mm}        
  \subfigure[Time snapshots of ESWs for $3.44<t<4.64 \ {\rm \mu s}$, shown by dashed white box in (a).]{\label{fig:time_snaps_3A}
        \includegraphics[trim = 9cm 0.14cm 10cm 0.14cm, clip,width = 0.48\textwidth]{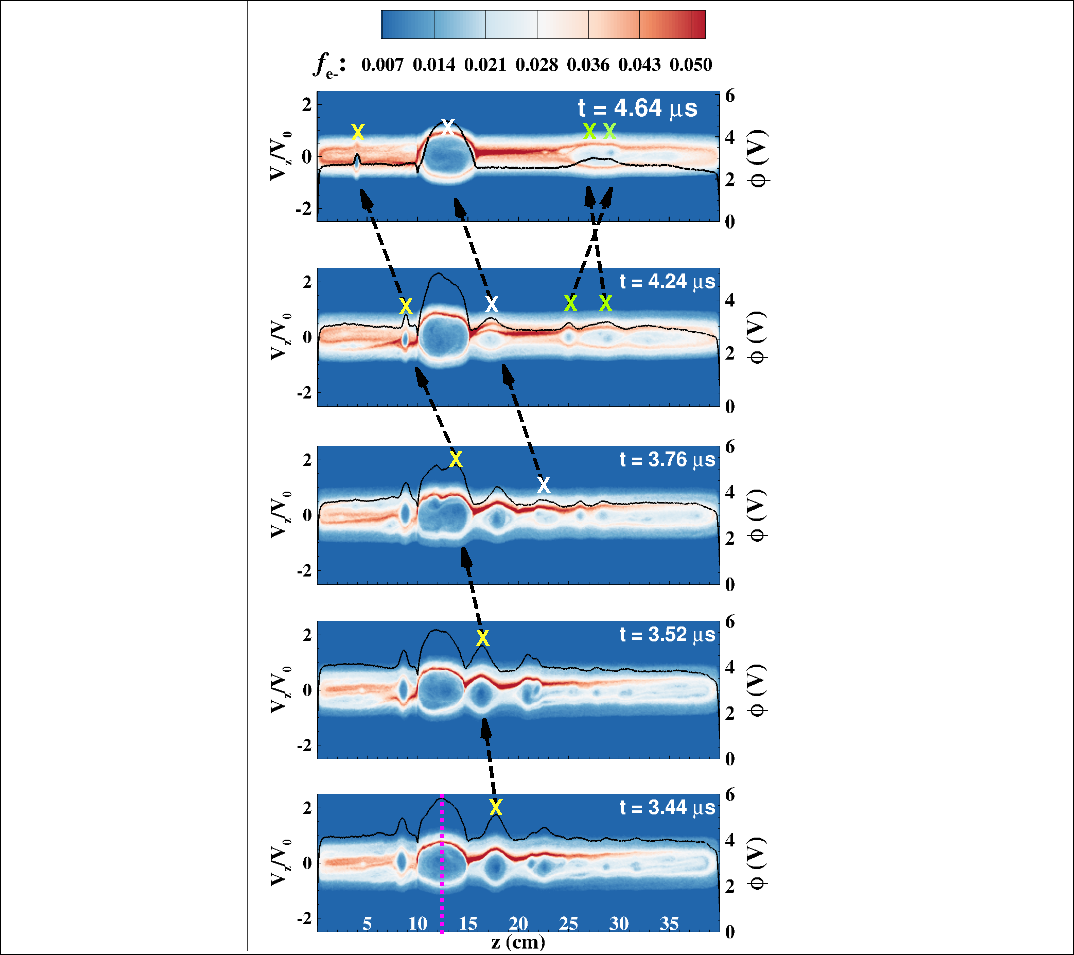}}
\caption{ Time evolution of electric potential and electron ESWs along the beam axis ($(x,y) = (1.5,1.5)$ cm) for 3D cylindrical beam with beam radius, $a= 2.5$ mm. In (a), the magenta colored line near $z = 12$ cm shows the electron hole attached to the electron source at $z = 10$ cm. In (b), the yellow and white ESWs move and merge with the attached electron hole, shown by the dashed magenta line in phase snapshot of $t = 3.44 \ \mu s$. The ESWs shown in green collide with each other and merge. In (b), the electron distribution, $f_{e-}$, is normalized by $n_0 = 1.75\times10^{14} \ {\rm m^{-3}}$.}
\label{fig:3D-A_soliton_movement_timeevol}
\end{figure}

 In the 3D cylindrical beam case with beam radius $a = 2.5$ mm and channel width $b = 15$~mm, the movement of ESW is shown in Fig. \ref{fig:3D-A_soliton_movement_timeevol}. Similar to the 2D planar beam case discussed above, the electron hole with the largest length, whose location is shown by the magenta dashed line in the time-snapshot of $t = 3.44 \ \mu s$ in Fig. \ref{fig:time_snaps_3A}, remains stationary and attached to the electron source at $z = 10$ cm, \textcolor{black}{as shown by the nearly vertical line of magenta color created by connecting the electron hole locations near $z \approx 12$ cm in Fig.~\ref{fig:Phi_z_t_3D-A}}. Also, similar to the 2D planar beam, we observe collisions and merging of ESWs in the 3D cylindrical beam case. An example of a collision can be seen in Fig. \ref{fig:time_snaps_3A}, where the ESW shown by a yellow \textbf{`X'} collides with the attached ESW at $z \approx 12$ cm and initiates the movement of the previously stationary ESW at $z = 8$ cm. Similarly, the ESW shown by the white \textbf{`X'} in Fig. \ref{fig:time_snaps_3A} merges with the attached electron hole at $z \approx 12$ cm. Finally, the two small ESWs, marked by green \textbf{`X'} in Figs. \ref{fig:Phi_z_t_3D-A} and \ref{fig:time_snaps_3A}, collide, pass-through, and finally merge with each other. For both 2D planar and 3D cylindrical cases with $a = 2.5$ mm, we find that most ESWs seem to move in the $-z$ direction with about the same range of ESW speeds between zero and $21$ cm/$\mu s$. 

\begin{figure}
\centering
\subfigure[$a = 5.0$ mm.]{\label{fig:Phi_z_t_3D-B}
        \includegraphics[trim = 0.14cm 0.14cm 0.14cm 0.14cm, clip,width = 0.48\textwidth]{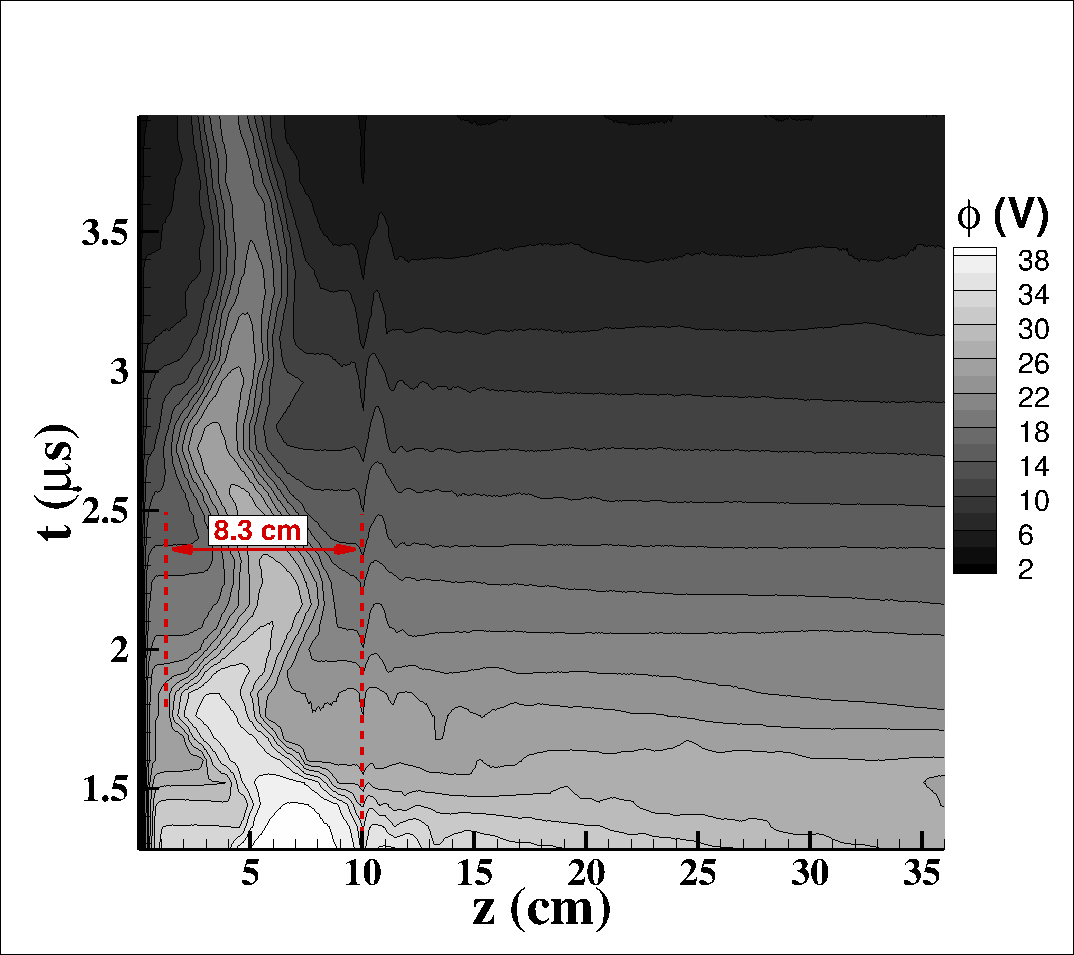}}
\hspace{0.3mm}        
  \subfigure[$a = 7.5$ mm.]{\label{fig:Phi_z_t_3D-C}
        \includegraphics[trim = 0.14cm 0.14cm 0.14cm 0.14cm, clip,width = 0.48\textwidth]{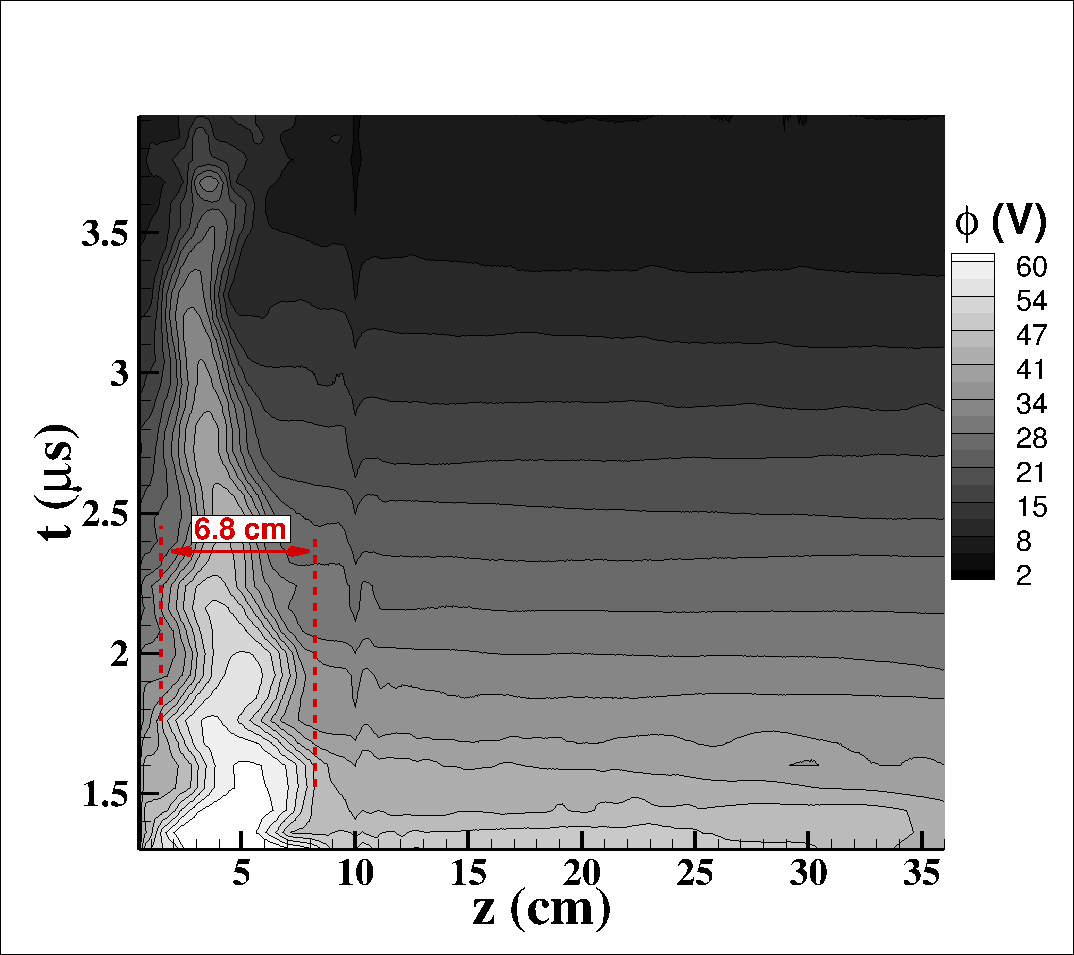}}
\caption{ Time evolution of electric potential along the beam axis ($(x,y) = (1.5,1.5)$ cm) for 3D cylindrical beam cases with $a = 5.0$ and $7.5$ mm.}
\label{fig:3D-BC_timeevol}
\end{figure}

 The nature of ESWs, however, dramatically changes for a 3D cylindrical beam case with a larger beam radius, $a = 5.0$ mm and the same channel width, $b = 15$ mm, as the previous cylindrical beam case. For this case, the widest ESW forms in the upstream direction of the electron source because of a higher electric potential near a larger ion source at $z = 0$ cm. This ESW oscillates in the region $0<z<10$ cm with an amplitude of about 8 cm, as shown in Fig. \ref{fig:Phi_z_t_3D-B}, and this amplitude of oscillation decays with time as the ESW shrinks in shape and size. Unlike the cases shown before, the widest electron hole for the $a = 5.0$ mm case is detached from the electron source. While an attached electron hole forms downstream of the electron source for the $a = 5.0$ mm case, it is much smaller in size compared to the upstream ESW. A similar ESW movement is seen for the case with $a = 7.5$ mm where we observe a detached ESW oscillating upstream of the electron source, as shown in Fig. \ref{fig:Phi_z_t_3D-C}. Also, the upstream ESW for the $a = 7.5$ mm case is of larger length than the $a = 5.0$~mm case because of the larger number of ions near $z = 0$ mm, which attract more beam electrons in the upstream region of the electron source. The ESWs formed downstream of the electron source, for both the $a = 5.0,$ and $7.5$ cm cases, are initially very small in size, and they soon disappear with time. This disparity in the formation of ESWs for different cases shows that the size and position of the ESWs formed in a 3D beam have a strong dependence on the beam radius and ion and electron currents. \textcolor{black}{The summary of locations of ESWs for different cases is listed in Tab. \ref{tab:Summary_of_all_case_results}. } The effect of trapped electron population on the shape and size of an ESW in 2D planar and 3D cylindrical beams will be discussed next.  

\begin{table}
		\begin{center}
		\caption{Position of ESWs and properties of surfaces waves in all PIC cases}
		\label{tab:Summary_of_all_case_results}
	
		\begin{tabular}{l c c c c c c c c c c c c}
            \hline \hline
             Case no.  & Type of beam &  $a^1$ (mm) & ESWs positions$^2$ & Surface waves $\left(\frac{\omega}{\omega_{pe}},\kappa a \right)$ \\
            \hline
     1. & 2D planar & 2.5 mm & All downstream & - \\
     2. & 3D cylindrical & 2.5 mm & All downstream & - \\
     3. & 3D cylindrical & 5.0 mm & One upstream, none downstream & (0.33,0.51) \\
     4. & 3D cylindrical & 7.5 mm & One upstream, none downstream & (0.36,0.79) \\
   \hline \hline
    \end{tabular}
\begin{tablenotes}
\item $^1$ $a = $beam half-width/radius for 2D planar/3D cylindrical beam.
\item $^2$ $z$ positions w.r.t. $e^-$ source, which lies at $z = 10$ cm, along the beam axis. 

\end{tablenotes}		
		\end{center}
\end{table}

\subsection{Shape of ESWs in a beam: trapped vs untrapped electrons} \label{sec:ESW_BGK}
Characterizing ESWs based on their amplitude and size is important in understanding their role in the neutralization of beams. While ESWs were found in the numerical study of a 2D beam by Lan et al.\cite{lan2020neutralization2}, a quantitative comparison of their amplitude and size with the BGK theory was not discussed. In this section, we isolate one ESW each from 2D planar and 3D cylindrical beams of $a = 2.5$ mm, and compare them with the 1D BGK theory\cite{bernstein1957exact}. As discussed in the previous section, ESWs move along the beam as independent entities with a low dissipation, and they have a population of trapped and untrapped electrons associated with them. \textcolor{black}{In the work of Bernstein-Greene-Kruskal\cite{bernstein1957exact} (BGK), an analytical relation between electric potential profile and electron velocity distribution was derived for a stationary solitary wave. In Hutchinson\cite{hutchinson2017electron}, it was shown that the analysis by BGK applies to a moving ESW as well if the total energy and electron velocity distribution is evaluated in the frame of the moving ESW. Similar to the symmetric electron holes of BGK and Hutchinson, the electric potential profiles of our ESWs are almost symmetric.  The electric potential profile of an ESW traps the low kinetic energy electrons, which are reflected from its edges and \textcolor{black}{keeps these electrons bouncing in the potential well.} These trapped electrons appear as a circular vortex-like structure in phase space, as was shown in Figs. \ref{fig:time_snaps_2D_merge}, \ref{fig:time_snaps_2D_coll} and \ref{fig:time_snaps_3A}. 
}

\begin{figure}[h]
\centering
  \subfigure[ Average total energy of electrons in a 2D planar beam. The RHS figure is the EVDF at the center of the ESW `2', $z = 22.50$ cm, at $t = 3.12 \ \mu s$.]{\label{fig:2D_soliton_t39000}
        \includegraphics[trim = 0.25cm 0.05cm 0.07cm 0.14cm, clip,width = 0.7\textwidth]{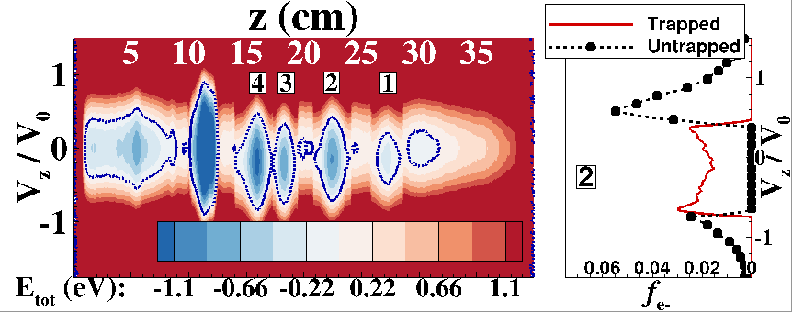}}
\subfigure[ESW `2' in a 2D planar beam at $t = 3.12 \ \mu s$.]{\label{fig:Soliton_2D}
        \includegraphics[trim = 0.14cm 0.14cm 0.14cm 0.14cm, clip,width = 0.48\textwidth]{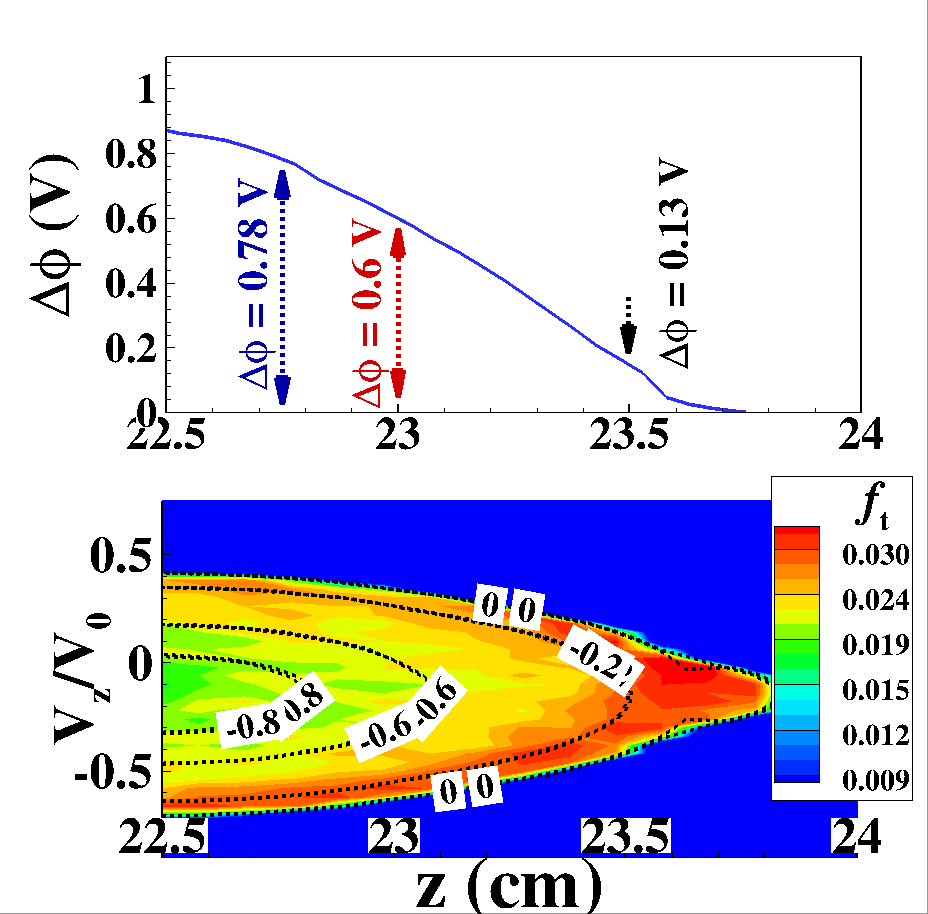}}
\hspace{0.3mm}        
  \subfigure[Trapped electron distribution from CHAOS (dashed) vs Eq.~\ref{eq:BGK_EVDF_selfsimilar} (solid) for 2D beam at $t = 3.12 \ \mu s$.]{\label{fig:BGK_2D}
        \includegraphics[trim = 0.14cm 0.14cm 0.14cm 0.14cm, clip,width = 0.48\textwidth]{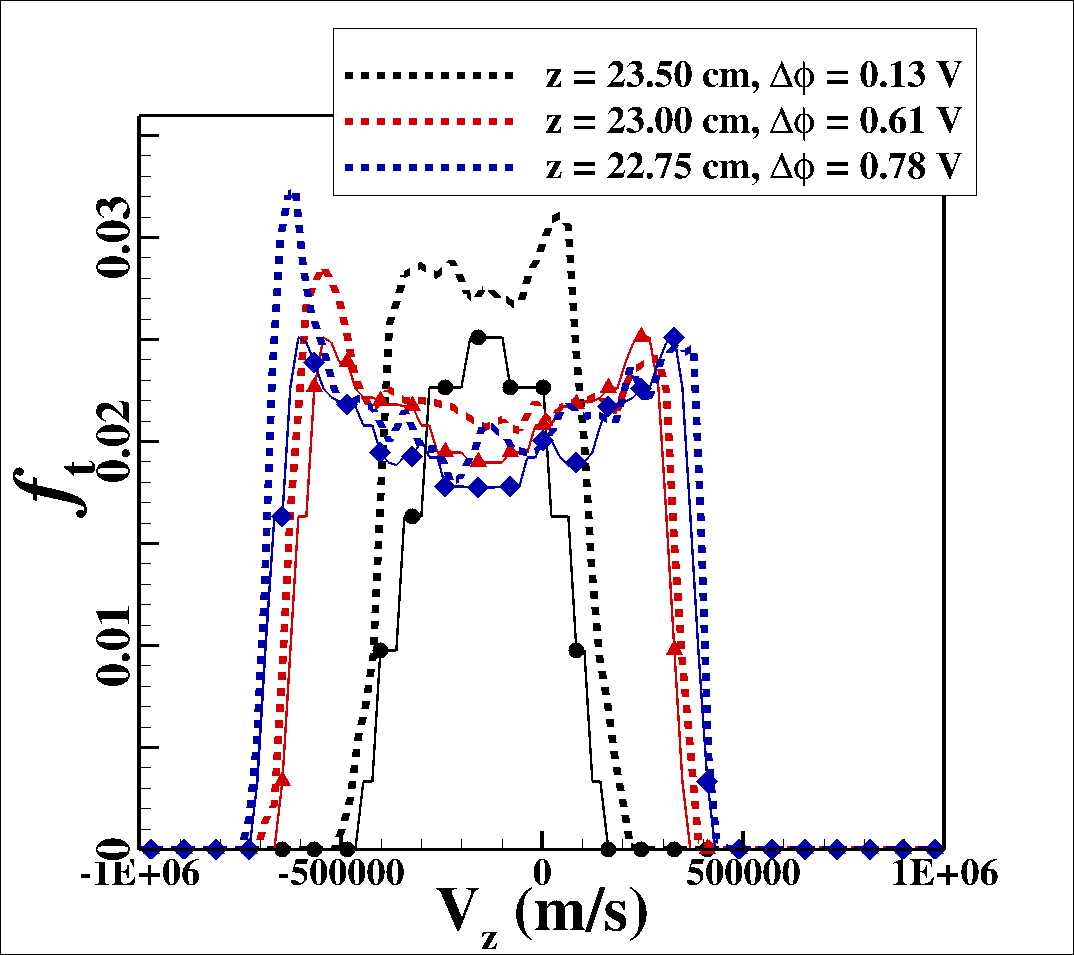}}
\caption{ Average electron energy distribution for different ESWs in a 2D planar beam at $t = 3.12 \ \mu s$. In (b), the electric potential profile and normalized trapped electron density in phase space are shown. The dashed black line in the bottom part of (b) are iso-contour lines of $E_{\rm tot}$, with their values labeled in white boxes. In (c), a comparison of predicted EVDF of trapped population based on Eq. \ref{eq:BGK_EVDF_selfsimilar}, shown by solid lines and CHAOS PIC result, shown by dashed lines. Here, $f_t$ and $f_{e-}$ are normalized by $n_0 = 1.75\times10^{14} \ {\rm m^{-3}}$. }
\label{fig:Soliton_structure_2D_all}
\end{figure}

Figure \ref{fig:2D_soliton_t39000} shows a contour plot of the average total electron energy, $E_{\rm tot}$, per electron for the 2D planar beam case at $t = 3.12 \ \mu s$, 
\begin{align}\label{eq:Etot}
    E_{\rm tot}(v_z,z) = \frac{\sum_{i=1}^{N_{\rm total}}\left[\frac{1}{2}m_e|(\vec{v_{e_i}}-\vec{v_s})|^2 - e\phi\right]\delta_{v_{e_i}v_z}\delta_{z_{e_i}z}}{\sum_{i=1}^{N_{\rm total}}\delta_{v_{e_i}v_z}\delta_{z_{e_i}z} },
\end{align}
where $\vec{v_{e_i}} = v_{e_{iz}}$ and $\vec{v_s}$ are the z-directional electron and ESW velocities, $\delta$ is the Kronecker delta function, $z_{e_i}$ is the $z$ position of the $i^{th}$ electron, and $N_{\rm total}$ is the total number of electron computational particles in the domain. For the phase-space points $(v_z,z)$ where there are no electrons, $E_{\rm tot} = 0$. For trapped electrons, the term in square brackets of Eq. \ref{eq:Etot} is less than zero. The average total energy per particle, $E_{\rm tot} = 0$ eV, shown by the dashed blue line in Fig. \ref{fig:2D_soliton_t39000}, marks the approximate boundary in phase space between the trapped and un-trapped electrons. \textcolor{black}{The discontinuities in the contour plot of Fig. \ref{fig:2D_soliton_t39000}, seen at the edges of ESWs at $z = 13,$ 19.4, 20.9, 23.9, 25.8, and 28.7 cm, are because of the difference in values of $v_s$ for each ESW, as estimated from Fig. \ref{fig:2D_soliton_movement_timeevol}.} Based on the average total energy criteria, the fractions of trapped ($N_{\rm t}$) and untrapped ($N_{\rm ut}$) computational electron particles for each of the ESWs numbered in Fig. \ref{fig:2D_soliton_t39000} are listed in Tab. \ref{tab:2D_solitions}. It can be seen that the length ($\lambda_{\rm ESW}$) and amplitude ($\Delta \phi$) of an ESW increases with a higher fraction of trapped electrons, which is consistent with a 1D BGK mode\cite{bernstein1957exact}. Among the four ESWs shown in Fig. \ref{fig:2D_soliton_t39000}, ESW `2' is of interest because it is relatively isolated from the other ESWs and it moves with a constant speed for a long duration (see yellow \textbf{`X'} in Fig. \ref{fig:2D_soliton_movement_timeevol}). The RHS of Fig. \ref{fig:2D_soliton_t39000} shows the electron velocity distribution function (EVDF) extracted at the center of ESW `2' ($z = $ 22.5 cm) in the phase plot on the LHS for the 2D planar beam case. For ESW `2', the un-trapped particle region of the EVDF has a peak at $v_z/v_0 \approx 0.5$, which corresponds to the electrons moving with a velocity close to the ion beam velocity of $v_z/v_0 = 0.427$, where $v_0 = 1\times10^6$ m/s. The trapped region of the EVDF in Fig. \ref{fig:2D_soliton_t39000} is symmetric about the ESW speed, $v_s=$ $14.5$ $ \rm{cm/\mu s}$. This symmetry is also consistent with a typical 1D BGK mode where continuous bouncing in the potential-well produces a symmetric distribution of trapped electrons\cite{bernstein1957exact}.

Furthermore, in a 1D BGK mode, all the electrons follow constant energy trajectories\cite{bernstein1957exact,hutchinson2017electron}. Therefore, if the electric potential profile of an ESW's potential-well, and the trapped particle EVDF at the center of the ESW are known, the trapped EVDF at different locations inside the ESW can be analytically evaluated\cite{bernstein1957exact,hutchinson2017electron} by,
\begin{align}\label{eq:BGK_EVDF_selfsimilar}
f((v_z-v_s),\Delta z) = f_0\left(\sqrt{0.5(v_z-v_s)^2 - \frac{e}{m_e}(\Delta \phi(\Delta z) - [\Delta\phi]_{\rm max}})\right),
\end{align}
where $\Delta \phi = \phi - \phi_{\rm base}$, $\phi_{\rm base}$ being the lowest electric potential value in the vicinity of the ESW, $\Delta z = z - z_{\rm center}$, $z_{\rm center}$ being the center of the ESW, and $f_0$ is the EVDF at the center of the ESW, shown on the RHS of Fig. \ref{fig:2D_soliton_t39000}, and $\Delta \phi(z = z_{\rm center}) = [\Delta \phi]_{\rm max}$\cite{hutchinson2017electron}. For our 2D planar beam, we obtain a 1D electric potential profile, from PIC, along the beam axis where the value of $\Delta \phi$ ranges from $[\Delta \phi]_{\rm max}$ at the center of the ESW to 0 V at its edge, as shown in the top part of Fig. \ref{fig:Soliton_2D}. Figure \ref{fig:Soliton_2D} also shows the distribution of trapped particles in the phase space (bottom) with iso-energy lines of $E_{\rm tot}$ for the ESW, labeled `2' in Fig. \ref{fig:2D_soliton_t39000}. Using the EVDF at the center of ESW `2', EVDFs at three different locations, shown in the top part of Fig. \ref{fig:Soliton_2D}, are evaluated using Eq. \ref{eq:BGK_EVDF_selfsimilar}. The EVDFs from Eq. \ref{eq:BGK_EVDF_selfsimilar}, shown by solid lines in Fig. \ref{fig:BGK_2D}, compare well with the PIC result from CHAOS, shown by dashed lines of the same colors in Fig. \ref{fig:BGK_2D}. The minor discrepancy is because of the error in estimating the ESW velocity, $v_s$, from Fig. \ref{fig:2D_soliton_movement_timeevol}. The predicted untrapped EVDFs using Eq. \ref{eq:BGK_EVDF_selfsimilar} also compared well with the PIC result. This comparison shows that the ESWs we see in the 2D beam neutralization case have electron trajectories that follow the 1D BGK mode type electron distribution \cite{bernstein1957exact}. 

\begin{figure}[h]
\centering
 \subfigure[ Electron density from PIC result for 2D planar beam at $t = 3.12 \ \mu s$.]{\label{fig:total_e_dens_CHAOS}
        \includegraphics[trim = 0.14cm 0.14cm 0.14cm 0.14cm, clip,width = 0.5\textwidth]{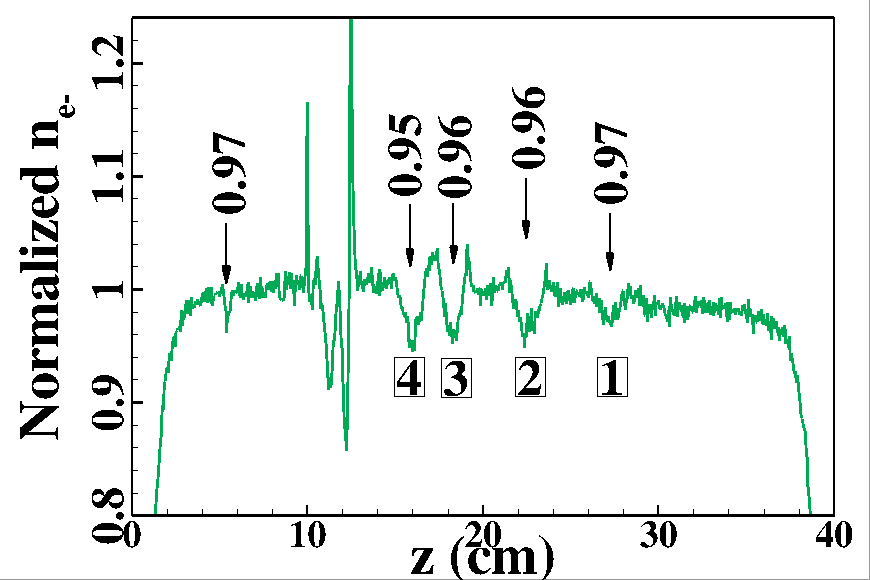}}
\subfigure[Fit of $\Delta\phi(\Delta z)$ with the PIC result]{\label{fig:Phi_profile_fit}
        \includegraphics[trim = 0.14cm 0.14cm 0.14cm 0.14cm, clip,width = 0.48\textwidth]{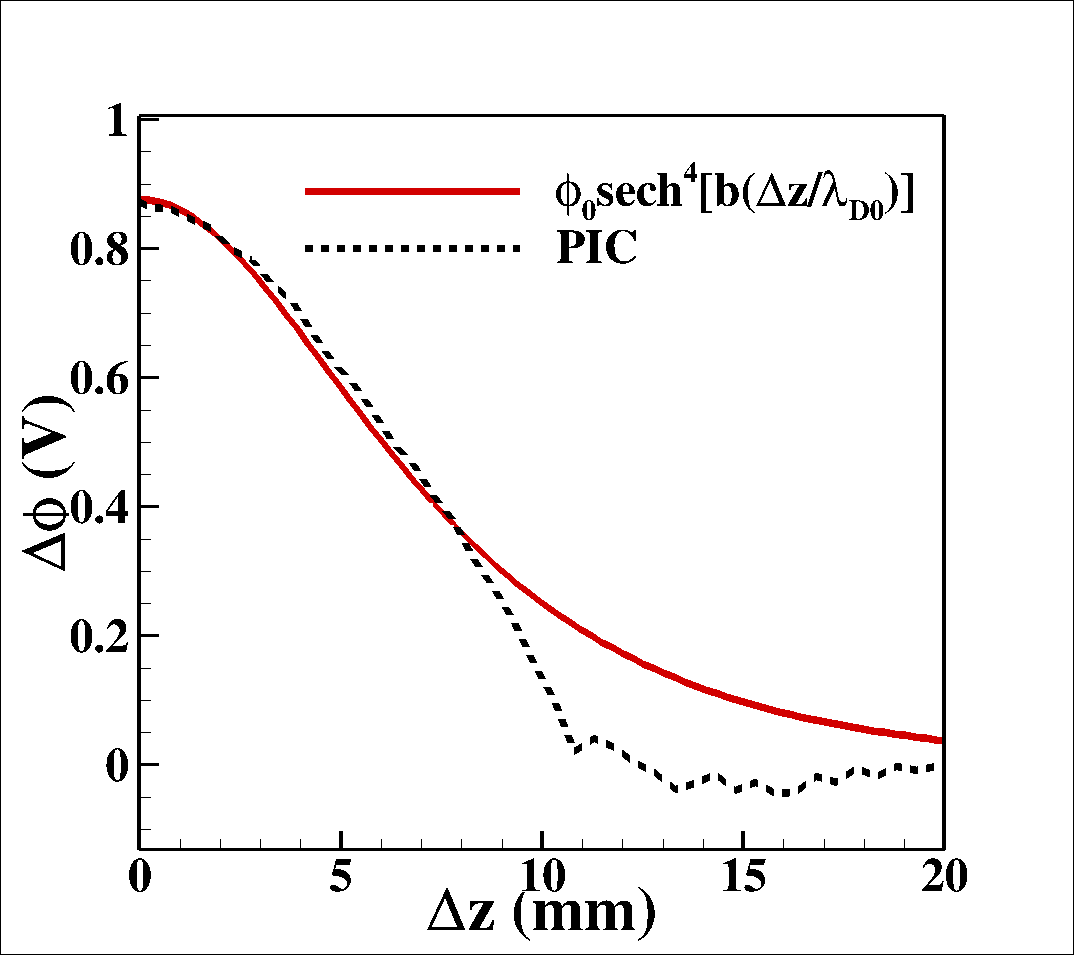}}  
\subfigure[Far-field EVDF, $f_{\rm FF}$, compared for Maxwellian and non-Maxwellian cases.]{\label{fig:EVDF_fit}
        \includegraphics[trim = 0.14cm 0.14cm 0.14cm 0.14cm, clip,width = 0.48\textwidth]{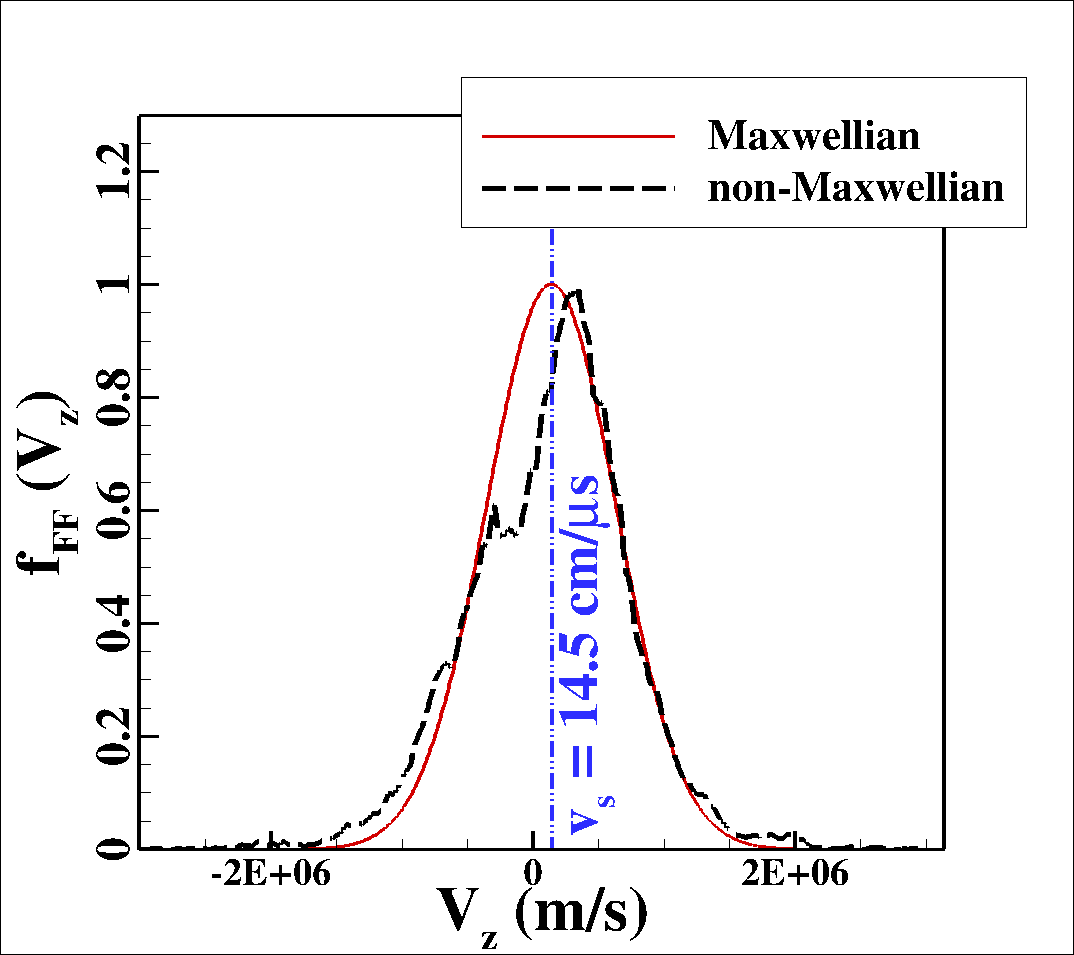}}        
\caption{ (a) Normalized electron density with perturbations at the locations of numbered ESWs from Fig. \ref{fig:2D_soliton_t39000}. Comparison of assumed and CHAOS(PIC) $\Delta\phi$ profile (b) and far-field EVDFs for the ESW `2' (c).}
\label{fig:BGK_theory_fit}
\end{figure}

\begin{figure}[h]
\centering
\subfigure[Untrapped particle population for $f_{\rm FF}$ as a non-Maxwellian EVDF obtained from PIC results.]{\label{fig:Non-Maxwellian_BGK}
        \includegraphics[trim = 0.14cm 0.14cm 0.14cm 0.14cm, clip,width = 0.48\textwidth]{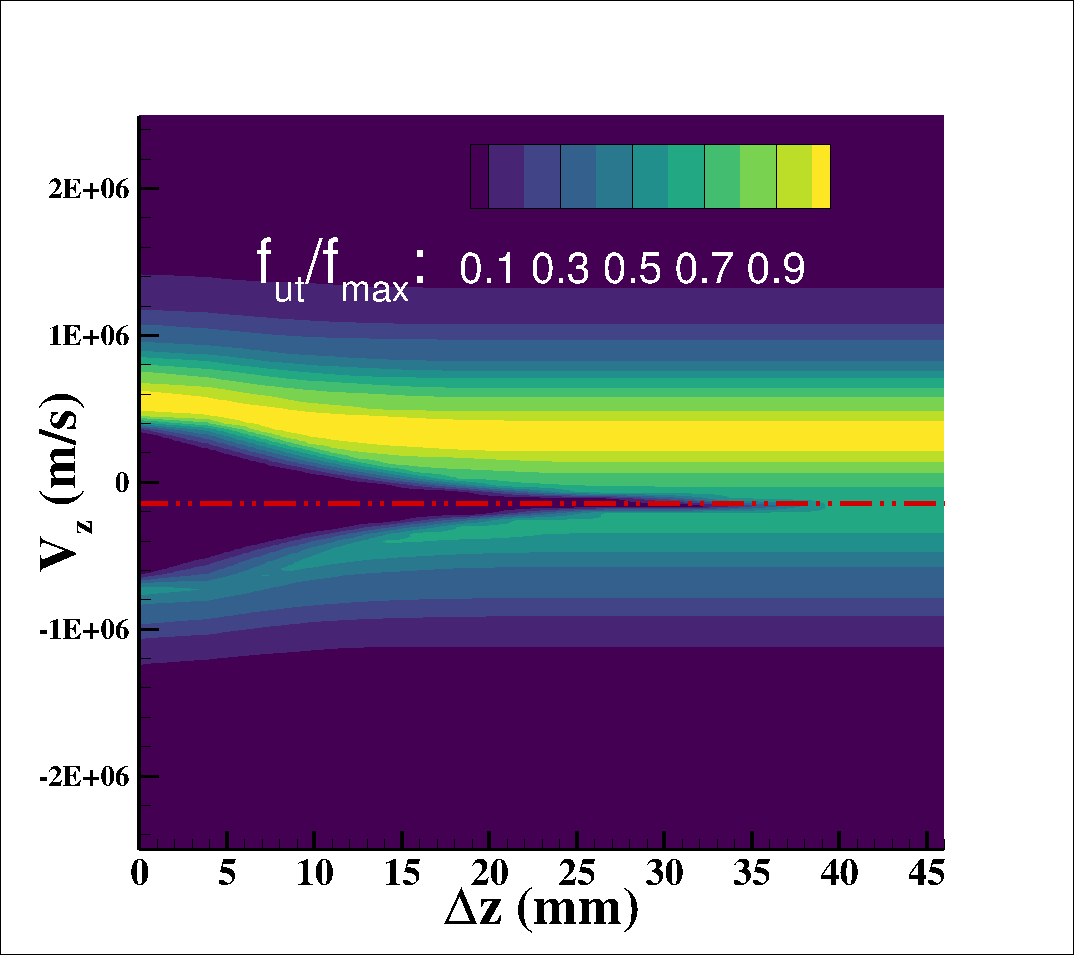}}
                \hspace{0.3mm}
  \subfigure[ Untrapped particle population for $f_{\rm FF}$ as a Maxwellian EVDF.]{\label{fig:Maxwellian_BGK}
        \includegraphics[trim = 0.14cm 0.14cm 0.14cm 0.14cm, clip,width = 0.48\textwidth]{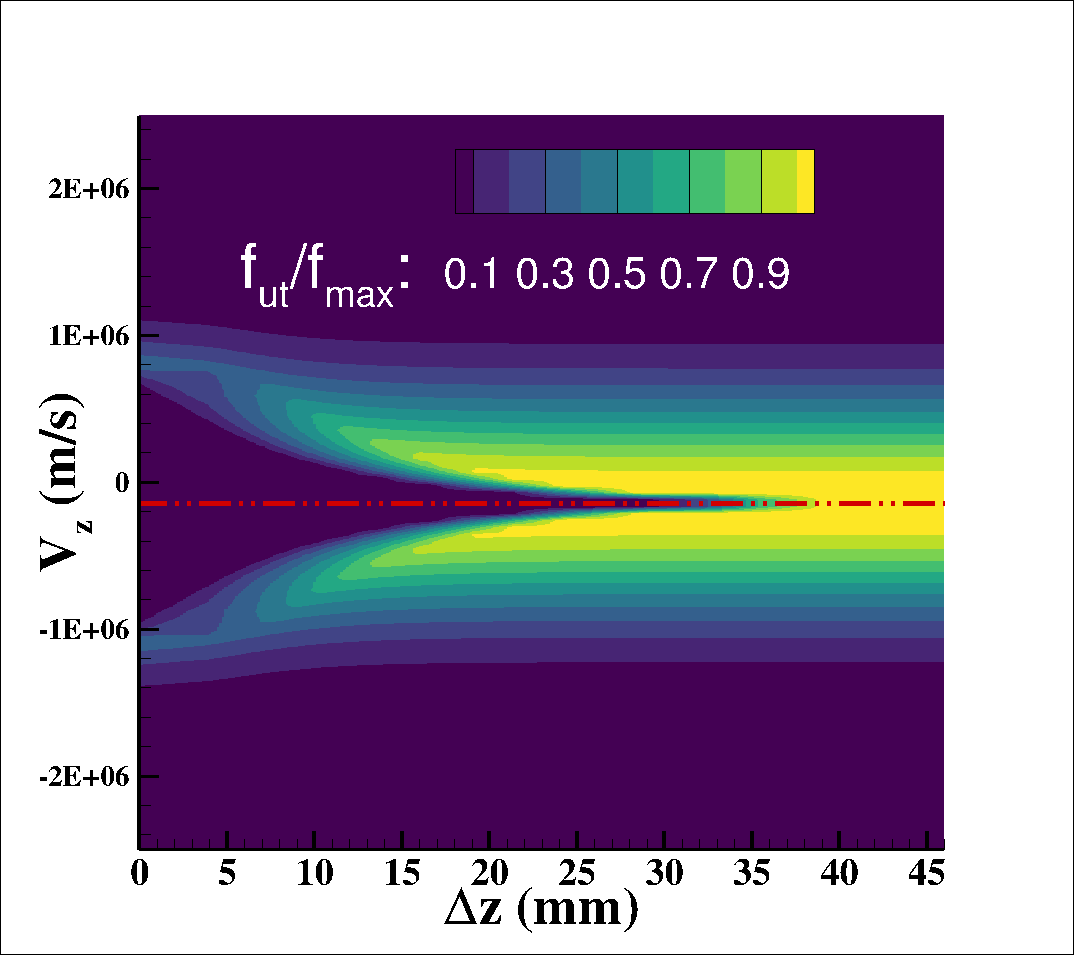}}        
\caption{ Comparison of normalized $f_{ut}$ (i.e. $f_{\rm ut}/f_{\rm max}$ here) for the cases of a Maxwellian vs non-Maxwellian far-field distribution. Here, $f_{\rm max}$ is the maximum of $f_{ut}(v_z,\Delta z)$, and the dashed red line represents the velocity of ESW `2', $v_s = 14.5 \ {\rm cm/\mu s}$. }
\label{fig:BGK_theory_Phase_two_cases}
\end{figure}

  We also observe that the ESWs listed in Tab. \ref{tab:2D_solitions} are about 30-45 Debye lengths long for the 2D planar beam case. While the typical ESWs analyzed in theory are about 2$\lambda_{\rm D}$ to 20$\lambda_{\rm D}$ in length and small in amplitude\cite{hutchinson2017electron,chen2002bgk2,lefebvre2010laboratory}, in our case, the ESWs move with low relative speeds, which allows them to merge with each other to create longer ESWs with larger trapped particle populations. This merging was also shown in Lan et al.\cite{lan2020neutralization2}, and it was identified as one of the properties of electron holes in Schamel\cite{schamel1979theory}. 
\textcolor{black}{Such long ESWs usually have a large electron density perturbation when they form in a Maxwellian far-field background plasma.} \textcolor{black}{However, Fig. \ref{fig:total_e_dens_CHAOS} shows that all the ESWs in the 2D planar beam PIC case have surprisingly low electron density perturbations between $3$ to $5\%$. The electron density profile in Fig. \ref{fig:total_e_dens_CHAOS} is computed by sampling all the electrons at multiple z planes along the beam axis at $t = 3.12 \ \mu s$.} With further study, we find that the non-Maxwellian EVDF, found in our beam, requires a smaller \textcolor{black}{electron density perturbation} to form an ESW of the same length than one where the EVDF is Maxwellian in the far-field. We demonstrate this by using the 1D BGK integral approach, described in Bernstein et al.\cite{bernstein1957exact} and Hutchinson\cite{hutchinson2017electron}, to predict the untrapped and trapped electron density profiles inside ESW `2'. In this approach, the background or far-field EVDF, $f_{\rm FF}$, of an ESW is known (\textcolor{black}{in our case from the PIC simulations}), and the ions are assumed to be of constant and uniform density in the ESW. This approach has the following four steps: (1) assume an electric potential profile for the ESW, (2) using conservation of total electron energy with the assumed potential profile of the ESW, calculate the untrapped EVDF, $f_{ut} (v_z,\Delta z)$, from the known far-field EVDF, $f_{\rm FF}$, (3) compute the untrapped electron density ($n_{ut} (\Delta z)$) of the ESW from $f_{ut} (v_z,\Delta z)$, and, (4) using Poisson equation, calculate the trapped electron density ($n_{t} (\Delta z)$). See Appendix \ref{sec:1DBGK_app} for details regarding how this approach is applied to the PIC particle data.



With respect to step (1), we fit the electric potential profile of ESW `2' from the PIC simulation to an analytical electric potential profile suggested by Schamel\cite{schamel1979theory}, 
\begin{align}\label{eq:phi_fit}
    \Delta \phi = {\phi_0} {\rm sech^4}\left(\frac{b_s\Delta z}{\lambda_{\rm D_0}}\right),
\end{align}
as shown in Fig. \ref{fig:Phi_profile_fit}, with $\phi_0 = 0.876$ V and $b_s = 0.05$ as the fit parameters, and $\lambda_{\rm D_0} = 0.78$ mm is the Debye length near the electron source where $n_e = 1.75\times10^{14} \ {\rm m^{-3}}$ and $T_e = 2$ eV. This analytically fit electric potential profile makes it easier for us to apply the 1D BGK approach. Since the total energy (i.e. kinetic + potential energy) is conserved along the electron trajectories in 1D, we can compute the untrapped electron phase distribution, $f_{ut}\left(v_z ,\Delta z\right)$, from a known far-field EVDF, $f_{\rm FF}$, i.e. step (2), 
\begin{align}\label{eq:untrapped_EVDF_selfsimilar}
    f_{ut}\left(v_z ,\Delta z\right) = \begin{cases}
    f_{\rm FF}\left( \sqrt{v_z^2-2\frac{e}{m_e}\Delta \phi(\Delta z)}\right) & \text{if $v_z > 0$} \\
    f_{\rm FF}\left( -\sqrt{v_z^2-2\frac{e}{m_e}\Delta \phi(\Delta z)}\right) & \text{if $v_z < 0$}.
    \end{cases}
\end{align}

\begin{figure}[h]
\centering
  \subfigure[ The untrapped and total electron density for two cases of $f_{\rm FF}$.]{\label{fig:Maxwellian_NonMax_trapped_BGK}
        \includegraphics[trim = 0.14cm 0.14cm 0.14cm 0.14cm, clip,width = 0.48\textwidth]{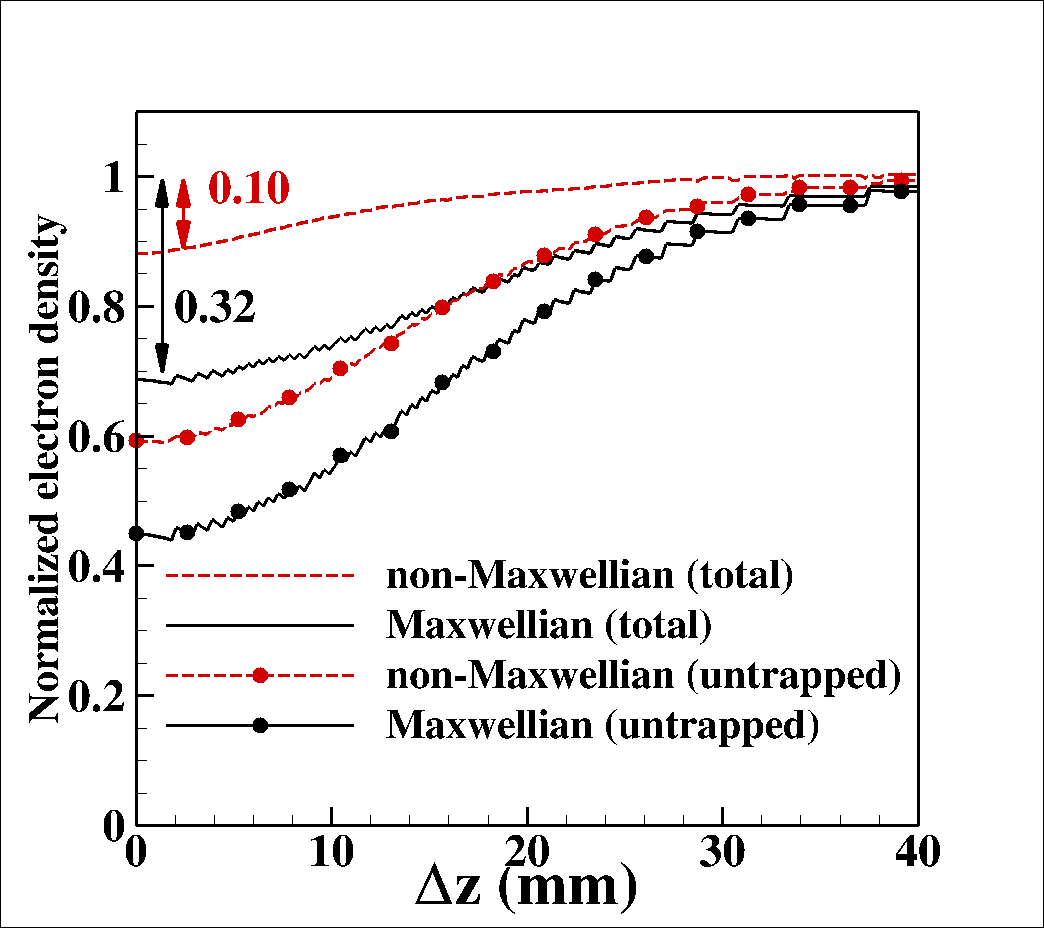}}
        \hspace{0.3mm}
\subfigure[Comparison of trapped and untrapped populations predicted by Eq. \ref{eq:trapped_population} and obtained from PIC results.]{\label{fig:trapped_untrapped_Anal_BGK}
        \includegraphics[trim = 0.14cm 0.14cm 0.14cm 0.14cm, clip,width = 0.48\textwidth]{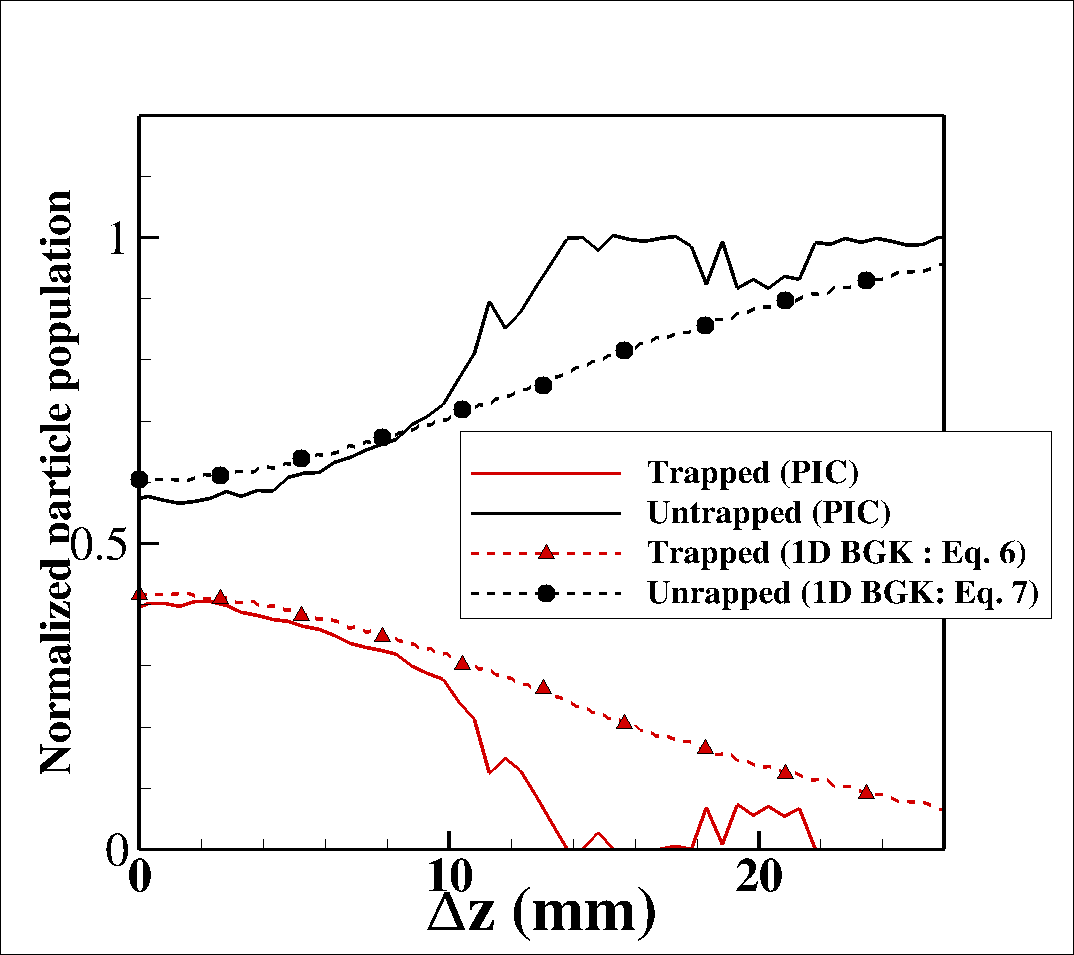}}        
\caption{ Comparison of untrapped and total electron density for a Maxwellian vs non-Maxwellian far-field distribution(PIC) using 1D BGK approach. In (b), the PIC result is compared with the 1D BGK approach and the densities are normalized by the electron density in the far-field of ESW `2', i.e. $1.5\times10^{14} \ {\rm m^{-3}}$. }
\label{fig:BGK_theory_trapped_untrapped_comp}
\end{figure}

The two electron streams in the beam results in a non-Maxwellian far-field EVDF, $f_{\rm FF}$, as shown in Fig. \ref{fig:EVDF_fit}, where the highest peak represents the electrons moving in the direction (+z) of the ion beam and the smaller peak represents the electrons reflected from the plasma sheaths at $z_{\rm min}$ and $z_{\rm max}$ boundaries and the virtual cathode that forms near the electron source at $z = 10$ mm. The electron phase distribution, $f_{ut}(v_z, \Delta z)$, computed using Eq. \ref{eq:untrapped_EVDF_selfsimilar} with the non-Maxwellian far-field EVDF, $f_{\rm FF}$, of Fig. \ref{fig:EVDF_fit}, is shown in Fig. \ref{fig:Non-Maxwellian_BGK}.  Figure \ref{fig:Non-Maxwellian_BGK} also shows that, similar to the far-field EVDF, $f_{\rm FF}$, the $f_{ut}(v_z, \Delta z)$ predicted by Eq. \ref{eq:untrapped_EVDF_selfsimilar} is asymmetric about the ESW's velocity, $v_s = -14.5 \ {\rm  cm /\mu s}$. In contrast, if we assume a Maxwellian EVDF for the far-field, \begin{align}\label{eq:Max_EVDF}
    f_{\rm FF}(v_z) = C \exp\left[{-\frac{v_z^2}{v_{te}^2}}\right],
\end{align}
where $C=350$ is a fit parameter based on the comparison with the PIC result shown in Fig. \ref{fig:EVDF_fit} and $v_{te}$ is the thermal speed, Eq. \ref{eq:untrapped_EVDF_selfsimilar} gives an untrapped electron phase distribution, $f_{ut} (v_z, \Delta z)$ that is symmetric about the ESW `2' velocity, $v_s =-14.5 \ {\rm  cm /\mu s}$, as shown in Fig. \ref{fig:Maxwellian_BGK}. In step (3), we use the untrapped electron phase distribution, $f_{ut}(v_z, \Delta z)$, to calculate the number density profile of untrapped electrons, $n_{ut}(\Delta z)$, given by, 
\begin{align}\label{eq:untrapped_population}
    n_{ut} (\Delta z) = \int_{-\infty}^{+\infty} f_{ut} (v_z,\Delta z) dv_z.
\end{align}
A comparison between untrapped electron density profiles, $n_{ut} (\Delta z)$, for a Maxwellian vs non-Maxwellian (PIC result) far-field EVDFs is shown in Fig. \ref{fig:Maxwellian_NonMax_trapped_BGK}, where a Maxwellian $f_{\rm FF}$ predicts a lower untrapped electron density than the non-Maxwellian $f_{\rm FF}$, obtained from PIC, inside the ESW. 
In the usual 1D BGK approach, we invoke Poisson equation to calculate the trapped electron density profile ($n_t(\Delta z)$) inside the ESW, i.e.,
\begin{align}\label{eq:trapped_population}
    n_t(\Delta z) &= n_i(\Delta z) + \frac{\epsilon_0}{e}\frac{d^2[\Delta\phi(\Delta z)]}{d[\Delta z]^2} - n_{ut}(\Delta z), 
\end{align}
where $n_i (\Delta z)$ is the ion number density profile, which is assumed to be constant and $n_i = n_{ut}(\Delta z\to \infty)$. \textcolor{black}{For the particular case of ESW `2', the second term on the RHS of Eq. \ref{eq:trapped_population} would be much smaller than the other two because of its large length, which also makes the trapped and untrapped electron density profiles nearly complementary and their difference is equivalent to $4\%$ electron density perturbation in the ESW.} Finally, using Eq. \ref{eq:trapped_population}, we calculate the trapped electron density profile using the non-Maxwellian $f_{\rm FF}$ that we obtained from our PIC simulation result, i.e. step (4).  The normalized $n_t$ and $n_{ut}$ profiles for the non-Maxwellian $f_{\rm FF}$ compare well with the PIC result in Fig. \ref{fig:trapped_untrapped_Anal_BGK}, which reaffirms our assertion that ESW `2' in the 2D planar beam case is a 1D BGK mode formed in a non-Maxwellian far-field plasma.

\textcolor{black}{However, using Eq. \ref{eq:trapped_population} for a Maxwellian $f_{\rm FF}$ may result in a fictitious trapped EVDF where it can have negative values for the number of electrons\cite{hutchinson2017electron}. To avoid that, we instead calculate the trapped electron phase distribution by assuming that the trapped EVDF at the center of the ESW ($\Delta z = 0$) is known from our PIC result. This treatment of trapped electrons results in identical trapped electron density profiles, $n_t(\Delta z)$, for both Maxwellian and non-Maxwellian $f_{\rm FF}$'s. So, the only difference in the total electron density is due to the untrapped electrons inside the ESW. Figure \ref{fig:Maxwellian_NonMax_trapped_BGK} shows a comparison of the total electron density drop in the Maxwellian and non-Maxwellian far-field cases, which shows that the Maxwellian $f_{\rm FF}$ would have a larger electron density perturbation, given that the trapped electron distribution at the center of the ESW and the potential profile is the same as it is for the non-Maxwellian $f_{\rm FF}$. This means that a Maxwellian background EVDF, $f_{\rm FF}$, would likely result in a larger electron density and electric potential perturbation than a non-Maxwellian one and this is the reason why such long ESWs with small density and potential perturbations appear in our ion beam, which has a non-Maxwellian $f_{\rm FF}$. }

\begin{table}
		\begin{center}
		\caption{Trapped and un-trapped computational particles of each ESW in the 2D planar beam at $t = 3.12 \ \mu s$, and 3D cylindrical beam at $t = 3.44 \ \mu s$, shown in Figs. \ref{fig:2D_soliton_t39000} and \ref{fig:Trapped_untrapped_Etot_3D_phase}, respectively, with $a = 2.5$~mm.}
		\label{tab:2D_solitions}
	
		\begin{tabular}{c c c c c c c c c c c c c}
            \hline \hline
           & ESW No.   & $N_{\rm t}^{\dagger}/N_0^\#$ & $N_{\rm ut}^{\dagger}/N_0$  & $\Delta \phi^*$ (V)& $[\lambda_{\rm D}]_{\rm FF}^d$ (mm) &$\lambda_{\rm ESW}^c/[\lambda_{\rm D}]_{\rm FF}$ & $v_{s} \ (\rm {cm/\mu s})$\\
            \hline
&	1. &  0.14 & 0.86  & 0.45 & 0.74 & 29.47 & 13.42\\
2D planar &   2. &  0.25 & 0.75 & 0.87 & 0.69 & 39.88 & 14.50\\
&	3. &  0.23 & 0.77   & 0.80 & 0.69 & 26.66 & 20.67\\
&	4. &  0.30 & 0.70 & 1.19 & 0.69 & 43.88 & 18.08\\

   \hline \hline\\
3D cylindrical &  	1. &  0.049 & 0.951 & 0.52 & 0.59 & 73.57 & 15.96\\
&	2. &  0.081 & 0.919   & 1.04 & 0.63 & 114.2  & 21.00\\
   \hline \hline
   
    \end{tabular}
\begin{tablenotes}
\item $^\dagger$ $N_{\rm t},$ $N_{\rm ut}$ are number of trapped and untrapped computational electron particles, respectively.
\item $^\#$ Total electron particles, $N_{0} = N_{\rm t} + N_{\rm ut}$.
\item $^*$ $\Delta \phi = $ amplitude of the ESW.
\item $^c$ $\lambda_{\rm ESW} =$ length of ESW based on $E_{\rm tot} = 0$ eV.
\item$^d$ $[\lambda_{\rm D}]_{\rm FF}$ is local Debye length in the far-field of the ESW.
\end{tablenotes}		
		\end{center}
\end{table}

\begin{figure}[H]
\centering
\subfigure[Iso potential surfaces of a 3D beam with $a = 2.5$ mm.]{\label{fig:3D_iso_phi_thinbeam}
        \includegraphics[trim = 0.14cm 0.14cm 1.1cm 0.14cm, clip,width = 0.60\textwidth]{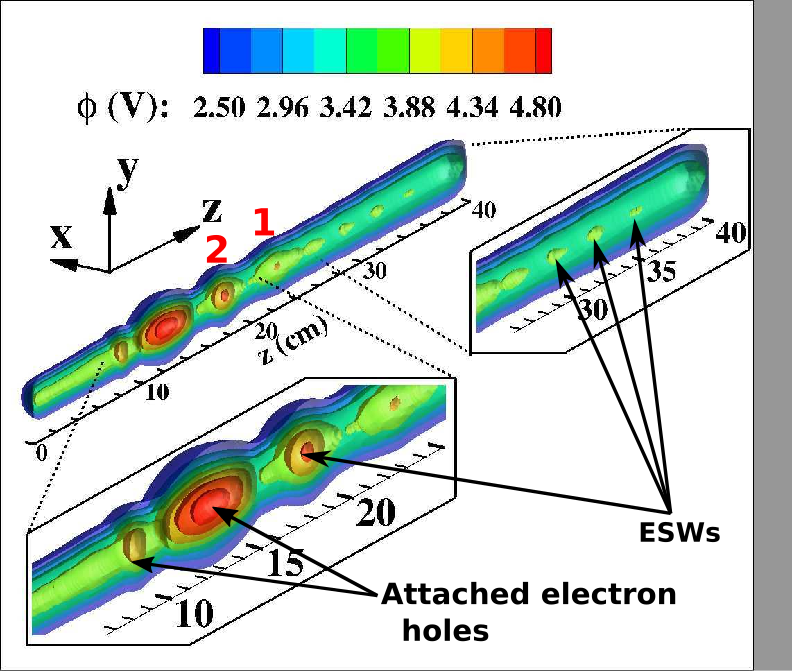}}
\subfigure[2D vs 3D beam potential profiles at $z = 20$ cm.]{\label{fig:3D_i2D_prof_comp}
        \includegraphics[trim = 0.14cm 0.14cm 0.14cm 0.14cm, clip,width = 0.45\textwidth]{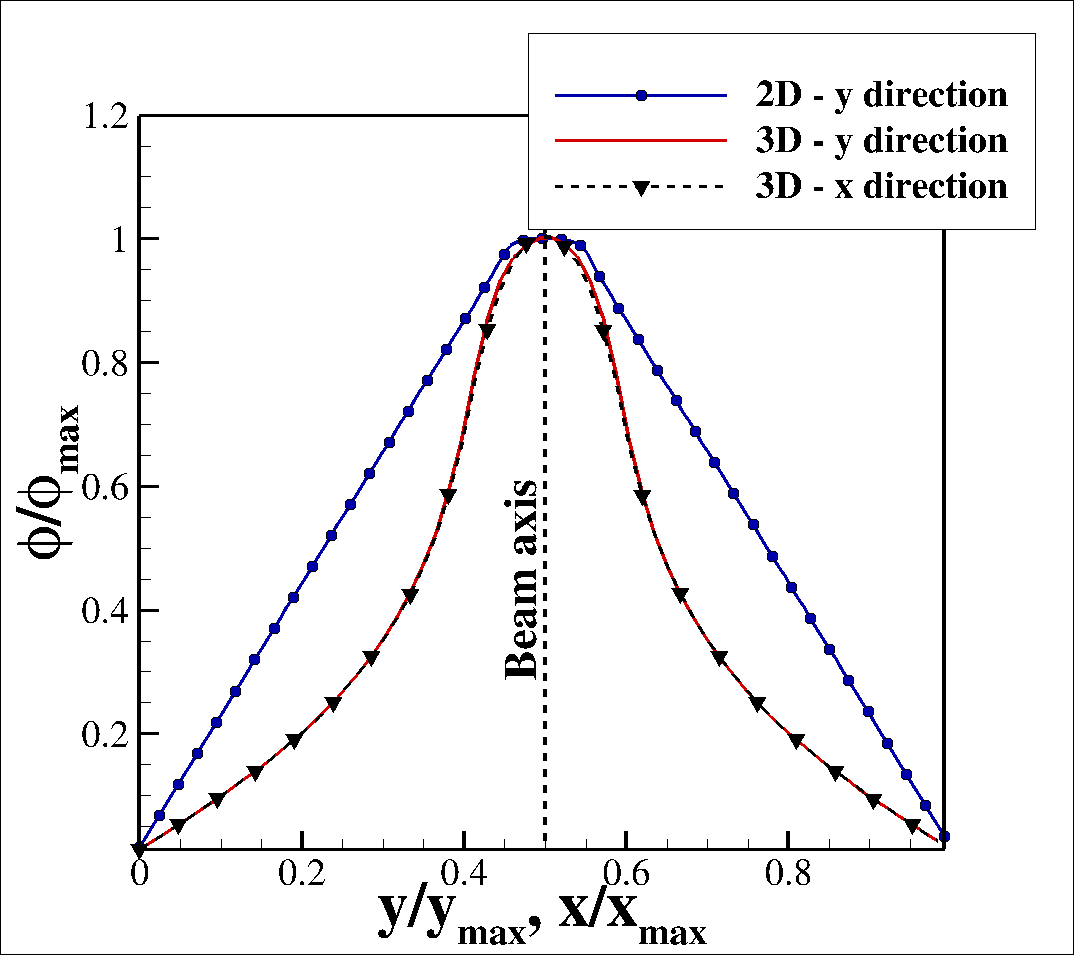}}        
\caption{Iso-potential surfaces of 3D beam in (a) show the oblong shaped electron holes next to the electron source and solitary waves at its downstream. In (b) the transverse direction $\phi$ profiles are compared for 2D planar and 3D cylindrical beams at $t = 3.12$ and $3.44 \ \mu s$, respectively.   }
\label{fig:3D_iso_and_r_profiles_phi}
\end{figure}

\begin{figure}[H]
\centering
        \includegraphics[trim = 0.14cm 0.14cm 0.14cm 0.14cm, clip,width = 0.8\textwidth]{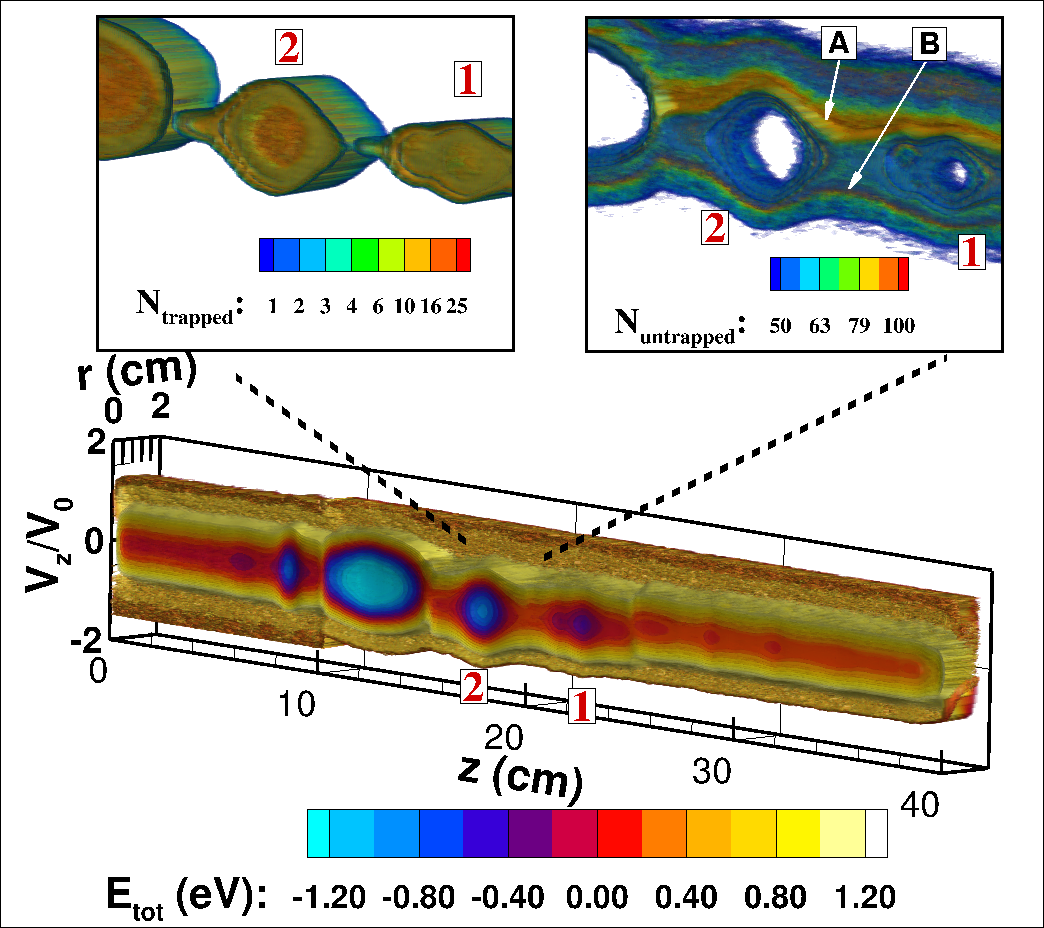}
\caption{ Trapped and untrapped number of particles in ($r,z,v_z$) phase space based on the average total energy per particle, $E_{\rm tot}$, of electrons. For trapped particles, $E_{\rm tot}<$0 and $v_0 = 1\times10^6$ m/s. In the inset images, the white space signifies the absence of computational particles in the phase space. \textcolor{black}{In the RHS inset figure, `A' and `B' indicate the difference in electron particle densities in the two opposing streams, which indicates an asymmetric non-Maxwellian EVDF.}  Note that the number of computational particles are divided into the 800 bins in $z$ direction, 30 bins in $r$ direction and 60 bins in $v_z$ direction.}
\label{fig:Trapped_untrapped_Etot_3D_phase}
\end{figure}


\textcolor{black}{ For the 3D cylindrical beam case witth $a = 2.5$ mm, however, both the electron holes near the electron source and the ESWs (i.e., solitary waves marked as `1' and `2') have 3D oblong shapes that are elongated along the beam axis, as shown in Fig. \ref{fig:3D_iso_phi_thinbeam}.} Similar to the 2D planar beam, the amplitude ($\Delta \phi$) of ESWs in the 3D beam is directly correlated with the fraction of electrons that are trapped. However, the electric potential structure is very different for the ESWs in the 3D cylindrical beam. Figure \ref{fig:3D_iso_phi_thinbeam} shows iso-potential surfaces in 3D where the beam has been cut in half along the $x = 1.5$ cm mid-plane. This electric potential is different from the 2D planar beam case where the electric potential profile was wide and flat near the beam axis (see the 2D curve in Fig. \ref{fig:3D_i2D_prof_comp}), thus enabling a 1D BGK analysis in that case. However, for the 3D cylindrical beam case, the electric potential rapidly declines with radial distance, as shown in Fig. \ref{fig:3D_i2D_prof_comp}. Since the electron source was placed along the axis of the ion beam, the beam structure in the 3D cylindrical beam case is perfectly axisymmetric, as shown by the comparison of the electric potential profiles at $z = 20$ cm along the $x$ and $y$ transverse directions in Fig. \ref{fig:3D_i2D_prof_comp}. Therefore the ESWs in the 3D cylindrical beam case can be analyzed in a $(r,z)$ cylindrical coordinates system.

 Similar to the 2D planar beam analysis discussed above, we can evaluate the trapped and untrapped electron particle populations along the beam by calculating the total energy of the electrons using all three velocity components. The average total electron energy, $E_{\rm tot}$, for an axisymmetric ESW in the 3D beam is given by, 
\begin{align}\label{eq:Etot_3D}
    E_{\rm tot}(r,z,v_z) = \frac{\sum_{i=1}^{N_{\rm total}}\left[\frac{1}{2}m_e|(\vec{v_{e_i}}-\vec{v_s})|^2 - e\phi\right]\delta_{v_{e_i}v_z}\delta_{z_{e_i}z} \delta_{r_{e_i}r} }{\sum_{i=1}^{N_{\rm total}}\delta_{v_{e_i}v_z}\delta_{z_{e_i}z} \delta_{r_{e_i}r} },\\
    \text{where } |(\vec{v_{e_i}}-\vec{v_s})| = \sqrt{v_{e_{ix}}^2+v_{e_{iy}}^2+(v_{e_{iz}}-v_s)^2}, \nonumber 
\end{align}
 and $\delta$ is the Kroeneker delta function and $r_{e_i}$ is the position of the $i^{th}$ electron in the domain.  
 
 Figure \ref{fig:Trapped_untrapped_Etot_3D_phase} shows the average total energy of electrons, $E_{\rm tot}$, per electron particle in $(r,z,v_z)$ phase space where the average electron energy is negative in the regions where an ESW is present, i.e. electrons are trapped. In the inset of Fig. \ref{fig:Trapped_untrapped_Etot_3D_phase}, the trapped and untrapped particle populations are shown in the same $(r,z,v_z)$ space based on the total electron energy, i.e. term in square brackets of Eq. \ref{eq:Etot_3D}, being negative and positive, respectively for the ESW labeled as `2'.   
\textcolor{black}{In the untrapped population shown in the RHS inset of Fig. \ref{fig:Trapped_untrapped_Etot_3D_phase}, we can see that a larger population of electrons have velocities $v_z > 0$ (shown by `A') than $v_z < 0$ (shown by `B'), because of the non-Maxwellian $z$-direction electron distribution, similar to the 2D planar beam EVDF that is shown in Fig. \ref{fig:2D_soliton_t39000}. }
The ESW `2' in Fig. \ref{fig:Trapped_untrapped_Etot_3D_phase} for the 3D beam is about $73\lambda_{\rm D}$ in length, which is slightly longer than the ESWs in 2D beam, as listed in Tab. \ref{tab:2D_solitions}. The ESW `1' in Fig. \ref{fig:Trapped_untrapped_Etot_3D_phase} has a \textcolor{black}{larger length} than ESW `2' because of the ongoing merging between the two ESWs shown in Fig. \ref{fig:time_snaps_3A} at $t = 3.44 \ \mu s$.

\begin{figure}[h]
\centering
\subfigure[EVDF at far-field region and at the center of ESW~2.]{\label{fig:EVDF_3D_CHAOS}
        \includegraphics[trim = 0.14cm 0.14cm 0.14cm 0.14cm, clip,width = 0.48\textwidth]{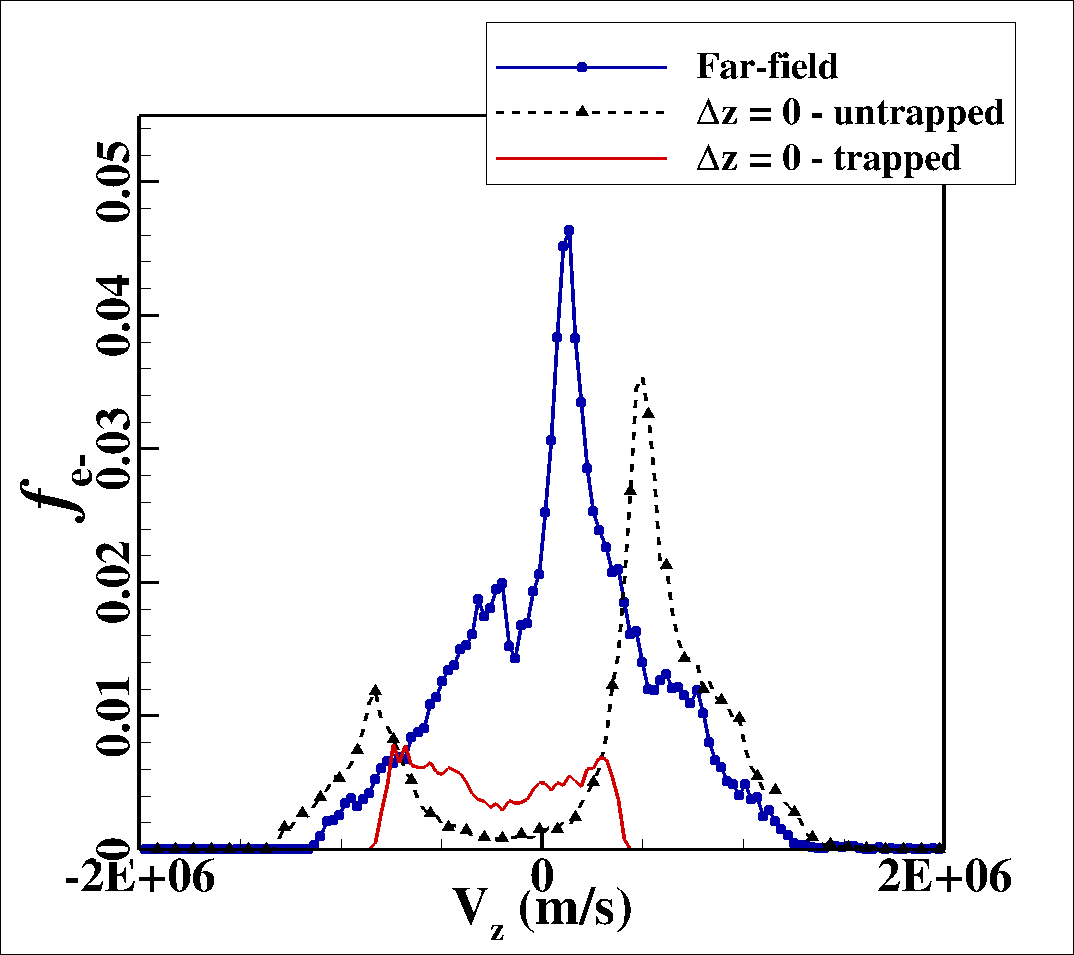}}
\subfigure[Trapped and untrapped electrons in ESW~2.]{\label{fig:BGK_CHAOS_3D}
        \includegraphics[trim = 0.14cm 0.14cm 0.14cm 0.14cm, clip,width = 0.48\textwidth]{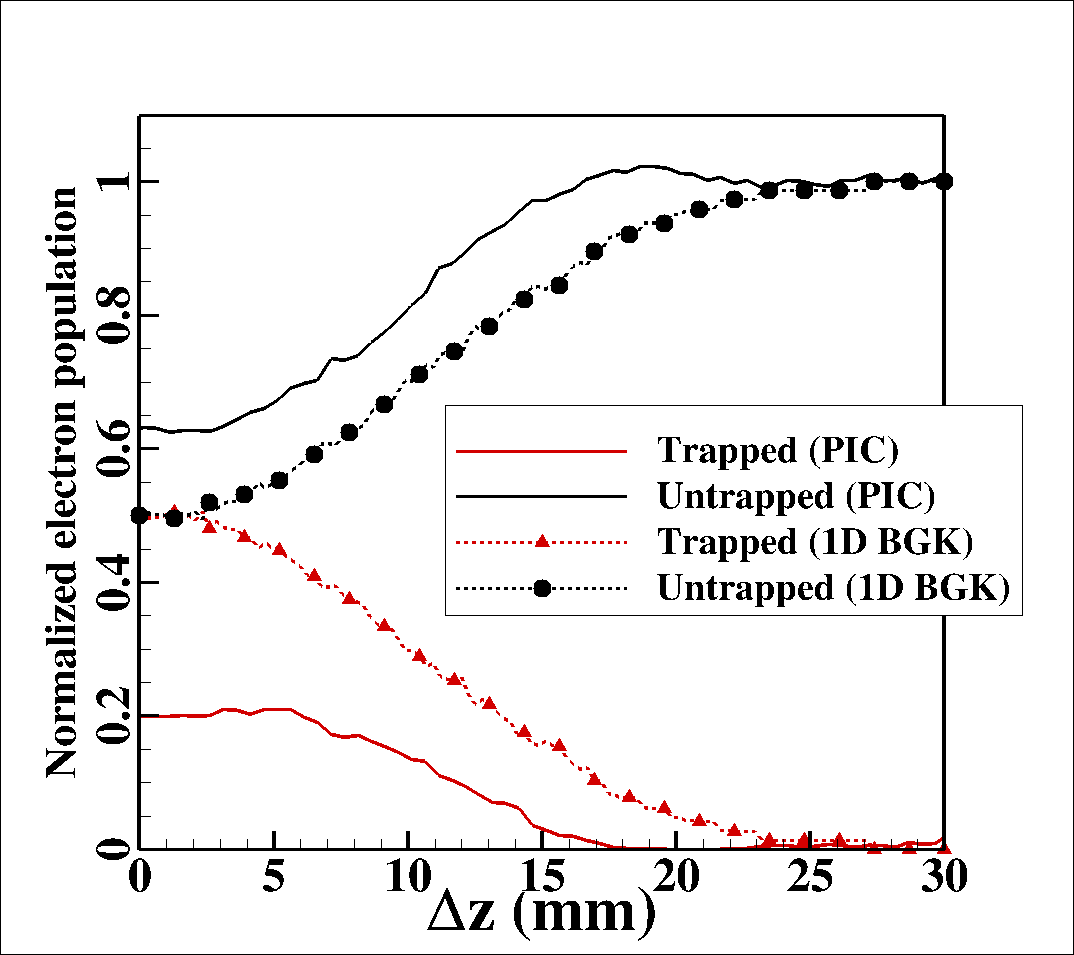}}        
\caption{(a) Electron velocity distribution of ESW 2 from PIC result and (b) trapped $vs$ untrapped electron population along the ESW. In (a), there is a non-zero population of untrapped electrons between $v_z = -1\times 10^6$ and $5\times10^5$ m/s as these electrons have high $v_x$ and $v_y$ values. In (a), the EVDF is normalized by $n_0 = 1.75\times10^{14} \ {\rm m^{-3}}$. In (b), the densities are normalized by the electron density in the far-field of ESW~`2' of 3D beam, i.e. $1.4\times10^{14} \ {\rm m^{-3}}$.  }
\label{fig:3D_BGK_analysis}
\end{figure}

\begin{figure}[h]
\centering
\subfigure[Trajectory of an electron trapped in ESW 2.]{\label{fig:trapped_traj}
        \includegraphics[trim = 0.14cm 0.14cm 0.14cm 0.14cm, clip,width = 0.48\textwidth]{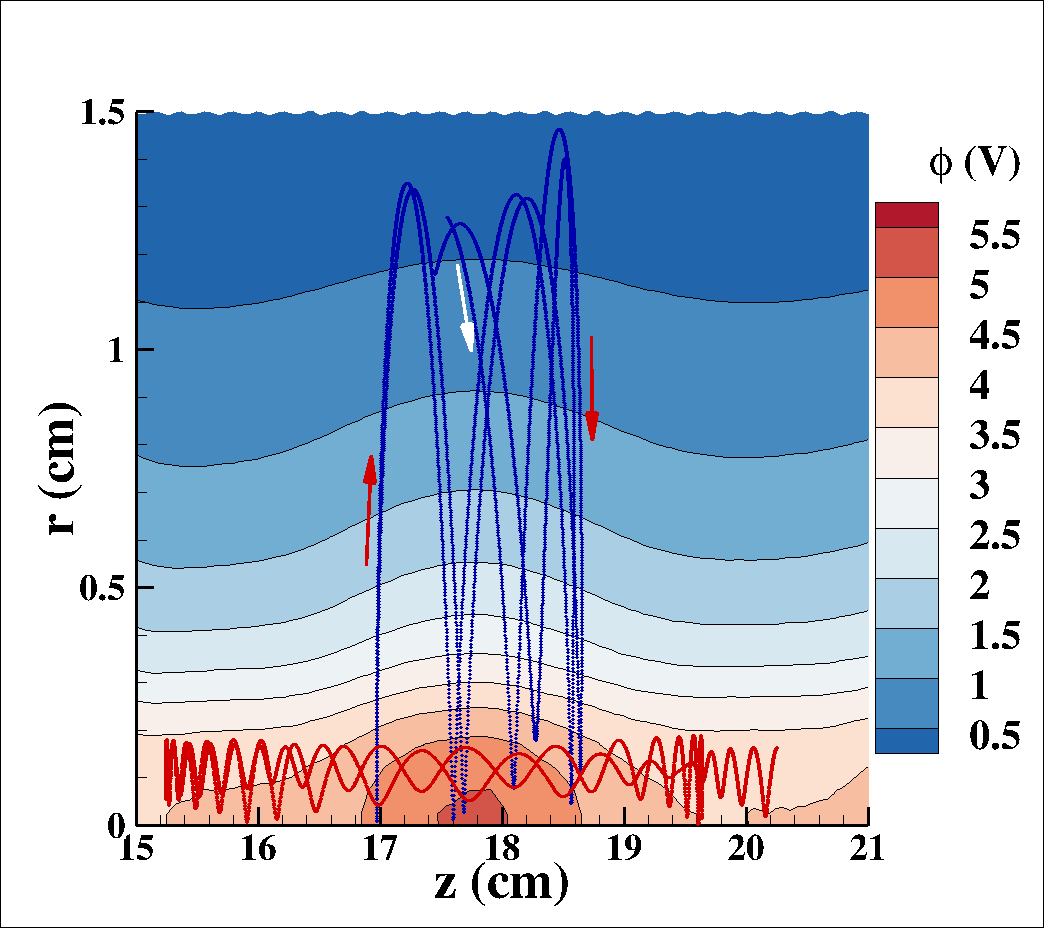}}
\subfigure[Trajectories of electrons not trapped in ESW 2]{\label{fig:untrapped_traj}
        \includegraphics[trim = 0.14cm 0.14cm 0.14cm 0.14cm, clip,width = 0.48\textwidth]{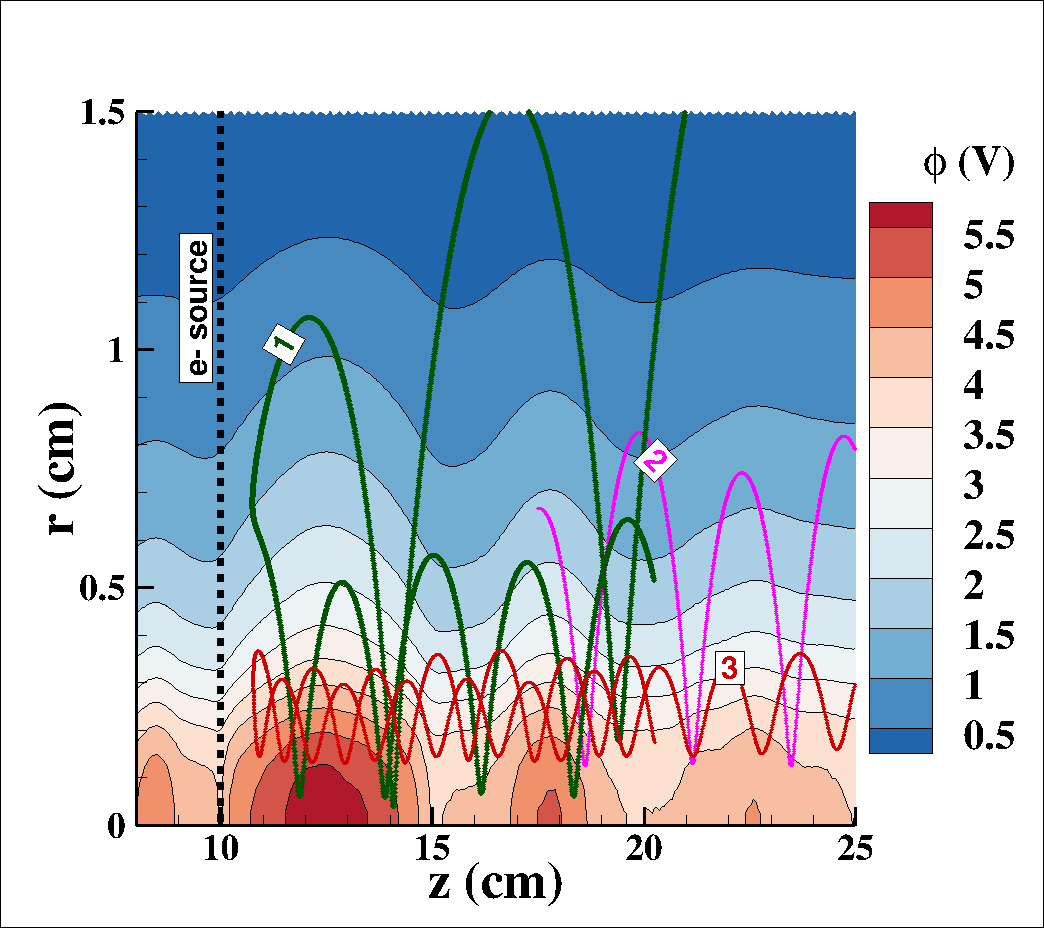}}        
\caption{Examples of electron trajectories which are (a) trapped and (b) untrapped in ESW 2 (15 $< z <$ 20 cm) superimposed over the electric potential in cylindrical coordinates at $t = 3.44 \ \mu s$ for the 3D cylindrical beam case with $a = 2.5$ mm. In (a) the white arrow indicates the starting of the traced electron trajectory and the red arrows indicate its direction when reflected from the ESW potential-well boundaries at $z = 17$ and 18.5 cm. In (b), traced electrons `1' and `3' are reflected from the virtual cathode, where `1' leaves from conducting boundary at $r = 1.5$ cm, while `3' keeps bouncing between virtual cathode and the ion sheath at $z = 40$ cm. Traced electron `2' leaves the domain from the $z = 40$ cm conducting boundary condition.      }
\label{fig:trapped_untrapped_traj}
\end{figure}

Applying the same 1D analysis to the electrons in the 3D-cylindrical beam, we use the non-Maxwellian far-field distribution, $f_{\rm FF}$, for the electrons, shown in Fig. \ref{fig:EVDF_3D_CHAOS}, obtained from the PIC result of the $a = 2.5$ mm 3D beam case. The $f_{\rm FF}$ (far-field) shown in Fig. \ref{fig:EVDF_3D_CHAOS} was obtained by binning all the electrons that lie in the $z = 20$ cm plane, i.e. outside the ESW `2' of Fig. \ref{fig:3D_iso_phi_thinbeam}. Using the electric potential profile along the beam axis, we obtain the untrapped electron population profile, $n_{ut}(\Delta z)$, using Eq. \ref{eq:untrapped_EVDF_selfsimilar}, as shown in Fig. \ref{fig:BGK_CHAOS_3D}, which can be seen to not agree with the PIC result. The trapped electron population also shows disagreement, and its value is over-predicted by the 1D BGK analysis when compared to the PIC result, as shown in Fig. \ref{fig:BGK_CHAOS_3D}. This difference is due to the fact that the 1D BGK analysis makes the assumption that for all the electron trajectories in the ESW, the electric potential profile along the trajectory path of an electron is the same as it is along the beam axis of the 3D cylindrical beam. However, even though the electric potential of ESW `2' is axisymmetric, the electron trajectories are quite complicated, as can be seen in Fig. \ref{fig:trapped_untrapped_traj}. For example, in Fig. \ref{fig:trapped_traj}, the trajectories of two electrons show that they cross several different iso-potential lines when trapped in an axisymmetric potential-well and are not parallel to the beam axis. The same is true for the untrapped electrons, shown by three example trajectories in Fig. \ref{fig:untrapped_traj}, where all three labeled electrons cross multiple iso-potential lines. In Fig. \ref{fig:untrapped_traj}, the electrons labeled `1', and `3' are reflected from the virtual cathode which forms at the electron source at $z = 10$ cm while electron `2' has enough kinetic energy to eventually leave from the $z = 40$ cm boundary. Note that since these trajectories in Fig. \ref{fig:trapped_untrapped_traj} are taken in a static potential field at $t = 3.44 \ \mu s$, they represent the constant total energy lines of electrons in the 3D beam at that time. Other approaches such as that of Chen et al.\cite{chen2002bgk} and Eliasson et al.\cite{eliasson2007theory}, which take into account the radial changes in electric potential profiles,  also assume that the electron trajectories are parallel to the beam axis, which, as we show in Fig. \ref{fig:trapped_untrapped_traj}, is not true for a non-magnetized beam.  For electrostatic plasma cases, such as ours, a fully 3D-3V BGK approach would have to be used to account for the 3D effects in a beam. Although spherical electrostatic ESWs have been studied theoretically in Ng et al.\cite{ng2005bernstein} for a canonical case with a Maxwellian far-field EVDF, a general theory for 3D is yet to be developed to understand the ESWs that are seen in our 3D beam PIC simulations.

\section{Excitation of surfaces waves in a plasma beam}\label{sec:surface_waves}
\subsection{Trivelpiece-Gould surface waves in 3D cylindrical beams}
\begin{figure}[h]
\centering
\subfigure[TG waves in a 3D beam with $a = 5.0$ mm.]{\label{fig:5mm_case_waves}
        \includegraphics[trim = 0.14cm 0.14cm 0.14cm 0.14cm, clip,width = 0.48\textwidth]{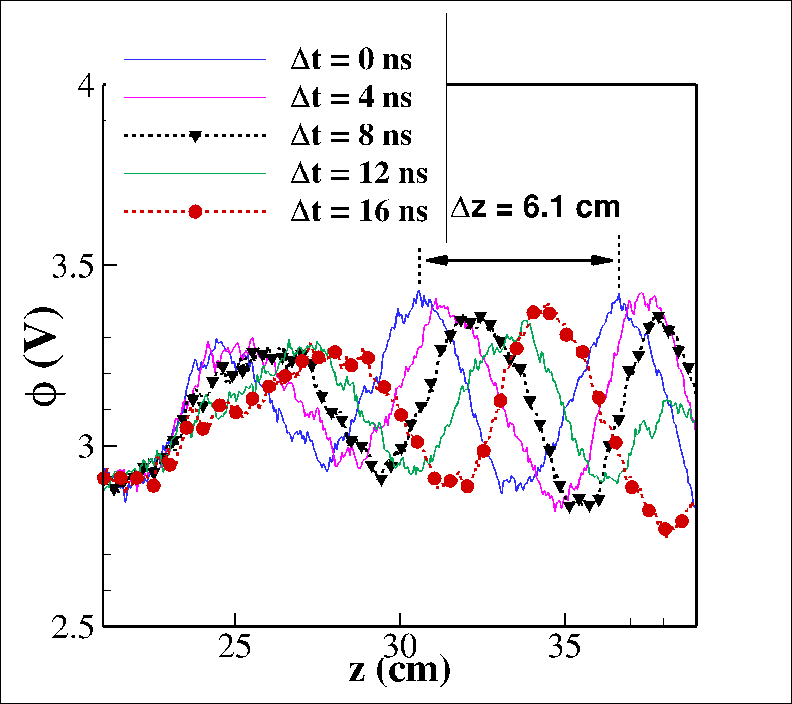}}
\subfigure[Dispersion curve for TG\cite{trivelpiece1959space} waves.]{\label{fig:TG_dispersion}
        \includegraphics[trim = 0.14cm 0.14cm 0.14cm 0.14cm, clip,width = 0.48\textwidth]{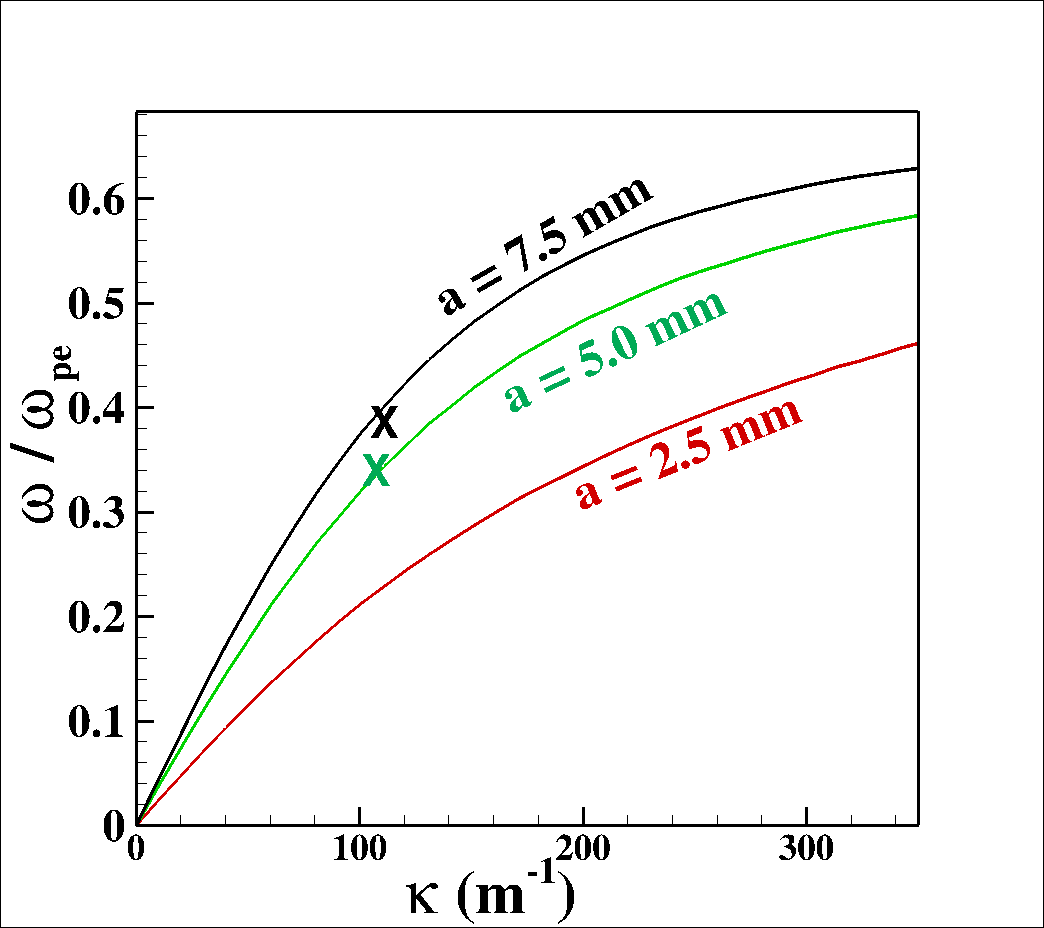}}        
\caption{Long wavelength surface waves traveling along the beam axis in a cylindrical beam. (a) The time evolution of the electric potential profile ($\Delta t = 0$ corresponds to $t = 4.72 \ \mu s$ from the start of the PIC simulation, at the end of rapid neutralization) shows waves with $v_\phi = 2.42\times10^6$ m/s, and wavelength, $\lambda_w$ = 6.1 cm. In (b), the curves show the theoretically calculated phase speed from dispersion relation of axisymmetric TG surface waves (Eq. \ref{eq:TG}); `X' marks shows the PIC simulation ($\omega-\kappa$) results for the phase velocity for different beam radii. }
\label{fig:TG_dispersion_all}
\end{figure}

As was shown in the previous section, the geometry of the plasma has important effects on the types of electron-induced instabilities that may arise in a plasma beam. For the 3D cylindrical beam cases with beam radius $a = 5.0$ and $7.5$ mm, as the beam potential decreases and the electrons fill in the cylindrical ion beam, long-wavelength waves appear, which travel along the beam axis with high phase speeds. Figure \ref{fig:5mm_case_waves} shows a wave of long wavelength, $\lambda_w = $ 6 cm, traveling at a phase speed of $v_{\phi} = 2.42\times 10^6$~m/s in the $+z$ direction with a frequency of $\omega = \kappa v_{\phi} = 2.49 \times 10^8$ rad/s, where $\kappa = 2\pi/\lambda_w$ is the wavenumber, observed in the simulations. The geometry of the beam, the low frequency ($\omega/\omega_{pe} = 0.33$), and the long wavelength of this wave are typical of a Trivelpiece-Gould (TG) surface wave\cite{trivelpiece1959space}, commonly found in electron beams in cylindrical cavities\cite{moisan1982experimental}.    

\begin{figure}[h]
\centering
\subfigure[Time evolution of electric potential at $z = 20$ cm along beam axis.]{\label{fig:Phi_z_waves}
        \includegraphics[trim = 0.14cm 0.14cm 0.14cm 0.14cm, clip,width = 0.48\textwidth]{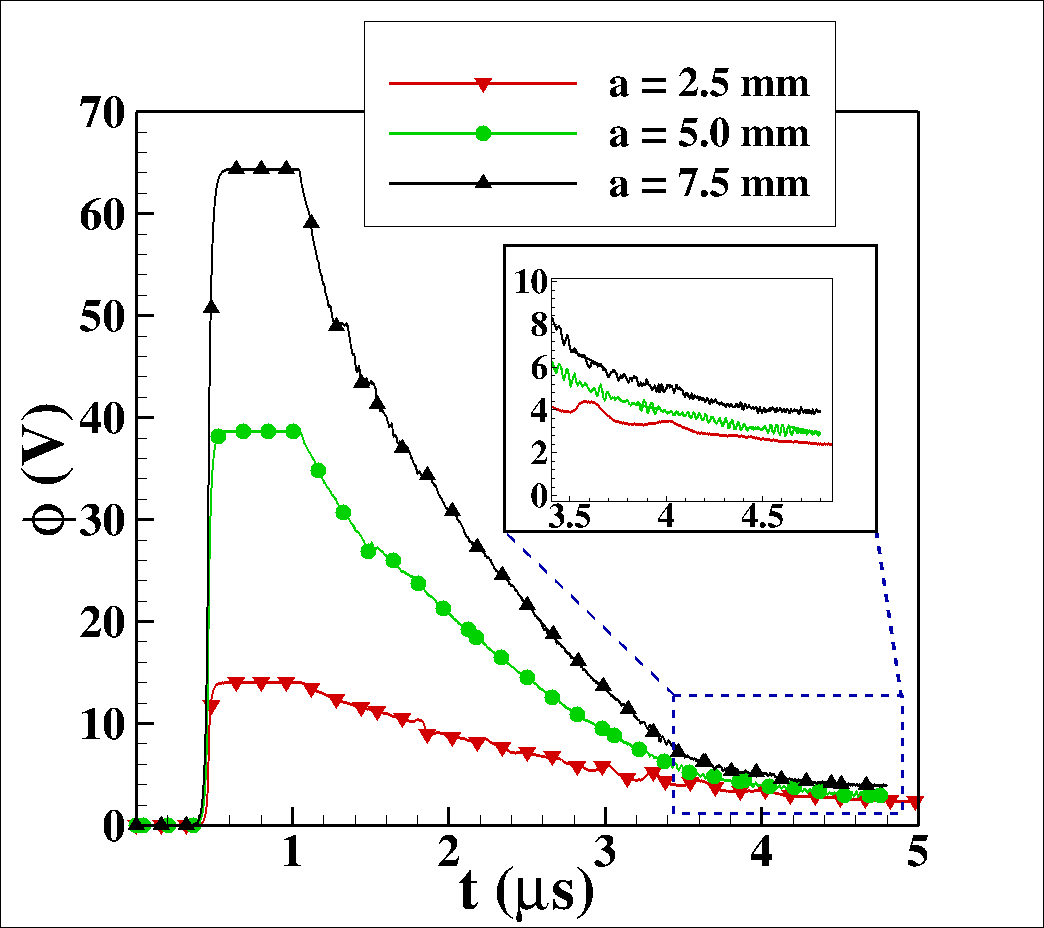}}
\hspace{0.3mm}        
  \subfigure[EVDF vs $v_\phi$ for different beam radius.]{\label{fig:EVDF_TG_vphi_comp}
        \includegraphics[trim = 0.14cm 0.14cm 0.14cm 0.14cm, clip,width = 0.48\textwidth]{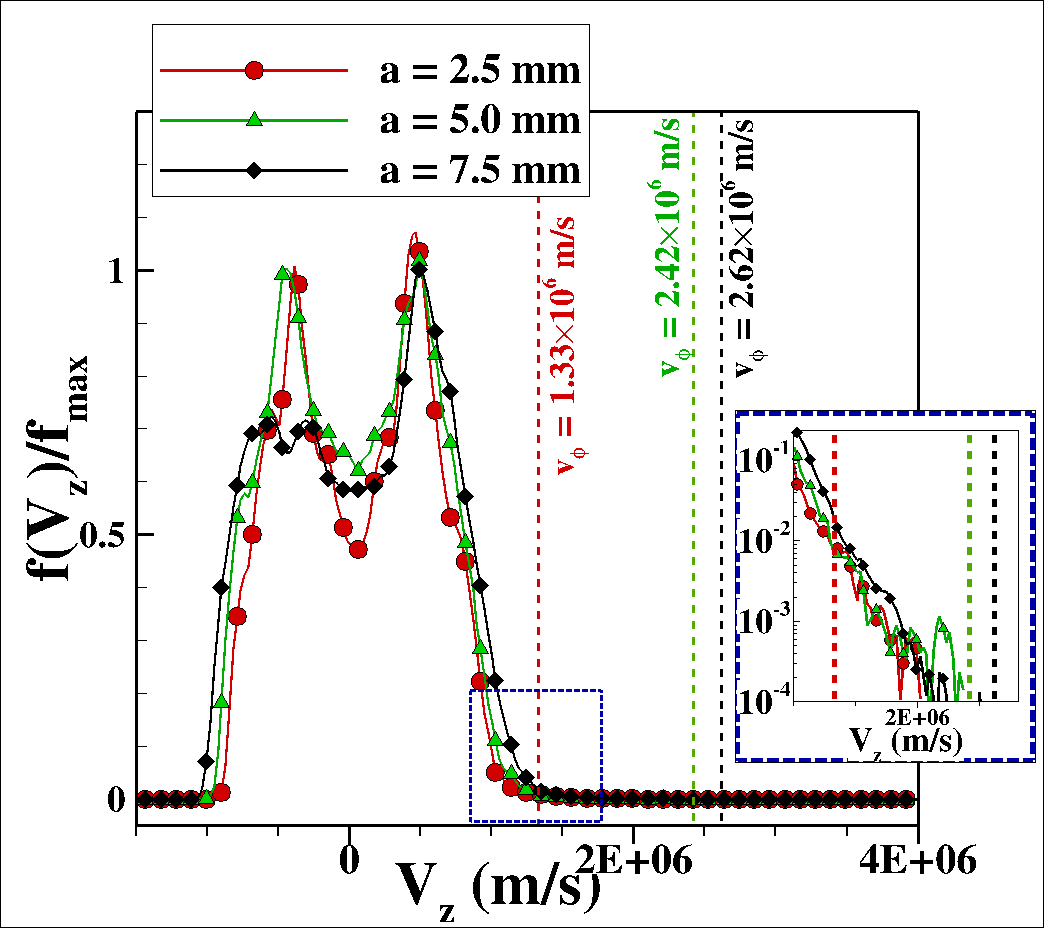}}
\caption{ (a) Beam neutralization vs beam radius, $a$, for the 3D cylindrical beam cases. In (b), EVDFs are shown at $z = 30$ cm and $t = 4.72 \ \mu s$ for all cases. In (b), the inset figure has a logarithmic $y$ axis. }
\label{fig:3D-B_timeevol}
\end{figure}

The $\omega-\kappa$ dispersion relation for a surface wave in a cylindrical cold electron beam with no external magnetic field, first derived by Trivelpiece and Gould (TG)\cite{trivelpiece1959space}, is given by,
\begin{align}\label{eq:TG}
    \epsilon_1  \frac{J_n'({\rm i} \kappa a)}{J_n({\rm i} \kappa a)} = \epsilon_0  \frac{I_n'(\kappa a)K_n(\kappa b) - I_n(\kappa b)K_n'(\kappa a)}{I_n(\kappa a)K_n(\kappa b) - I_n(\kappa b)K_n(\kappa a)},
\end{align}
where $J_n$ is the $ n^{\rm th}$ order Bessel function of the first kind and $I_n$ and $K_n$ are the modified Bessel functions of the first and second kind, respectively, $\epsilon_1 = \epsilon_0(1-\omega_{pe}^2/\omega^2)$ is the dielectric constant of the plasma medium, $a$ and $b$ are radii of the plasma beam and the cylindrical conductor channel, respectively, and $n$ is the azimuthal ($\theta$ direction) index of the electric potential perturbation\cite{trivelpiece1959space}. Figure \ref{fig:TG_dispersion} shows a comparison of the axisymmetrical ($n=0$) TG wave dispersion relation of Eq. \ref{eq:TG} with the long-wavelength wave for the $a=5.0$ mm case. Similarly for the case with $a = 7.5$ mm, these long-wavelength waves show up near $t = 4 \ \mu s $, as shown by high frequency fluctuations in Fig. \ref{fig:Phi_z_waves}. For the $a = 7.5$ mm case, the long-wavelength waves appear at $\omega/\omega_{pe} = 0.37$ and $\lambda_w = 5.94$ cm (see blue `X' in Fig.~\ref{fig:TG_dispersion}). The surface wave properties of these two cases are listed in Tab. \ref{tab:Summary_of_all_case_results}. The TG waves have also been observed in experiments by Lynov et al.\cite{lynov1979observations}, and Moisan et al.\cite{moisan1982experimental}, where these waves were excited by pulsing the plasma column in a cylindrical TG wave-guide cavity.  However, no experimental evidence which shows their spontaneous excitation during beam neutralization has been reported in the literature, to the best of our knowledge. 

\begin{figure}[h]
\centering
\subfigure[$a = 5.0$ mm.]{\label{fig:TG_excite_mid}
        \includegraphics[trim = 9cm 0.14cm 10cm 0.14cm, clip,width = 0.475\textwidth]{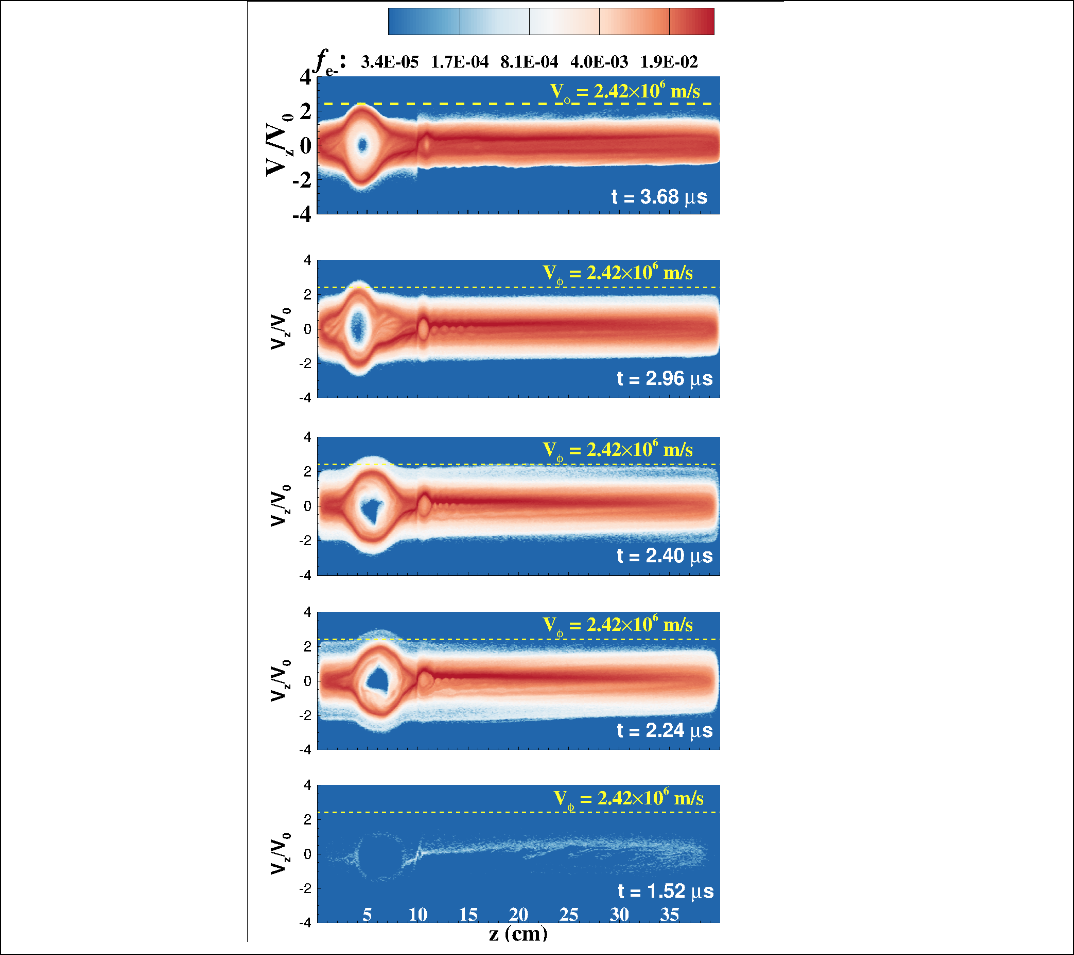}}
\hspace{0.3mm}        
  \subfigure[$a = 7.5$ mm.]{\label{fig:TG_excite_thick}
        \includegraphics[trim = 9cm 0.14cm 10cm 0.2cm, clip,width = 0.48\textwidth]{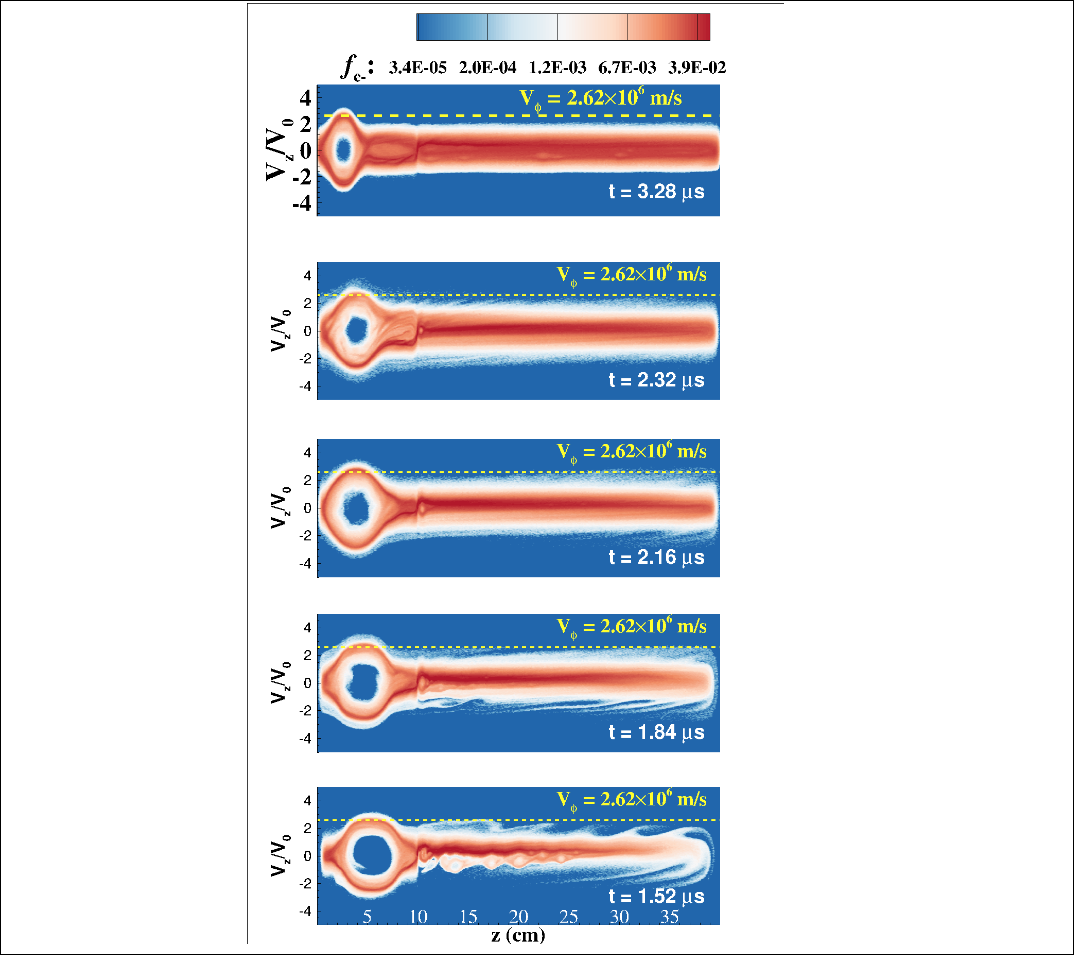}}
\caption{ Time evolution of electrons in phase-space extracted in the $x = 1.5$ cm mid-section plane of a 3D cylindrical beam with $a = 5.0$ and $7.5$ mm. In (a) and (b), the yellow dashed line represents the phase speed of the surface waves found in their respective cases. Here, $v_0 = 1\times10^6$ m/s.}
\label{fig:TG-excitation}
\end{figure}

The surface wave phase speed, $v_{\phi}$, is higher than the local electron velocities, as shown in Fig. \ref{fig:EVDF_TG_vphi_comp} by the EVDF in the region $z>25$ cm for the cases with $a = 5.0$, and 7.5 mm. From the EVDFs shown in Fig. \ref{fig:EVDF_TG_vphi_comp}, it appears that these waves become excited by the high energy electrons that are present at the beginning of beam neutralization, starting from $t = 1.0 \ \mu s$, when the beam potential is high, causing the electrons to accelerate to high velocities. \textcolor{black}{This acceleration is enhanced by the large sized ESWs, which oscillate upstream of the electron source for both the $a = 5.0$ and 7.5 mm cases, as was shown in Fig. \ref{fig:3D-BC_timeevol}. } \textcolor{black}{This oscillation of upstream ESWs is also shown in phase space in Figs. \ref{fig:TG_excite_mid} and \ref{fig:TG_excite_thick} between $t = 1.52$ and $3.68 \ \mu s$ and $t = 1.52$ and $3.28 \ \mu s$ for the $a = 5.0$ mm and 7.5 mm cases, respectively. For both cases, we can see that the large ESWs accelerate the electrons coming from the electron source at $z = 10$ cm. These electrons then gain velocities close to $2.42\times10^6$ and $2.62\times10^6$ m/s in the downstream region of the electron source at $t = 2.40 \ \mu s$ and $2.16 \ \mu s$, respectively for the $a = 5.0$ and 7.5 mm cases, as shown in Figs. \ref{fig:TG_excite_mid} and \ref{fig:TG_excite_thick}. These high-speed electrons then excite the surface waves seen in both these cases at later times with the respective phase speeds of $2.42\times10^6$ and $2.62\times10^6$ m/s for the $a = 5.0$ and 7.5 mm cases.} \textcolor{black}{After the high energy electrons create surface waves, these electrons leave the beam as the beam potential decreases with time. This leaves the beam with insufficient number of electrons required to resonate and dampen these surface waves by Landau damping, as shown by the EVDF and $v_\phi$ comparisons in Fig. \ref{fig:EVDF_TG_vphi_comp}.} Therefore, these waves survive and can be seen as the beam potential falls below 3.5 and 4.0 V, respectively, for the $a = 5.0$ and 7.5 mm cases. However, these waves do not seem to be excited for the case with $a = 2.5$ mm because the phase speed of the TG wave corresponding to a wavelength of $\lambda_w = 6$ cm is about $1.33\times10^6$ m/s, which is sufficiently small to be Landau damped by the particles in the tail of the electron distribution function, as shown by the EVDF comparison with $v_\phi$ in the inset of Fig. \ref{fig:EVDF_TG_vphi_comp}. 

\subsection{Dispersion relation of surface waves in 2D planar {\it vs} 3D cylindrical electron beams}

\begin{figure}[h]
\centering
\subfigure[2D plasma beam of width $2a$ in a channel of width $2b$.]{\label{fig:2D_schematic_a_b}
        \includegraphics[trim = 0.14cm 0.14cm 0.14cm 0.14cm, clip,width = 0.48\textwidth]{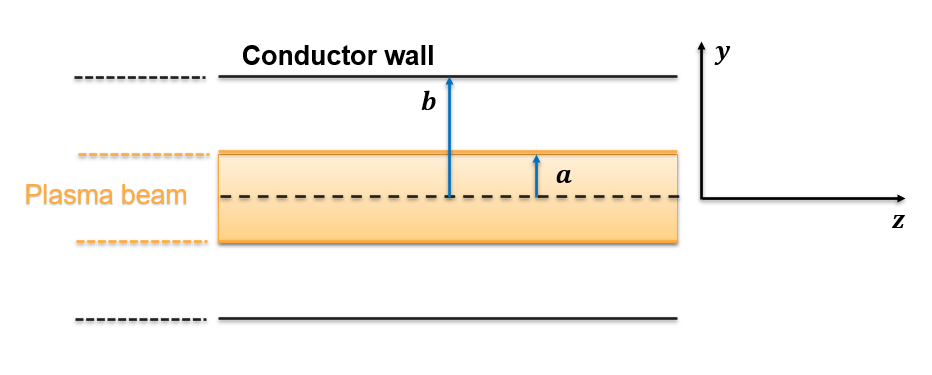}}
    \hspace{0.3mm}         
\subfigure[2D Slab Dispersion relation for different values of $a$ and $b$.]{\label{fig:2D_diap_a_b}
        \includegraphics[trim = 0.2cm 0.2cm 0.2cm 0.2cm, clip,width = 0.48\textwidth]{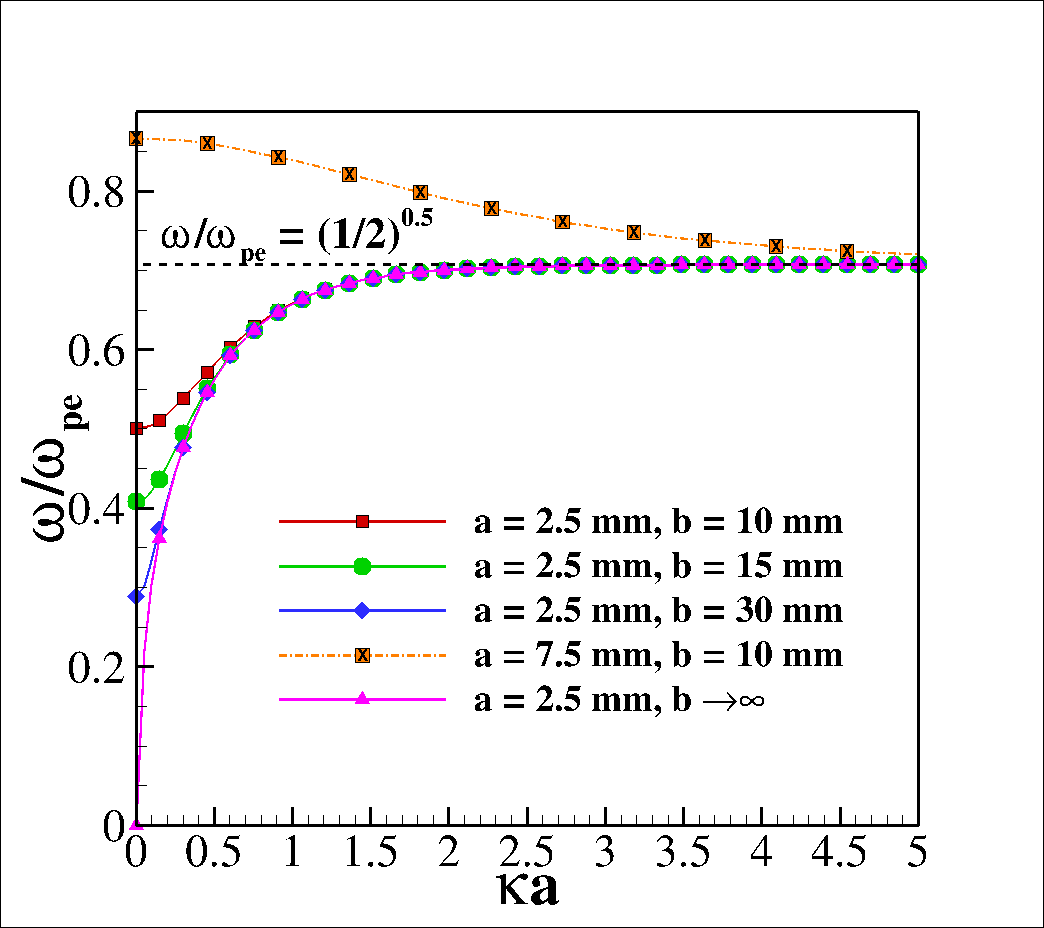}}
\caption{ 2D surface wave dispersion relation for plasma beam with different plasma widths, $2a$, and channel widths, $2b$.}
\label{fig:2D-dispersions}
\end{figure}

The question arises why these long wavelength surface waves do not appear in the 2D planar beam simulation. A fully neutralized 2D planar electron beam in a channel has a geometry similar to a plasma slab which lies between two conducting surfaces, as shown in Fig. \ref{fig:2D_schematic_a_b}, where, $a$ and $b$ are the half-widths of the plasma beam and the channel, respectively. Following a procedure similar to Vedenov\cite{vedenov1965solid}, Krall\cite{krall1973principles}, and Trivelpiece-Gould\cite{trivelpiece1959space}, it can be shown that a cold-electron beam assumption gives, 
\begin{align}\label{eq:2D_dispersion}
    \frac{\omega}{\omega_{pe}} = \frac{1}{\left[1+ \frac{\tanh{[(b-a)\kappa]}}{\tanh{a\kappa}}\right]^{1/2}},
\end{align}
where $\omega_{pe} = \sqrt{n_ee^2/m_e\epsilon_0}$ is the electron plasma frequency of the beam. Further details are discussed in Appendix \ref{sec:2D_surface_app}. As seen from Eq. \ref{eq:2D_dispersion}, the 2D surface waves depend on the width of the beam and channel, where all the geometric parameters converge to $\omega/\omega_{pe} = 1/\sqrt{2}$ for high values of wavenumber, $\kappa$, as shown in Fig. \ref{fig:2D_diap_a_b}. Most notably, $\omega/\omega_{pe} \propto \sqrt{\kappa}$ for small values of $\kappa$ and $a << b$, i.e. a thin film surface wave, and $\omega/\omega_{pe} \to 1/\sqrt{2}$ for $a\to\infty, \ b\to \infty$, and $a<<b$, i.e. surface wave with no conducting walls\cite{vedenov1965solid}. 

A comparison of the 2D planar and 3D cylindrical surface wave dispersion relation is shown in Fig. \ref{fig:2D_3D_disp_comp}, where it can be seen that the 2D surface waves have a much higher frequency than the 3D cylindrical waves of the same wavenumber. Hence, for surface waves to be excited in a 2D planar beam, the electrons would need to reach much higher velocity values than they would be required to excite a 3D cylindrical surface wave. For example, to excite a surface wave of $\lambda_w = 6$ cm in a 2D planar beam of beam half-width $a = 2.5$ mm, the required phase speed is $v_{\phi} = 3.23\times10^6$ m/s, which is about 2.5 times that required for the corresponding 3D cylindrical beam of beam radius $a = 2.5$ mm. Since electrons are injected with $T_e = 2$ eV they do not reach such high velocities in our 2D planar beam case at the time when they acquire the shape of a planar beam, and therefore we do not observe surface waves in our 2D numerical simulation results.

\subsection{Excitation of surface plasma waves by electrons in the tail of the EVDF}

\begin{figure}[h]
\centering
\subfigure[2D $vs$ 3D dispersion relation.]{\label{fig:2D_3D_disp_comp}
        \includegraphics[trim = 0.8cm 0.14cm 0.14cm 0.14cm, clip,width = 0.6\textwidth]{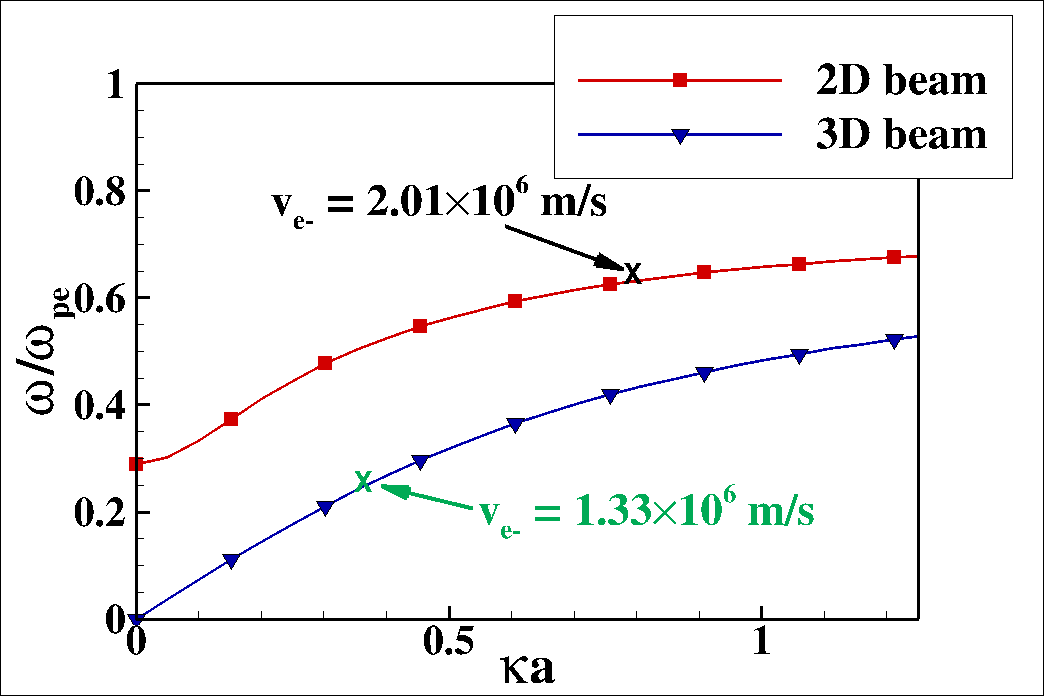}}
\hspace{0.3mm}             
  \subfigure[Artificial excitation in 2D planar beam.]{\label{fig:2D_excitation}
        \includegraphics[trim = 0.2cm 0.2cm 0.2cm 0.2cm, clip,width = 0.48\textwidth]{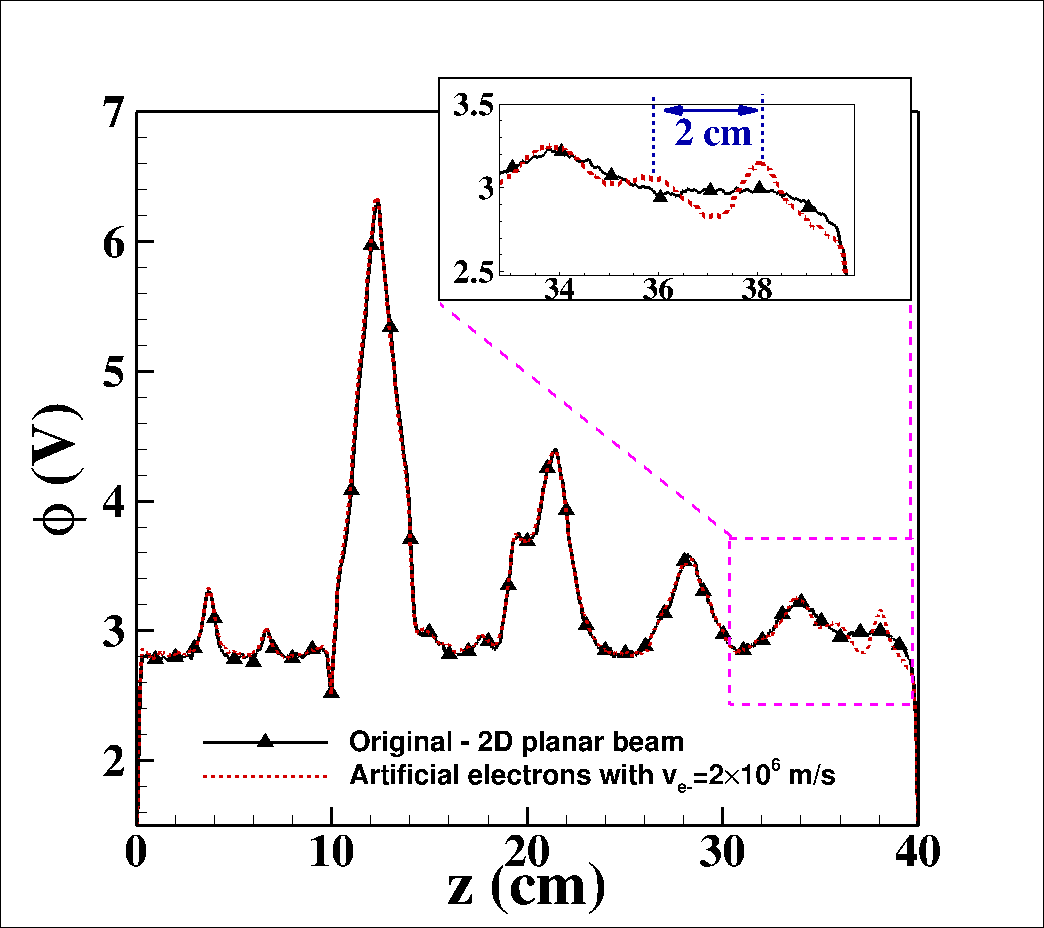}}   
    \subfigure[Artificial excitation in 3D cylindrical.]{\label{fig:3D_excitation}
        \includegraphics[trim = 0.2cm 0.2cm 0.2cm 0.2cm, clip,width = 0.48\textwidth]{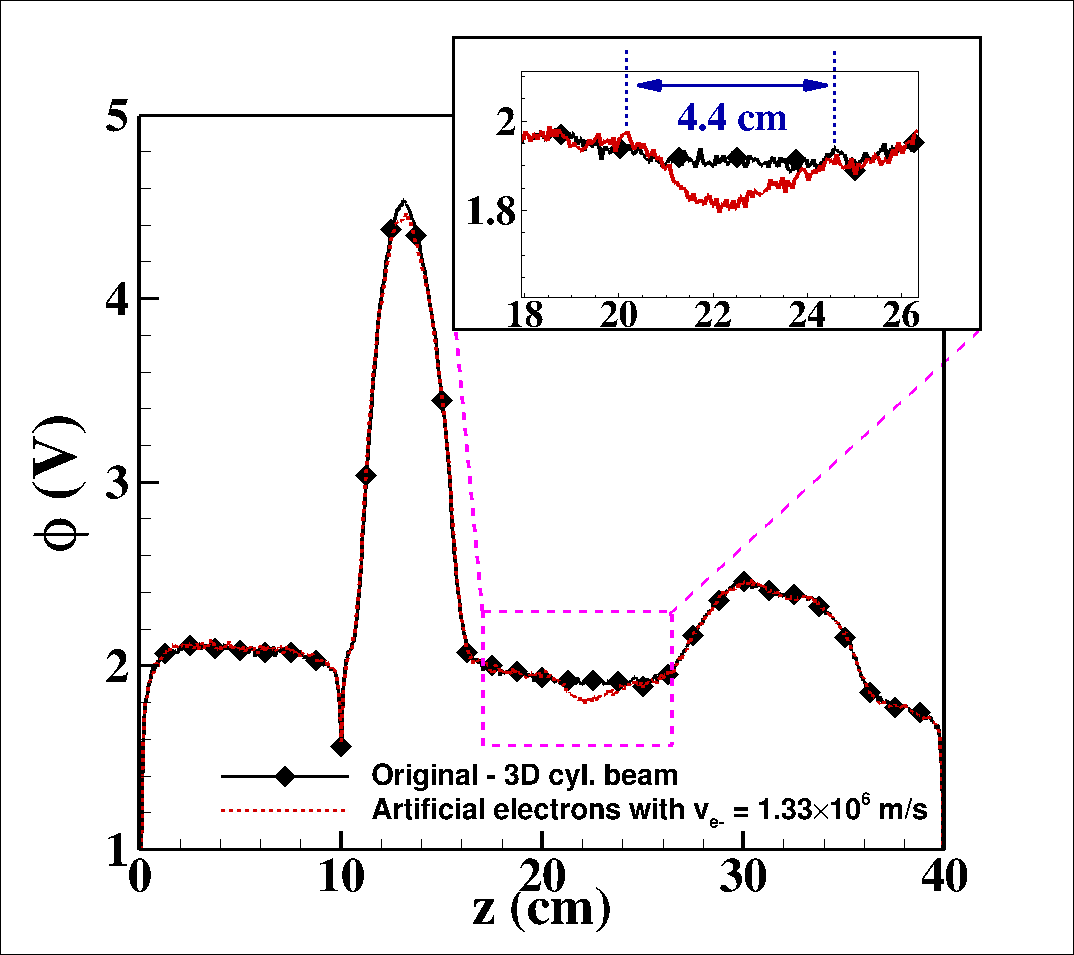}}  
\caption{ (a) 2D planar (Eq. \ref{eq:2D_dispersion}) vs 3D cylindrical (Eq.~\ref{eq:TG}) surface wave dispersion relation and excitation of surface waves by tail electrons in (b) 2D and (c) 3D beams with $a = 2.5$ mm. In (b) and (c), original result is shown in black and the artificially perturbed result in red line.}
\label{fig:2D-dispersions_2}
\end{figure}

Since the surface waves for the 3D cylindrical beams were found to be excited by high energy electrons, we were curious to see whether we could excite them in the 2D planar beam case by introducing a small fraction of 3\% of electrons with a specific velocity. To do this, we restart our 2D planar beam PIC simulation from $3.92 \ \mu s$, when the beam is near the end of the rapid neutralization phase (see Fig. \ref{fig:Probe_z20cm})  and randomly initialize 3\% of the electrons entering from the source at $z = 10$ cm with a specific high velocity of $2.01 \times 10^6$ m/s. Figure \ref{fig:2D_excitation} shows a comparison of the electric potential profiles of the 2D planar beam cases with and without the artificial excitation of surface waves by the high energy electron tail of $v_e = 2.01\times10^6$ m/s at $t = 4.08 \ \mu s$. The excited surface wave (dashed red curve) has a wavelength of 2.0 cm and a phase speed of $v_\phi \approx 2.01\times10^6$ m/s. From this, $\omega = \kappa v_\phi = 6.31\times10^8$ rad/s, and $(\omega,\kappa)$ of the wave compares well with the 2D surface wave dispersion curve, shown by the black `X' in Fig. \ref{fig:2D_3D_disp_comp}. Similarly, the surface waves, which were not originally self-excited in the 3D cylindrical beam case with $a = 2.5$ mm, were excited by the introduction of 10\% high energy electrons with $v_e = 1.33\times10^6$ m/s at $t = 5.84 \ \mu s$, as shown by the comparison of the original and artificially excited potential profiles in Fig. \ref{fig:3D_excitation} and a green `X' in Fig. \ref{fig:2D_3D_disp_comp}. Note that the surface waves are only seen in the region of the beam where there are no ESWs that can dominate the local electric potential and phase space, as they would make it hard to identify the small amplitude surface waves. This artificial excitation shows that a small number of high energy electrons can excite a surface wave in either a planar or cylindrical beam. This also confirms our assertion that the TG surface waves found in the $a = 5.0$ and $7.5$ mm 3D cylindrical beam cases are excited by a small number of high energy electrons in the first phase of beam neutralization.     

\section{Conclusions}\label{sec:conclusions}
In this work, we demonstrated the formation of ESWs in the rapid neutralization phase of 2D planar and 3D cylindrical beams by an external electron emission source. Similar to Lan et al.\cite{lan2020neutralization,lan2020neutralization2}, ESWs are formed due to the two streams of electrons\cite{omura1996electron} created by the electron reflection from the sheaths at the axial boundaries (i.e., $z_{\rm min}$ and $z_{\rm max}$) and the virtual cathode at $z = 10$ cm. We demonstrated the tracking of ESWs along the beam and their collisions, which results in their merging together to form slow moving ESWs of larger lengths. Since the slow moving (or nearly stationary) ESWs have a lower probability of colliding with the domain walls or other ESWs, they will dissipate slowly and hence, will slow down the process of beam neutralization. By comparing the individual ESW structures, we identified that the ESWs in the 2D planar beam case corresponded to those of a 1D BGK mode, while the ESWs in the 3D cylindrical beam case did not. Since the electron trajectories in the latter case were found to be highly three dimensional, a BGK analysis with higher dimensionality would be required to analyze 3D ESWs.  The 1D BGK analysis of the 2D beam ESWs found in our simulation led to the conclusion that they were of a larger length than usual because of the non-Maxwellian nature of the beam electrons. This shows that such ESWs are possible with a smaller total electron density perturbation as compared to the case where the beam electrons are Maxwellian.  

Furthermore, we demonstrated the spontaneous excitation of Trivepiece-Gould\cite{trivelpiece1959space} surface waves in a 3D cylindrical beam with beam radius, $a\ge 5.0$ mm, in which the initial high energy electrons were found to be responsible for the excitation of the wave. By deriving the theoretical dispersion relation for a cold 2D planar beam, we showed that the high phase velocity requirement was why surface waves did not appear in our 2D planar beam results. Finally, we demonstrated that by introducing a small fraction of electrons with a velocity higher than the usual Maxwellian of $T_e = 2$~eV, surface waves could be excited even in the 2D planar beam case.

\section*{Acknowledgments}
This project at the University of Illinois Urbana-Champaign is supported by the Department of Energy Fusion Energy Sciences DOE - Award DE-SC00021348. Dr. I. D. Kaganovich is supported by the US DOE under contract DE-AC02-09CH11466 as a part of the Princeton Collaborative Research Facility (\textcolor{black}{PCRF}). We would also like to thank NCSA for providing us with the computational resources on Bluewaters petascale computing facility\cite{Bode2013,Kramer2015}. We would also like to thank XSEDE for providing us with GPU computing resources on San Diego Supercomputing Cluster (SDSC). 

\section*{Data availability statement}
The data that support the findings of this study are available from the corresponding author upon reasonable request.


\appendix
\section*{Appendix}

\section{Numerical convergence of PIC simulations and method of ESW position location}\label{sec:Num_conv}

\begin{figure}[H]
\centering
        \includegraphics[trim = 0.14cm 0.14cm 0.14cm 0.14cm, clip,width = 0.6\textwidth]{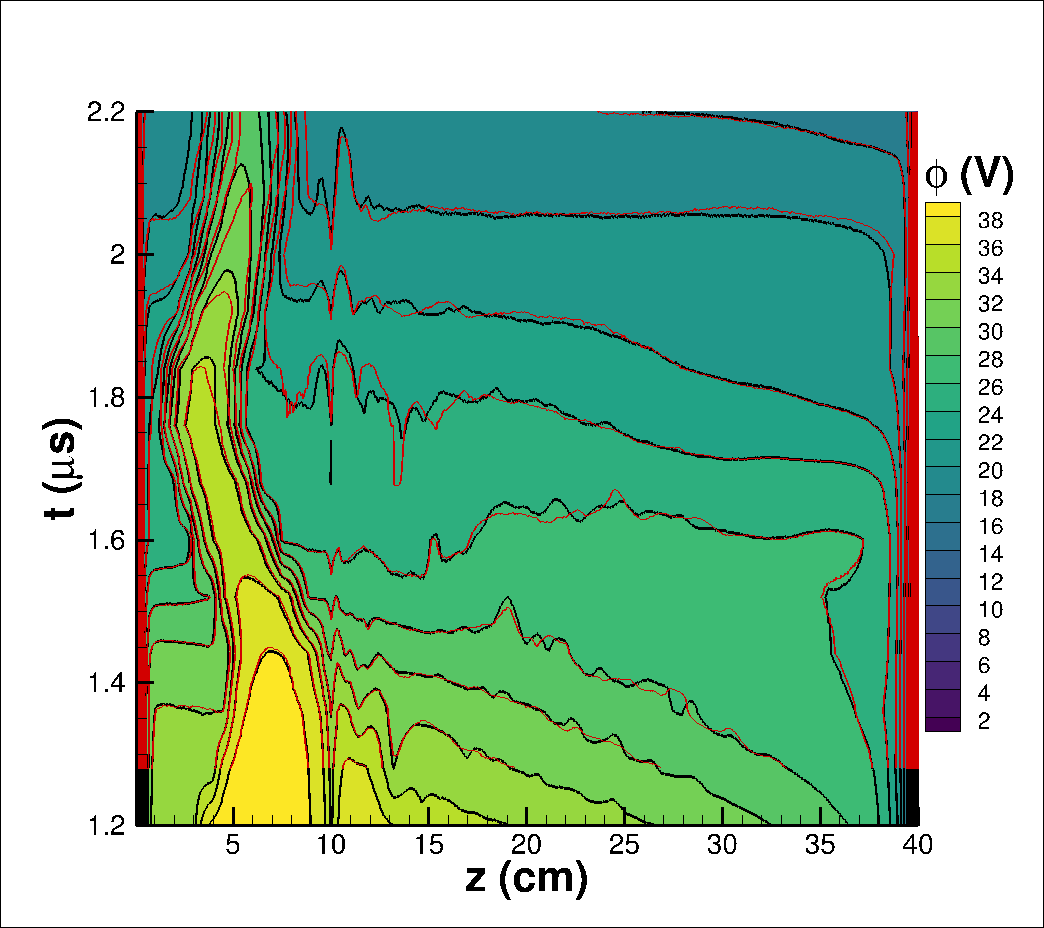}
       
\caption{ Numerical convergence of the cylindrical beam electric potential for $a = 5.0$ mm case. Here, comparison of iso-potential lines with $F_{\rm num} = 60$(black), and $30$(red) is shown.}
\label{fig:Num_conv_ctr}
\end{figure}

To show numerical convergence, we performed a case with $a = 5.0$ mm with twice as many charged particles per cell as in the original case in which $F_{\rm num}$ was 60. We performed this case with 128 GPUs until $t = 2.2 \ \mu s$ to show convergence in the ion only ($t < 1.0 \ \mu s $) and ion-electron ($t > 1.0   \ \mu s$) regime of the beam neutralization. The electric potential at the $z = 20$ cm probe along the beam axis was found to agree within 2\% for cases with $F_{\rm num} = 30$ and 60. Also, the time evolution of the electric potential profile along the beam is shown in Fig. \ref{fig:Num_conv_ctr}, and it can be seen that the iso-potential lines compare well for both cases. 

\begin{figure}[h]
\centering
        \includegraphics[trim = 0.14cm 0.1cm 0.14cm 0.14cm, clip,width = 0.8\textwidth]{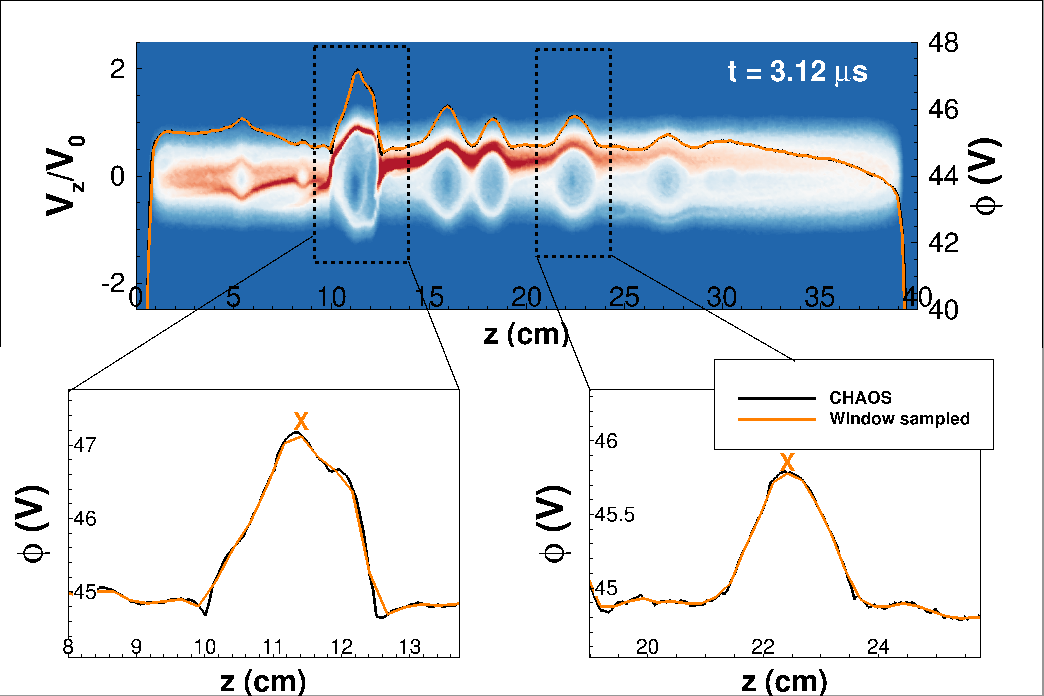}
\caption{ Window averaging of the electric potential along the 2D planar beam axis at $t = 3.12 \ \mu s$. The orange \textbf{X} shows the location of the local maxima of electric potential peak.}
\label{fig:Window_sampling}
\end{figure}

To locate the ESWs along the beam axis, we use the fact that the position of the ESW is accompanied with a local maxima in the electric potential ($\phi$) profile, as shown in Fig. \ref{fig:Window_sampling}. We extract 800 points spanned over the range $z \in (0,40)$ cm along the beam axis.  To smooth the $\phi (z)$, we subdivide 800 points along the beam axis into 160 $z$-location points,
\begin{align}
    z_{\rm window} = \frac{1}{N_w} \sum_{n=1}^{N_w} z_n,
\end{align}
where $z_{\rm window}$ is the average $z$ location of the window of size $N_w$. For all cases in this work, we use $N_w=5$. The five points around those 160 points, $N_w$, are then averaged to assign the electric potential value to those points, i.e., 
\begin{align}
    \phi(z_{\rm window}) = \frac{1}{N_w} \sum_{n=1}^{N_w} \phi(z_n),
\end{align}
This results in a smoother electric potential profile, $\phi (z_{\rm window})$, as shown by orange line in Fig. \ref{fig:Window_sampling}. The profile, $\phi (z_{\rm window})$, is then differentiated twice to satisfy, $d\phi (z_{\rm window})/dz_{\rm window} \approx 0$ and $d^2\phi (z_{\rm window})/dz_{\rm window}^2 < 0$ to obtain local maxima, shown by orange \textbf{X} in Fig. \ref{fig:Window_sampling}. This process is performed for various time profiles to generate the locations of ESWs, shown in Figs.  \ref{fig:2D_soliton_movement_timeevol} and \ref{fig:Phi_z_t_3D-A}.  

\section{Applying 1D BGK theory to obtain $f_{ut}$ from the PIC EVDF data}\label{sec:1DBGK_app}
\begin{figure}[h]
\centering
\subfigure[]{\label{fig:phi_fit_schematic}
        \includegraphics[trim = 0.14cm 0.14cm 0.14cm 0.14cm, clip,width = 0.40\textwidth]{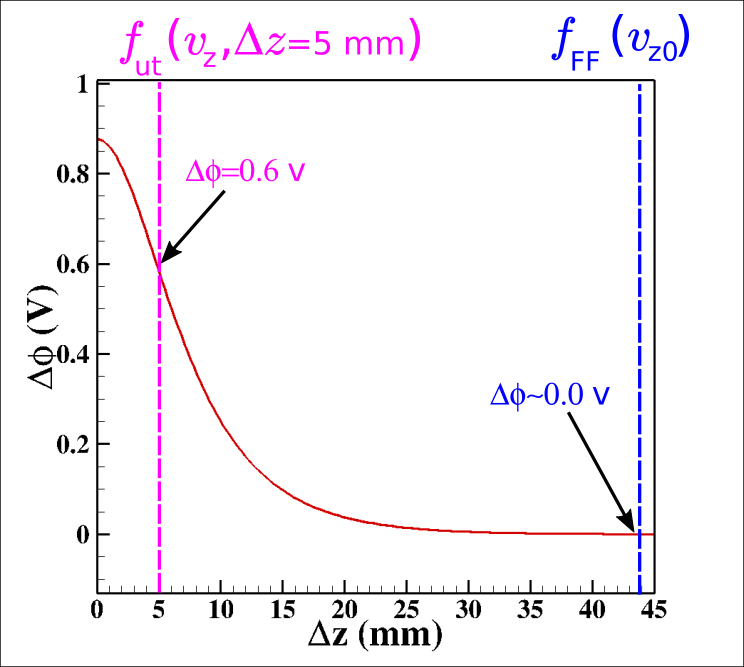}}
\subfigure[]{\label{fig:Vz_z_phase_schematic}
        \includegraphics[trim = 0.14cm 0.14cm 0.14cm 0.14cm, clip,width = 0.40\textwidth]{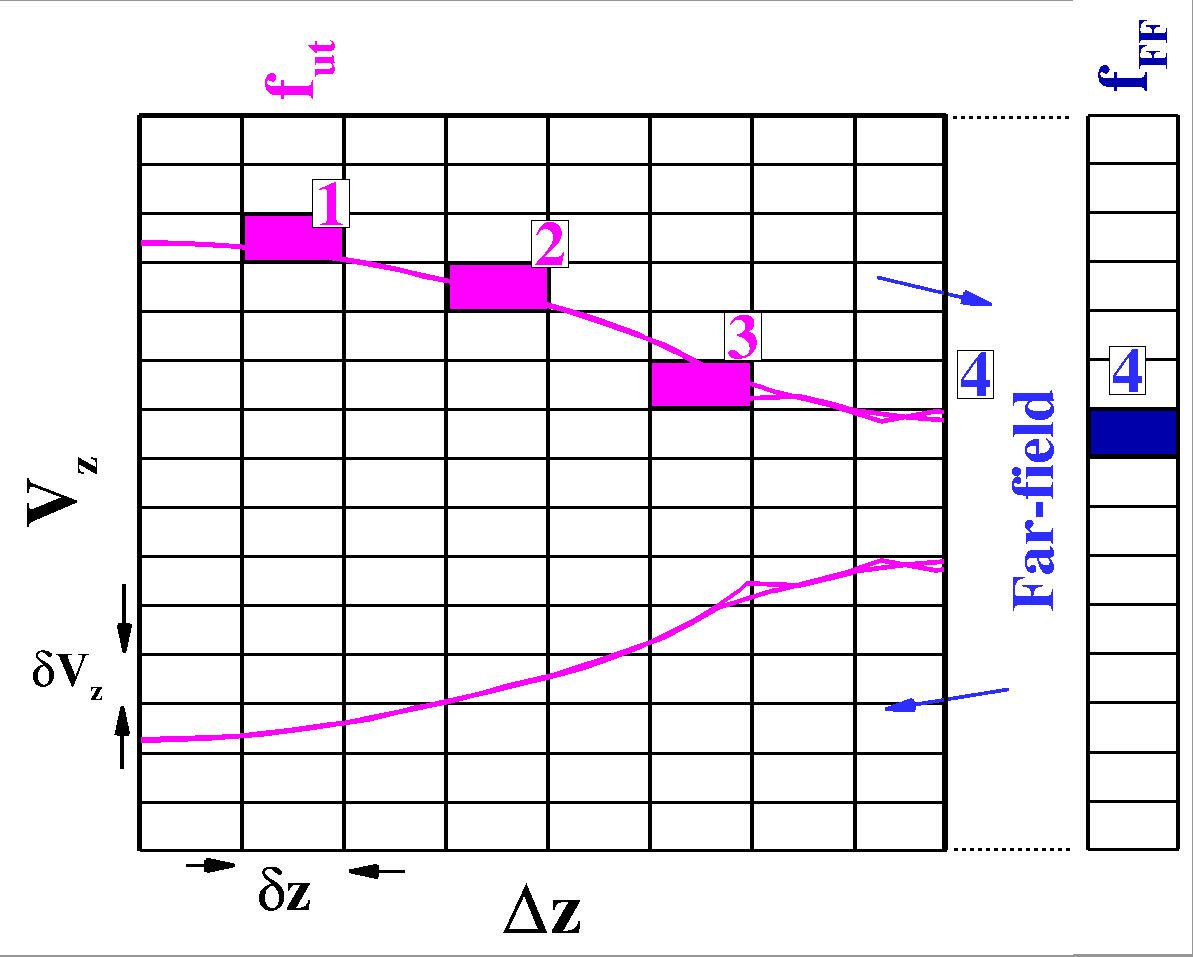}}      
\subfigure[]{\label{fig:ESW_trajs}
        \includegraphics[trim = 0.14cm 0.14cm 0.14cm 0.14cm, clip,width = 0.40\textwidth]{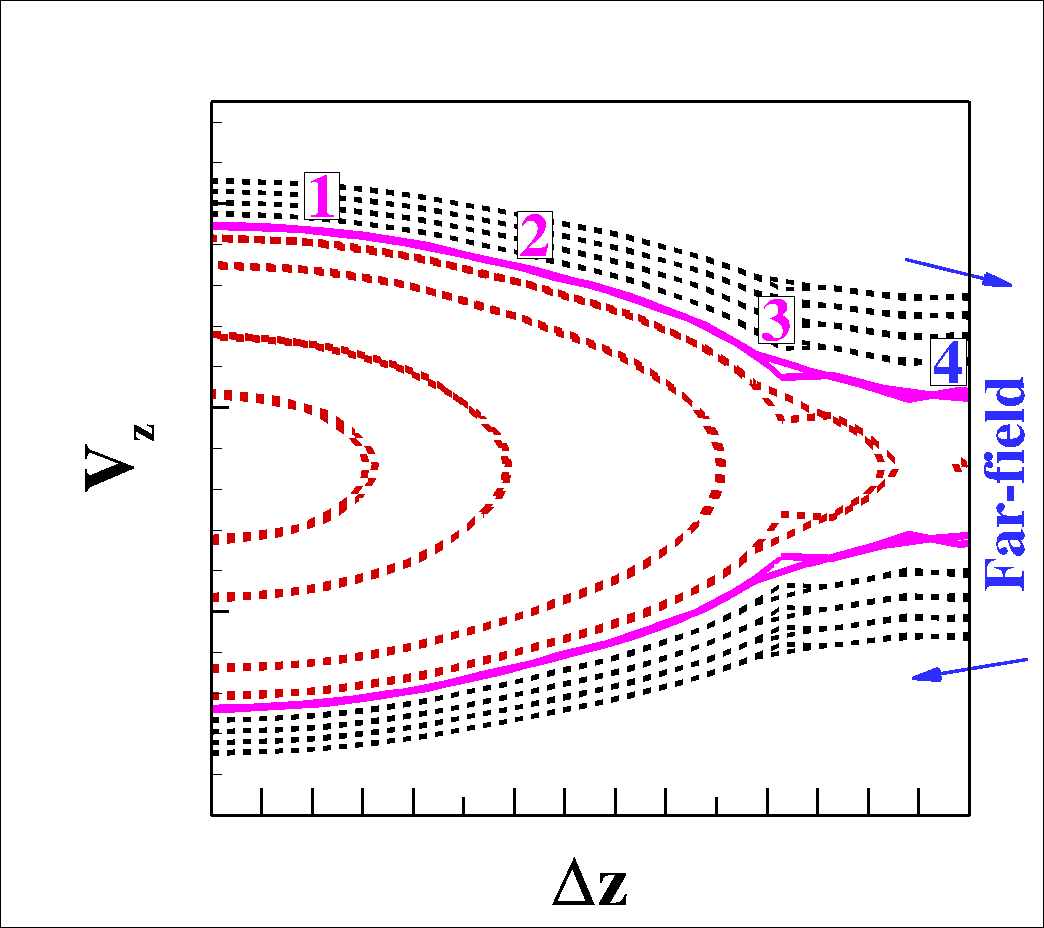}}           
\caption{ (a) Electric potential profile of the ESW. (b) Schematic of discretized phase grid of $(v_z,\Delta z)$ space. In (b), $\delta z = 0.15$ mm and $\delta v_z = 12,000$ m/s. (c) A schematic of trapped (red) and untrapped (black) electron trajectories (i.e. constant total energy lines) for an ESW in ($v_z, \Delta z$) phase space. In (b), since the magenta colored constant $E_{\rm tot}$ line (also shown in (c)) passes through points 1, 2, and 3 in the $(v_z,\Delta z)$ phase space, the number of electrons are identical in the highlighted magenta colored grid-boxes in (b).   }
\label{fig:BGK_theory_fit}
\end{figure}

Here, we discuss how we computed the untrapped electron phase distribution, $f_{ut}$, from the far-field EVDF $f_{\rm FF}$ obtained from the PIC result. The analytical fit profile from Eq. \ref{eq:phi_fit}, shown in Fig. \ref{fig:phi_fit_schematic}, makes it easier for us to perform the step 2 of 1D BGK approach. Note that the far-field in our PIC result occurs near $\Delta z \approx 15$ mm, however, for this theoretical analysis, we assume the far-field to be at $\Delta z = 45$ mm to ensure that the Eq. \ref{eq:phi_fit} fit decreases to nearly 0 V. From this point on, we will consider  Eq.~\ref{eq:phi_fit} and Fig.~\ref{fig:phi_fit_schematic} as the electric potential profile of the ESW.  

Since all electrons outside of the ESW (i.e. $\Delta z\geq 45$ mm in Fig. \ref{fig:phi_fit_schematic}) are untrapped electrons, the untrapped distribution, $f_{ut}(v_z,\Delta z\to\infty) = f_{\rm FF}$. Numerically, we discretize the z-velocity space ($v_z$) from $-3\times10^6$ to $3\times10^6$ m/s into 500 bins ($\delta v_z = 12000$ m/s), where all the electrons in a bin are considered to be of the same velocity. We also discretize the $\Delta z$ space into 300 points between 0 to 45 mm. A schematic of the generated ($v_z,\Delta z$) phase-space grid is shown in Fig. \ref{fig:Vz_z_phase_schematic}. Now, we want to fill each cell of this discretized grid with the value of number of electrons. We first fill the $\Delta z = 45$ mm column (i.e. column designated to far-field) of the phase-space grid with the electron particle velocity data from PIC divided into 500 bins of discrete velocity space to generate the far-field EVDF, $f_{\rm FF}$, that was shown in Fig. \ref{fig:EVDF_fit}. Although the distribution is initially shown in units of number of computational particles, it could easily be changed to the units of number density, $\rm m^{-3}$ by multiplication by the factor $F_{\rm num} / {\rm Vol}$, where $\rm Vol$ is the volume of cell of size $\delta z$.  

The electrons in the rest of the phase-space of ESW are either trapped or untrapped. At steady state of the ESW (i.e. in the reference frame of the moving ESW), the electron trajectories in the phase space look like those shown in Fig. \ref{fig:ESW_trajs}, where along each electron trajectory, the total energy, i.e. kinetic + potential energy, is conserved. This means that the number of electrons are the same along each contour line shown in Fig. \ref{fig:ESW_trajs} and, therefore, if we know the EVDF along any of the $\Delta z$ locations along with the electric potential profile of ESW, we can compute the rest of the phase space distribution. Moreover, since we know the EVDF in the far-field, $f_{\rm FF}$, we can directly compute the untrapped electron phase distribution from that, by following the dashed black lines in Fig. \ref{fig:ESW_trajs} from the far-field EVDF, $f_{\rm FF}$, to any $\Delta z$ location inside the ESW. Because the trapped electrons never leave the ESW and they have no electron trajectories, shown by dashed red lines in Fig. \ref{fig:ESW_trajs}, entering the far-field, we cannot estimate the trapped EVDF from $f_{\rm FF}$ alone.

We will now apply the conservation of total energy to compute the phase-space distribution of the untrapped electron particles based on the known $f_{\rm FF}$. The electrons at the location $\Delta z$ and in the velocity bin $v_z$, have a total energy of $\frac{1}{2}m_ev_z^2 - e\Delta \phi$, and will reach the far-field with a kinetic energy of $\frac{1}{2}m_ev_{z0}^2$ such that,
\begin{align}\label{eq:vz0_pos}
    \frac{1}{2}m_ev_{z0}^2 = \frac{1}{2}m_ev_z^2 - e\Delta \phi, \nonumber \\
    v_{z0} = \text{sgn($v_z$)}\sqrt{v_z^2 - 2\frac{e}{m_e}\Delta \phi}.
\end{align}
An example of such a trajectory is shown in Figs. \ref{fig:Vz_z_phase_schematic} and \ref{fig:ESW_trajs}, where the magenta colored iso-contour line travels from the center ($\Delta z=0$) to the far-field of the ESW. Since all the electrons which had velocity $v_z$ at $\Delta z$ location reach the far-field with the velocity $v_{z0}$, the number of electrons with velocity $v_z$ at $\Delta z$, i.e. $f_{ut} (v_z,\Delta z)$, should be equal to the number of electrons in the far-field distribution velocity bin $v_{z0}$. In other words,
\begin{align}\label{eq:untrapped_EVDF_selfsimilar_appx}
    f_{ut}\left(v_z ,\Delta z\right) = f_{\rm FF}\left( v_{z0}\right).
\end{align}
If we apply this to multiple points along the $\Delta z$ direction, we can fill the phase grid of $f_{ut}$. After computing the EVDFs for all 300 $\Delta z$ locations, the untrapped electron distribution in the phase-space was shown in Fig. \ref{fig:Non-Maxwellian_BGK}.

\section{Dispersion relation for 2D surface waves}\label{sec:2D_surface_app}

\begin{figure}
\centering
  \subfigure[Schematic for 2D slab and 2D surface.]{\label{fig:2D_b_vals}
        \includegraphics[trim = 0.14cm 0.14cm 0.14cm 0.14cm, clip,width = 0.4\textwidth]{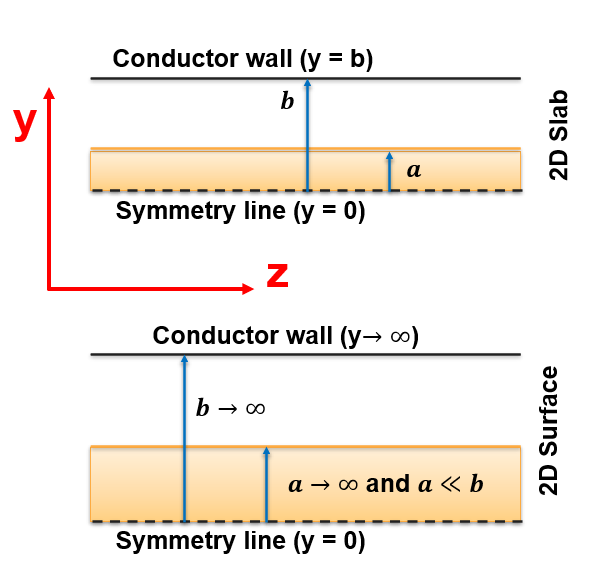}}
    \hspace{0.3mm}     
\subfigure[Comparison of 2D Slab dispersion relation with cylindrical and planar surface waves.]{\label{fig:2D_cyl_planar}
        \includegraphics[trim = 0.14cm 0.14cm 0.14cm 0.14cm, clip,width = 0.48\textwidth]{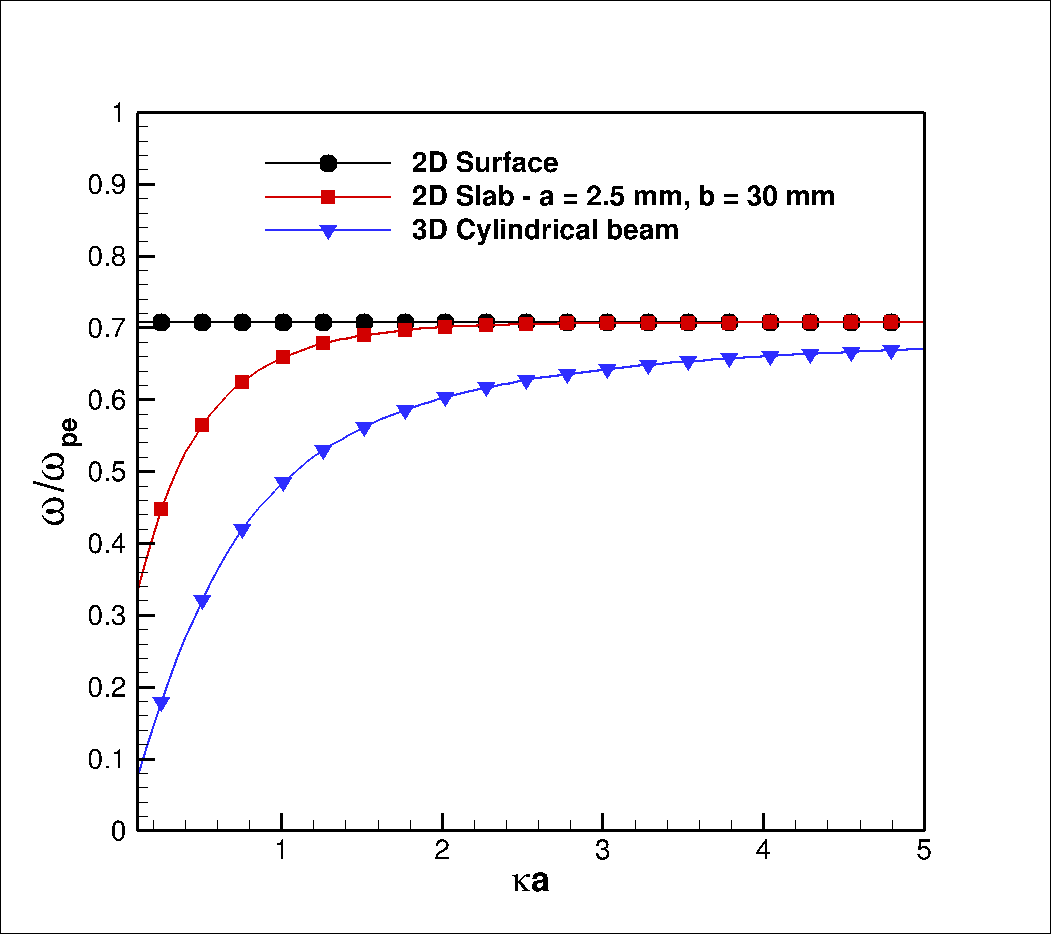}}
       
\caption{ 2D surface wave dispersion relation for plasma beam with different plasma widths, and channel widths.}
\label{fig:2D-dispersions}
\end{figure}

Equation (4.8.6) of Krall et all.\cite{krall1973principles} is the general wave equation, 
\begin{align}\label{eq:wave_equation}
    \nabla(\nabla . \vec{E_1}) - \nabla^2\vec{E_1} = \frac{\omega^2}{c^2} \epsilon.\vec{E_1},
\end{align}
used to derive a differential equation for determining the perturbed electric field, $\vec{E_1}$, in the plasma medium of dielectric tensor, $\epsilon$.  However, Trivelpiece et al.\cite{trivelpiece1959space} used a simpler approach in which, $\vec{E_1} = -\nabla\phi_1$, where $\phi_1$ is the perturbed scalar electric potential. Then, Poisson's equation in a dielectric medium is given by, 
\begin{align}\label{eq:possion_perterb}
\nabla . \epsilon .\vec{E_1} = 0, \\
\nabla . (\epsilon . (\nabla\phi_1)) & = 0 \nonumber,
\end{align}
where the dielectric tensor, $\epsilon$, in 2D Cartesian coordinates is given by, 
\begin{align}\label{eq:dielec_tensor}
    \epsilon &=\begin{bmatrix}\left(1 - \frac{\omega_{pe}^2}{\omega^2}\right) & 0 \\ 0 & \left(1 - \frac{\omega_{pe}^2}{\omega^2}\right) \end{bmatrix},
\end{align}
and $\omega_{pe} = \sqrt{\frac{e^2n_0}{m\epsilon_0}}$, is the electron plasma frequency. After inserting Eq. \ref{eq:dielec_tensor} into Eq. \ref{eq:possion_perterb}, we obtain,
\begin{align}\label{eq:poisson_pert_insidebeam}
    \left[1 - \frac{\omega_{pe}^2}{\omega^2} \right]\frac{\partial^2 \phi_1}{\partial y^2} + \left[1 - \frac{\omega_{pe}^2}{\omega^2} \right]\frac{\partial^2 \phi_1}{\partial z^2} & = 0, \text{for $-a<y<a$} 
\end{align}
inside the plasma beam. Outside the plasma beam,  
\begin{align}\label{eq:poisson_pert_outsidebeam}
    \frac{\partial^2 \phi_1}{\partial y^2} + \frac{\partial^2 \phi_1}{\partial z^2} & = 0, \text{ for $|y|>a$}
\end{align}
because the dielectric tensor is the identity matrix in a vacuum. Both Eqs. \ref{eq:poisson_pert_insidebeam} and \ref{eq:poisson_pert_outsidebeam} have to be solved to obtain the perturbation field $\phi_1$ and the corresponding dispersion relation.
These solutions should satisfy the following boundary conditions, 
\begin{align}
    \phi_1|_{y\to a^-} & = \phi_1|_{y\to a^+}, \label{eq:phi_cont}\\
    \left[\frac{\partial \phi_1}{\partial y}\right]_{y = 0} &= 0, \label{eq:phi_symm}\\
    \phi_1|_{y = b} & = 0,\label{eq:phi_cond} \\
    \left[1 - \frac{\omega_{pe}^2}{\omega^2} \right]\left[\frac{\partial \phi_1}{\partial y}\right]_{y\to a^-} &= \left[\frac{\partial \phi_1}{\partial y}\right]_{y\to a^+}.\label{eq:D_cont}
\end{align}

The analytical solution of Eq. \ref{eq:poisson_pert_insidebeam}, obtained by separation of variables, is of the form, 
\begin{align}\label{eq:phi_inside}
    \phi_1 &= \left[A_y \exp{(\kappa y)} + B_y \exp{(-\kappa y)}\right]\exp{\left[i(\kappa z - \omega t)\right]},
\end{align}
where $A_y$ and $B_y$ are the constants of integration, $\kappa$ is the wave-number in the $z$ direction, and $\omega$ is the frequency for the wave along the axial (i.e. $+z$) direction. Similarly, the analytical solution of Eq. \ref{eq:poisson_pert_outsidebeam} is,   
\begin{align}\label{eq:phi_outside}
    \phi_1 &= \left[C_y \exp{(\kappa' y)} + D_y \exp{(-\kappa' y)}\right]\exp{\left[i(\kappa' z - \omega t)\right]},
\end{align}
where $C_y$ and $D_y$ are also the constants of integration, and $\kappa'$ is the wave-number along the axial beam direction. Since the surface wave produced has to be continuous at the plasma-vacuum interface, $\kappa' = \kappa$. Finally, when we substitute the boundary conditions in Eqs. \ref{eq:phi_cont}-\ref{eq:D_cont} into Eqs.~\ref{eq:phi_inside} and \ref{eq:phi_outside}, we obtain, 
\begin{align}\label{eq:2D-disp}
    \frac{\omega}{\omega_{pe}} = \frac{1}{\left[1+ \frac{\tanh{[(b-a)\kappa]}}{\tanh{a\kappa}}\right]^{1/2}},
\end{align}
which is the dispersion relation for the surface waves in a 2D planar beam of half-width $a$ inside a channel of half-width $b$. Equation \ref{eq:2D-disp} converges to $\omega/\omega_{pe} = 1/\sqrt{2}$ for large values of $\kappa$, as shown in Fig. \ref{fig:2D_diap_a_b}, which is the dispersion relation for an infinite cold-plasma and vacuum interface surface wave. The value of $\omega/\omega_{pe}$ varies from 0.3 to 0.707 for long-wavelength waves (i.e. low values of $\kappa$) for $a = 2.5$, $5.0$, and $7.5$ mm, and $b = 30$ mm, as shown in Fig. \ref{fig:2D_diap_a_b}. Also, the 2D dispersion relation changes its shape when the beam width, $2a$ is nearly equal to the channel width $2b$, in which the value of $\omega/\omega_{pe}$ decreases to $1/\sqrt{2}$, shown in Fig. \ref{fig:2D_diap_a_b}, with increase in wavenumber, $\kappa$.

\bibliography{sample}
\bibliographystyle{aiaa}
\end{document}